\documentclass[fleqn,usenatbib,useAMS]{mnras}

\usepackage{graphicx}	% Including figure files
\usepackage{amsmath}	% Advanced maths commands
\usepackage{amssymb}	% Extra maths symbols
\usepackage{multicol}        % Multi-column entries in tables
\usepackage{bm}		% Bold maths symbols, including upright Greek
\usepackage{pdflscape}	% Landscape pages
\usepackage{media9} % for playing mp4 

\newcommand{\kms}{\,km\,s$^{-1}$} % kilometres per second
\newcommand{\angstrom}{\AA} % kilometres per second

\usepackage[T1]{fontenc}
\usepackage{ae,aecompl}

\usepackage{newtxtext,newtxmath}
\title[Bayesian AGN Decomposition Analysis for SDSS Spectra]{Bayesian AGN Decomposition Analysis for SDSS Spectra: A Correlation Analysis of [\ion{O}{iii}]$\lambda5007$ Outflow Kinematics with AGN and Host Galaxy Properties}

\author[R. O. Sexton]{Remington O. Sexton$^{1}$ \thanks{Contact e-mail: \href{mailto:remington.sexton@email.ucr.edu}{remington.sexton@email.ucr.edu}}, William Matzko$^{2}$, Nicholas Darden$^{1}$, Gabriela Canalizo$^{1}$, \newauthor Varoujan Gorjian$^{3}$
\\
$^{1}$Department of Physics and Astronomy, University of California, Riverside, 900 University Avenue, Riverside, CA 92521, USA\\
$^{2}$George Mason University, Department of Physics and Astronomy, MS3F3, 4400 University Drive, Fairfax, VA 22030, USA\\
$^{3}$Jet Propulsion Laboratory, M/S 169-327, 4800 Oak Grove Drive, Pasadena, CA 91109, USA
}
\date{}

\begin{document}
\label{firstpage}
\pagerange{\pageref{firstpage}--\pageref{lastpage}}
\maketitle

\begin{abstract}
We present Bayesian AGN Decomposition Analysis for SDSS Spectra (BADASS), an open source spectral analysis code designed for automatic detailed deconvolution of AGN and host galaxy spectra, implemented in Python, and designed for the next generation of large scale surveys.  BADASS simultaneously fits all spectral components, including power-law continuum, stellar line-of-sight velocity distribution, \ion{Fe}{ii} emission, as well as forbidden (narrow), permitted (broad), and outflow emission line features, all performed using Markov Chain Monte Carlo to obtain robust uncertainties and autocorrelation analysis to assess parameter convergence. BADASS utilizes multiprocessing for batch fitting large samples of spectra while efficiently managing memory and computation resources and is currently being used in a cluster environment to fit thousands of SDSS spectra. \\
\indent We use BADASS to perform a correlation analysis of 63 SDSS type 1 AGNs with evidence of strong non-gravitational outflow kinematics in the [\ion{O}{iii}]$\lambda5007$ emission feature.  We confirm findings from previous studies that show the core of the [\ion{O}{iii}] profile is a suitable surrogate for stellar velocity dispersion $\sigma_*$, however there is evidence that the core experiences broadening that scales with outflow velocity.  We find sufficient evidence that $\sigma_*$, [\ion{O}{iii}] core dispersion, and the non-gravitational outflow dispersion of the [\ion{O}{iii}] profile form a plane whose fit results in a scatter of $\sim0.1$ dex.  Finally, we discuss the implications, caveats, and recommendations when using the [\ion{O}{iii}] dispersion as a surrogate for $\sigma_*$ for the $M_{\rm{BH}}-\sigma_*$ relation.

\end{abstract}

\begin{keywords}
methods: data analysis, galaxies: active, quasars: absorption lines, emission lines
\end{keywords}

%%%%%%%%%%%%%%%%%%%%%%%%%%%%%%%%%%%%%%%%%%%%%%%%%%

%%%%%%%%%%%%%%%%% BODY OF PAPER %%%%%%%%%%%%%%%%%%

% The MNRAS class isn't designed to include a table of contents, but for this document one is useful.
% I therefore have to do some kludging to make it work without masses of blank space.
% \begingroup
% \let\clearpage\relax
% \tableofcontents
% (This table of contents is for easy navigation purposes. It will be removed before submission.)
% \endgroup
% \newpage

\section{Introduction}

Data analysis codes, recipes, and software distributions for spectroscopic data analysis have become commonplace in the astronomy and astrophysics community, especially in the advent of large all-sky surveys, such as the Sloan Digital Sky Survey \citep[SDSS;][]{SDSS} and the highly-anticipated Large Synoptic Survey Telescope \citep[LSST;][]{LSST}.  Despite their widespread use, many of the computational methods used to perform these analyses are either (1) not shared by authors for various reasons, or (2) not open source and cannot be accessed without the purchasing of proprietary software.  In addition to this, many data analysis pipelines designed for large-scale surveys are written with the intent of fitting as many objects as possible in the shortest amount of time, while other analysis recipes may be suited for more-detailed analyses.  Finally, many software packages are suited for fitting for specific types of objects, usually either galaxies or active galactic nuclei (AGNs), with no general means of fitting for both or other types of objects.  As astronomy advances through the 21st century, sacrificing quality for speed will no longer be necessary given the increasingly widespread use and availability of supercomputing resources in astronomy.  Likewise, software designed for fitting specific astronomical objects will yield to more general fitting algorithms which can fit a diverse set of objects autonomously and in great detail.\\
\indent Some notable existing software packages and codes have attempted to address the aforementioned issues.  The Gas AND Absorption Line Fitting (GANDALF; \citet{Sarzi2006}) code was one of the first large-scale algorithms to fully decompose gas emission from stellar absorption features, using penalized pixel-fitting (pPXF; \citet{ppxf1}; \citet{ppxf2}) to measure the stellar line-of-sight velocity distribution (LOSVD) with stellar templates.  In the context of AGN studies, GANDALF was ill-suited for the complexities of fitting type 1 AGNs, which contain additional features such as broad lines, \ion{Fe}{ii} emission, power-law continuum, and possible ``blue-wing'' components  indicative of outflowing narrow-line gas.  The Quasar Spectral Fitting package (QSFit; \citet{qsfit}) allowed for fitting of type 1 AGNs with a variety of optional features, making it ideal for large-scale surveys of many thousands of objects.  More recently, the release of PyQSOFit \citep{PyQSOFit} includes many similar features of QSFit with the added functionality of Python.  However, since QSFit and PyQSOFit use a library of galaxy templates to model the host galaxy component instead of attempting to model the LOSVD using stellar templates, AGNs with a strong stellar continuum component, such as type 2 AGNs, suffer from poor continuum modelling.  Furthermore, while both GANDALF and QSFit are technically open source, they are implemented using proprietary software and language (namely IDL).  While certain licensed software may have once been prevalent in the astronomical community, there is a growing push toward open source software that can be easily shared, modified, and used among the research community.  Among these open source languages is Python, which is one of the fastest growing programming languages for data analysis today, and its widespread use makes it ideal for research in the astronomical community.\\
\indent We address the limitations of current spectral fitting codes with a comprehensive fitting package implemented in Python and utilizing a Markov-Chain Monte Carlo (MCMC) fitting approach for accurate estimation of parameters and uncertainties, which we call \emph{Bayesian AGN Decomposition Analysis for SDSS Spectra} (BADASS).  In its current version, BADASS is written for the SDSS spectra data model in the optical (specifically 3460 \angstrom\; to 9463 \angstrom, based on choice of stellar template library), however, because it is written in Python and is open source\footnotemark, it can be easily modified to accommodate other instruments, wavelength ranges, stellar libraries, templates, and has already been used successfully to perform decomposition on 22 type 1 AGNs observed with the Keck-I LRIS instrument \citep{Sexton2019}.  \\
\footnotetext{\url{https://github.com/remingtonsexton/BADASS3}}
\indent The BADASS software attempts to address some notable and relevant problems with spectral fitting software available today.  Because BADASS was designed for detailed decomposition of type 1 AGNs, which contain various components such as forbidden ``narrow'' (typical FWHM $<500$ \kms) and permitted ``broad'' (typical FWHM $>500$ \kms) emission lines, broad and narrow \ion{Fe}{ii} emission, AGN power-law continuum, ``blue-wing'' outflow components, and the host galaxy stellar continuum, these components can be optionally turned on or off to fit less-complex objects such as type 2 AGNs or non-AGN host galaxies altogether, with all of these options easily configured through the Jupyter Notebook \citep{Jupyter} interface.  As a result, BADASS can be deployed for fitting a diverse range of astronomical objects and customized to the user's needs.  Additionally, the choice of Python as the programming language of BADASS follows suit with a number of other software packages, such as Astropy \citep{astropy}, which aim to replace antiquated software such as IRAF \citep{Valdes1984} or proprietary languages such as IDL, for astronomers now entering the field and/or adopting the Python programming language for their analyses.  If anything, the open source nature of the BADASS software will serve as a template for developing various implementations of the software for individual specific needs.\\
\indent To our knowledge, the BADASS algorithm is the first of its kind to address a number of issues specific to the fitting of AGN spectra that other algorithms have yet to implement.  First, BADASS was initially designed to fit all spectral components simultaneously, as opposed to masking regions of spectrum and fitting components separately.  This is specifically advantageous for the decomposition of the stellar continuum and \ion{Fe}{ii} emission from other components for studies of AGN and host galaxy relations such as the the $M_{\rm{BH}}-\sigma_*$ relation.  As noted in \citet{Sexton2019}, stellar kinematics remain the single-most difficult quantity to measure in type 1 AGN, and obtaining reliable values and uncertainties for stellar quantities is a non-trivial effort that includes a number caveats and systematics which can be difficult to account for \citep{Greene2006b}.  Simultaneous fitting with \ion{Fe}{ii} emission templates also allows for detailed study of \ion{Fe}{ii} emission properties of type 1 AGN while taking into account the underlying stellar continuum.  Finally, BADASS is the first software of its kind to use specific criteria for the automated detection and decomposition of outflow components in forbidden emission lines, which have recently become a topic of much study in the context of AGN and host galaxy evolution (see Section \ref{sec:motivation} and references therein).\\
\indent The Bayesian MCMC approach used by BADASS for fitting spectral parameters is unique in that it provides an easily-extensible framework for the user to modify the fitting model, free parameters, and convergence criteria.  Many fitting software packages typically utilize a simpler least-squares minimization approach, however, it is recommended (almost universally) to perform Monte Carlo resampling of the data and re-fitting (also known as ``bootstrapping'') to ensure accurate estimation of uncertainties.  While the least-squares approach is typically faster, an MCMC approach allows the user to estimate robust uncertainties, visualize possible degeneracies, and assess how well individual parameters are constrained or if they have properly converged on a solution.  While fitting algorithms that utilize random-sampling techniques admittedly suffer from slower runtimes, modern personal computers capable of multi-processing to decrease runtimes are becoming commonplace.  Since BADASS utilizes the affine invariant MCMC sampler \textit{emcee} \citep{emcee}, multi-processing is also an available option for fitting large samples of objects.  The use of powerful Bayesian and computational techniques, open source framework, and diverse fitting options together make help achieve the ultimate goal of BADASS, which is to provide the most detailed and versatile fitting software for optical spectra in future sky surveys.  \\
\indent We describe the BADASS model construction, fitting procedure, and autocorrelation analysis used to assess parameter convergence in Section \ref{sec:BADASS}.  In Section \ref{sec:Application} we discuss the significant correlations between stellar velocity dispersion $\sigma_*$, the decomposed [\ion{O}{iii}] core and outflow dispersions of the [\ion{O}{iii}]$\lambda5007$ emission line found using BADASS. \\
\indent Throughout this work, we assume a standard cosmology of $\Omega_m=0.27$, $\Omega_\Lambda=0.73$, and $H_0=71$ km s$^{-1}$ Mpc$^{-1}$.

\section{The BADASS Algorithm}\label{sec:BADASS}

In the following subsections, we discuss the BADASS model construction, spectral components, fitting procedure, and autocorrelation analysis used to assess parameter convergence.  We also discuss the results of benchmarking tests for the recovery of stellar velocity dispersion using BADASS.

\subsection{Model Construction}

BADASS constructs a model for each spectrum under the assumption that all AGN components (i.e., the AGN power-law continuum, emission lines, possible outflows, and \ion{Fe}{ii} emission) reside atop a  host galaxy component whose stellar contribution we wish to measure.  A general overview of the model construction is as follows.\\
\indent First, each spectral component is initialized using reasonable assumptions from the data.  For example, the stellar continuum and power-law continuum are initialized at amplitudes that are each half of the total galaxy continuum level, which is estimated using the median flux of the fitting region.  As another example, emission line amplitudes are initialized at the maximum flux value within fixed wavelength regions centered at the expected rest frame locations of the emission line.  These initial parameter values need not be exact, as BADASS will iteratively improve on their estimated values with each fitting iteration.\\
\indent Models of each spectral component are then sequentially subtracted from the original data, and any remaining continuum is assumed to be the stellar continuum contribution, which is then fit with a predefined host galaxy template or empirical stellar templates to estimate the LOSVD.  Once all parameters have been estimated and all model components have been constructed, their sum-total is used to assess the quality of the fit to the original data.  This process is repeated for each iteration of the algorithm until a best-fit is achieved.  \\
\indent We describe each of the spectral components used for constructing the model below.

\subsubsection{AGN Power-Law Continuum}

\begin{figure}
 \includegraphics[width=\columnwidth]{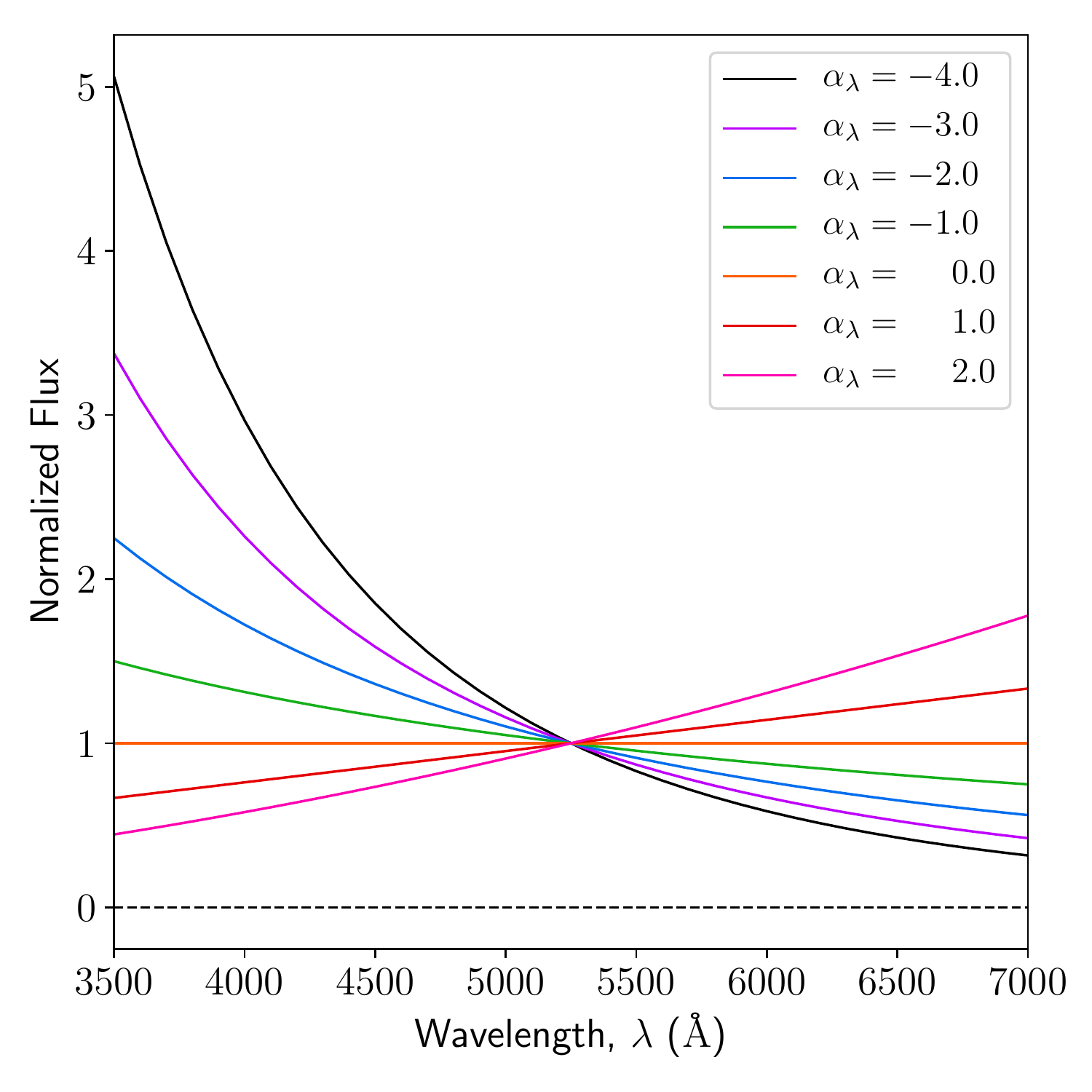}
 \caption{The AGN simple power-law model adopted from \citet{qsfit}.  Colors represent different values of $\alpha_\lambda$ in the range [-4,2] for the wavelength range [3500,7000].  The reference wavelength for this wavelength range, $\lambda_b=5250$ \angstrom, is the locus of different models for $\alpha_\lambda$ and is held fixed by default to be the center of the fitting range.}
 \label{fig:BADASS_agn_cont_model}
\end{figure}

In the simplest construction, the non-stellar thermal continuum in type 1 AGNs can be modeled as the sum of different temperature blackbodies at various radii within the AGN accretion disk \citep{Malkan1983}.  This manifests itself in the UV and optical as a ``big blue bump'', which flattens out at longer wavelengths towards the near-IR, resembling a power-law continuum.  We adopt the QSFit simple power-law implementation from \citet{qsfit} given by 

\begin{equation}\label{eq:agn_cont}
    p(\lambda) = A\left(\frac{\lambda}{\lambda_b}\right)^{\alpha_\lambda},
\end{equation}

where $A$ is the power-law amplitude, $\alpha_\lambda$ is the power-law index (or spectral slope), and $\lambda_b$ which is a reference wavelength chosen to be the central wavelength value of the fitting region and determines the break in the power-law model.  The power-law amplitude $A$ and slope index $\alpha_\lambda$ are free parameters throughout the fitting process.  The flat priors we set on these parameters dictate that $A$ must be non-negative and no greater than the maximum flux density value of the data, and $\alpha_\lambda$ can vary in the range [-4,2].  As in QSFit, the reference break wavelength $\lambda_b$ is fixed by default to be the center wavelength value of the fitting region (i.e., ($\lambda_{\rm{max}}$-$\lambda_{\rm{min}}$)/2), since the power-law slope is poorly constrained at optical wavelengths, however, this constraint can be relaxed if there is sufficient wavelength coverage in the near-UV.  We show different values of the power-law slope in Figure \ref{fig:BADASS_agn_cont_model}.\\
\indent We find that the simple power-law model adequately describes the AGN continuum in the optical, especially if the object fitting region is limited to rest-frame $\lambda_{\rm{rest}}>3460$ \angstrom, which is the lower limit of the wavelength range of the Indo-US Stellar Library \citep{IndoUS} used for fitting the stellar LOSVD.  To better model the true shape of the power-law continuum, a large fitting region at $\lambda_{\rm{rest}}<3500$ \angstrom\; is necessary to better constrain the power-law index $\alpha_\lambda$.  For fitting regions $\lambda_{\rm{rest}}>3500$ \angstrom, the AGN continuum can become highly degenerate with the host galaxy stellar continuum, especially if the shape of the power-law continuum is relatively flat.  We nevertheless include a power-law continuum in our fitting model because its inclusion does not affect the overall fitting process.

\subsubsection{Broad and Narrow Emission Lines}

\begin{figure}
 \includegraphics[width=\columnwidth]{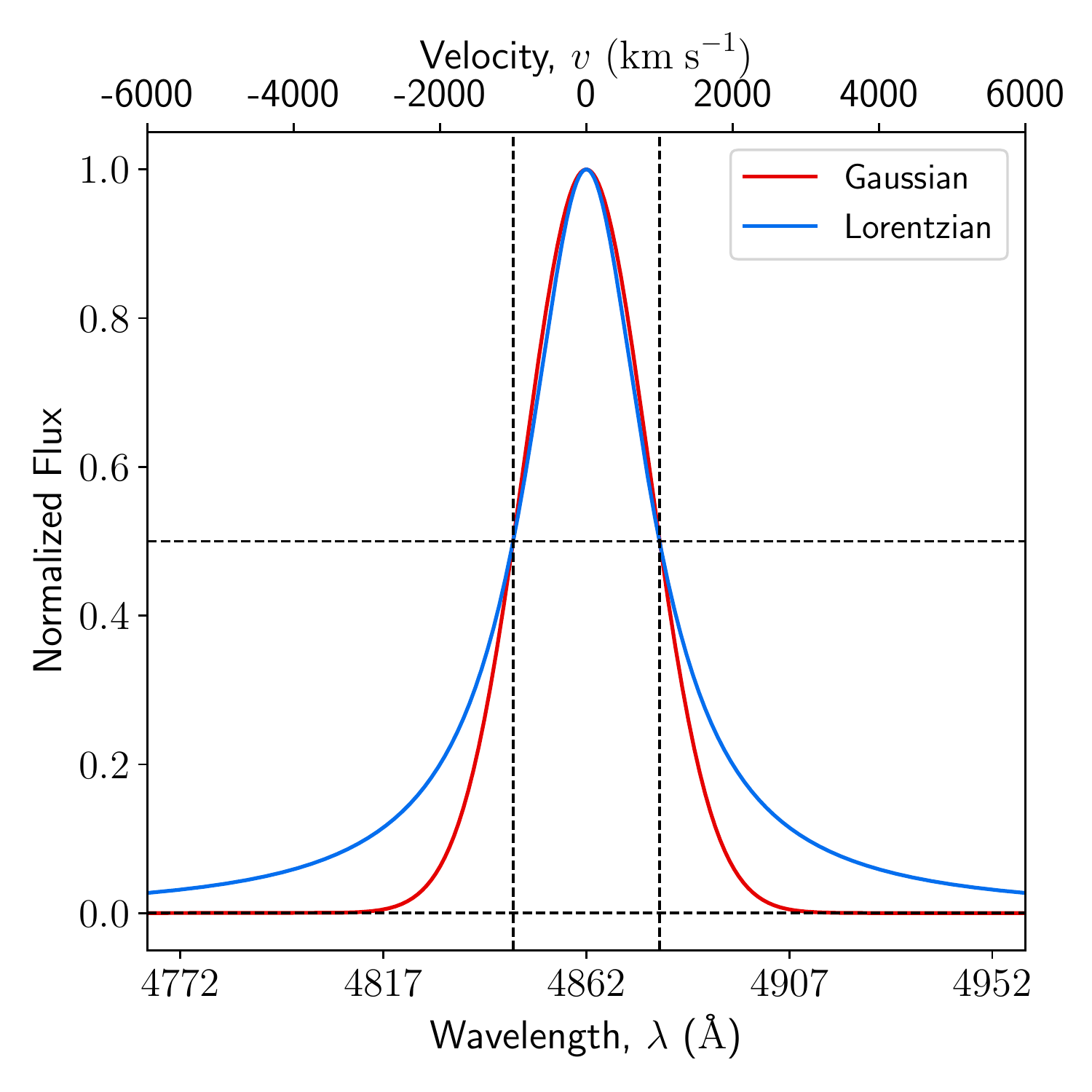}
 \caption{A comparison of the Gaussian and Lorentzian emission line models centered on the location rest frame H$\beta$ with a FWHM of 2000 km s$^{-1}$.  Dashed lines indicate the location and width of the FWHM. }
 \label{fig:BADASS_emline_model}
\end{figure}

All broad and narrow emission line features are, by default, modeled as a simple Gaussian function given by 
\begin{equation}\label{eq:gaussian}
    g(\lambda)= A\exp\left[-\frac{1}{2}\left(\frac{(\lambda-v)^2}{\sigma^2}\right)\right],
\end{equation}
where $A$ is the line amplitude, $\sigma$ is the Gaussian dispersion, and $v$ is the velocity offset of the Gaussian profile from the rest frame wavelength of the line.  Some types of objects, such as NLS1s, exhibit broad lines with extended wings \citep{Moran1996,Leighly1999,Veron-Cetty2001,Berton2020}, for which BADASS can optionally model the emission line with a Lorentzian function given by 
\begin{equation}\label{eq:lorentzian}
    \ell(\lambda) = \frac{A\gamma^2}{\gamma^2+(\lambda-v)^2},
\end{equation}
where $\gamma=\textrm{FWHM}/2$.  A comparison of the two emission line models is shown in Figure \ref{fig:BADASS_emline_model}.\\

\indent Because SDSS spectra are logarithmically-rebinned, each pixel represents a constant velocity scale measured in \kms.  This convenience allows us to initialize width and velocity parameters in units of \kms and fitting performed in units of pixels without the conversion from \angstrom, which is wavelength dependent.  All narrow and broad line widths are corrected for the  wavelength-dependent instrumental dispersion of the SDSS spectrograph during the fitting process so that final reported widths do not have to be corrected by the user.  However, we still place a minimum value for all measured emission line widths to be the velocity scale of our spectra (in units of km s$^{-1}$ pixel$^{-1}$), to ensure that emission line widths are at least greater than a single pixel in width to avoid the fitting the noise spikes. \\
\indent Given the modest resolution ($\sigma\sim 69$ \kms) and signal-to-noise (S/N) of SDSS spectra , we find that in most cases a simple Gaussian function is sufficient to model the full shape emission lines.  Other fitting algorithms attempt to model emission lines in higher detail, using Gauss-Hermite polynomials or additional higher-order moments, in order to account of line asymmetries, which are especially obvious in broad line emission.  \citet{Sexton2019} however showed that some line asymmetries can be attributed to strong absorption near Balmer features, and that emission line asymmetries are generally resolved with simple Gaussian models as long as the underlying stellar population is modeled.  Narrow lines that exhibit a ``blue wing'' outflow component, as typically seen in the [\ion{O}{iii}] emission lines, can be fit as an additional Gaussian component and is a standard feature of BADASS (see Section \ref{sec:outflow_tests}).  Higher resolution spectra of nearby objects with strong emission lines can exhibit further complex non-Gaussian profiles even after outflow components are accounted for.  These non-Gaussian profiles are best modeled iteratively using multiple Gaussian components until a optimal fit is achieved, and adding additional components can be easily achieved by modification of the BADASS code. \\
\indent Many line fitting algorithms tie the widths of narrow lines to be the same across the entire fitting region, however, we leave this as an optional constraint in BADASS.  The advantage of tying the widths only decreases the number of free parameters, while the disadvantage of tying all narrow line widths to each other can lead to a worse fit.  Instead, BADASS ties widths of lines that are nearest to each other in groups.  For example, the narrow [\ion{N}{ii}]/H$\alpha$/[\ion{S}{ii}] line group's widths are tied, and the narrow H$\beta$/[\ion{O}{iii}] line group's widths are tied and fit separately. The H$\alpha$/[\ion{N}{ii}]/[\ion{S}{ii}] line widths can be biased due to line blending, and/or the presence of outflows and broad lines, and since these lines tend to have larger fluxes than most other lines, they carry greater statistical weight in determining widths if they are tied to other lines in the spectrum.  A similar argument can be made for the H$\beta$/[\ion{O}{iii}] line group.  Ideally, one would model each individual line separate from the rest, however, tying widths is still required for narrow forbidden lines obscured by broad permitted lines.  Thus tying widths of groups of lines both reduces fitting bias within each group while also reducing the number of free fitting parameters. 

\subsubsection{\ion{Fe}{ii} Templates}

\begin{figure*}
 \includegraphics[width=\textwidth]{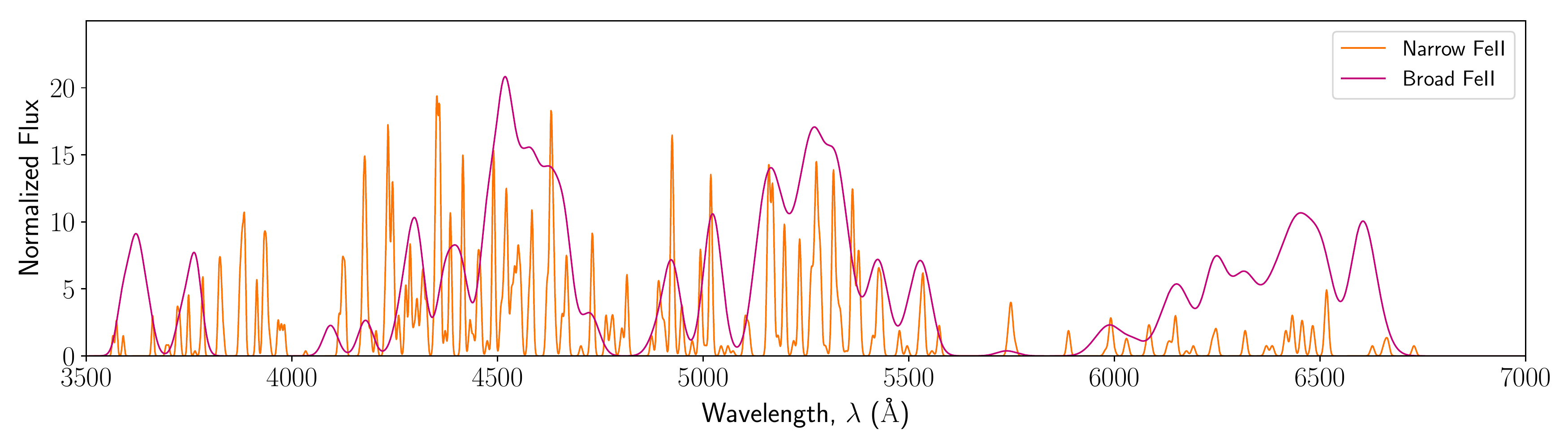}
 \caption{Broad and narrow \ion{Fe}{ii} templates from \citet{Veron-Cetty2004}, which (by default) have fixed zero velocity offset and fixed width in BADASS.  Broad \ion{Fe}{ii} is initialized with a FWHM of 3000 \kms, and narrow \ion{Fe}{ii} with a FWHM of 500 \kms, and held constant throughout the fitting process, following the implementation of QSFit \citep{qsfit}. }
 \label{fig:BADASS_VC04_feii_templates}
\end{figure*}

To account for \ion{Fe}{ii} emission typically present in the spectra type 1 AGNs, BADASS uses the broad and narrow \ion{Fe}{ii} templates from \citet{Veron-Cetty2004}, which are optimal for subtraction since they include emission features that are commonly found in many Seyfert 1 galaxies, as opposed to a single template based solely on I Zw 1 \citep{Barth2013}.  Figure \ref{fig:BADASS_VC04_feii_templates} shows the narrow the broad \ion{Fe}{2} emission features from the \citet{Veron-Cetty2004} template.\\
\indent All \ion{Fe}{ii} lines from \citet{Veron-Cetty2004} are modeled as Gaussian functions using Equation \ref{eq:gaussian} and are summed together into two separate broad and narrow templates, each of which can be scaled by a multiplicative free-parameter amplitude $A$ during the fit.  Following QSFit, the default FWHM of broad and narrow \ion{Fe}{ii} lines are fixed at 3000 \kms and 500 \kms, respectively, which are adequate given the resolution and typical S/N of SDSS data.  The velocity offset of each line is also fixed by default.  We justify holding the FWHM and velocity offsets fixed due to the fact that broad and narrow emission are blended together and superimposed atop one another, usually at varying amplitudes.  This leads to a strong degeneracy in both the FWHM and velocity offsets in these features. However, the FWHM and velocity offset constraints can be optionally turned off or adjusted to particular values for each the broad and narrow templates for more-detailed fitting.\\
\indent In addition to the template from \citet{Veron-Cetty2004}, BADASS can alternatively use the temperature-dependent \ion{Fe}{ii} model from \citet{Kovacevic2010}, which independently models each of the $F$, $S$, and $G$ atomic transitions of \ion{Fe}{ii}, as well as some strong lines from I Zw 1, in the region between 4400 \angstrom~and 5500 \angstrom, with amplitude, FWHM, velocity offset, and temperature as free parameters.  The template from \citet{Kovacevic2010}, while slightly smaller in wavelength coverage compared to the \citet{Veron-Cetty2004} template, can more accurately model objects with particularly strong \ion{Fe}{ii} such as NLS1s \citep{Veron-Cetty2001,Xu2012,Rakshit2017}. Individual transition amplitudes, widths, and velocity offsets can be optionally fixed during the fitting process as well.

\begin{figure*}
 \includegraphics[width=\textwidth]{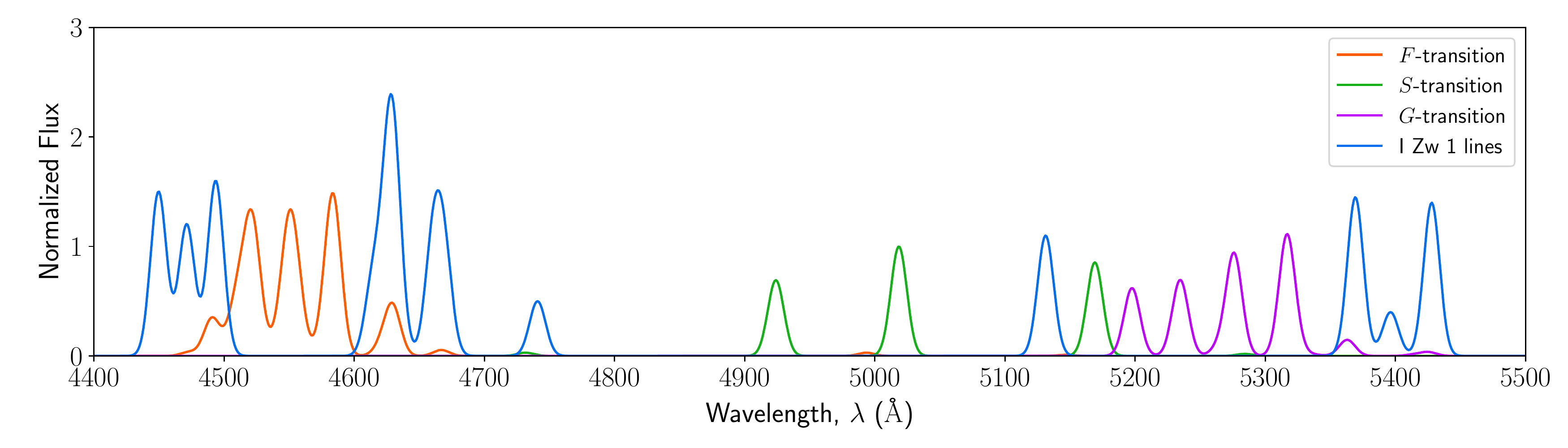}
 \caption{The temperature-dependent \ion{Fe}{ii} template from \citet{Kovacevic2010} for a more-detailed analysis of optical \ion{Fe}{ii} emission.  BADASS independently models each of the $F$, $S$, and $G$ atomic transitions of \ion{Fe}{ii}, as well as some strong lines from I Zw 1, in the region between 4400 \angstrom and 5500 \angstrom, with amplitude, FWHM, velocity offset, and temperature as a free-parameters. }
 \label{fig:BADASS_K10_feii_templates}
\end{figure*}

\subsubsection{Host galaxy \& Stellar Absorption Features}

\begin{figure*}
 \includegraphics[width=\textwidth]{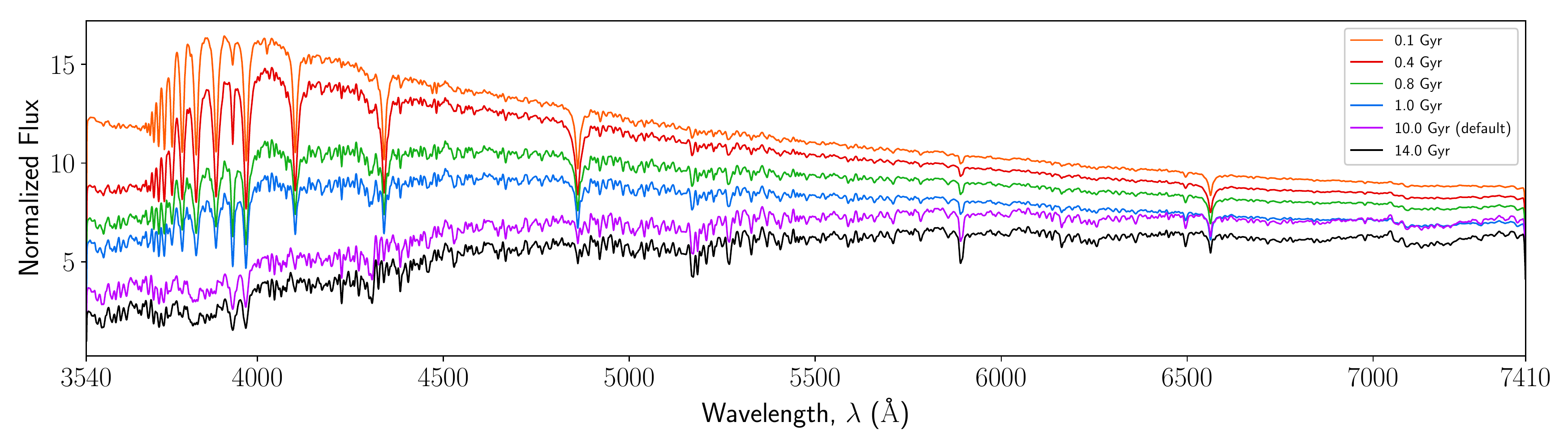}
 \caption{Single stellar population models from \citet{Vazdekis2010}  generated using the MILES Tune Stellar Libraries webtool used for fitting the host galaxy contribution if the LOSVD (stellar template) fitting is not performed.}
 \label{fig:BADASS_MILES_ssp_models}
\end{figure*}

The original purpose of BADASS was to extract the host galaxy contribution in type 1 AGN spectra, and in particular, estimate the stellar LOSVD to obtain stellar velocity $v_*$ and stellar velocity dispersion $\sigma_*$.  In this regard, BADASS serves as a wrapper for the stellar-template fitting code pPXF \citep{ppxf1,ppxf2}, allowing the user to optionally fit the underlying stellar population to extract stellar kinematics.  After subtracting off all the aforementioned components from the original data, BADASS models the stellar population using 50 empirical stellar templates from the Indo-US Library of Coud\'e Feed Stellar Spectra \citep{IndoUS}.  The Indo-US Library was chosen for its high-resolution (FWHM resolution of 1.35 \angstrom; \citet{Beifiori2011}) as well as its wide wavelength coverage from 3460 \angstrom\;to 9464 \angstrom.  The 50 chosen templates include the full range of spectral types from O to M, and were specifically chosen for minimal gaps in coverage.  We find that using 40-50 stellar templates strikes an optimal balance between reducing the chances of template mismatch and large computation times, since the non-negative least squares routine used by pPXF for choosing templates and calculating weights carries the largest computational overhead for BADASS and scales with the number of templates used in the fit.\\
\indent By default, only $v_*$ and $\sigma_*$ are fit for SDSS spectra, however, if given higher-resolution spectra, it is still possible for pPXF to estimate the higher-order Gauss-Hermite moments of the LOSVD.  One caveat to fitting the LOSVD via stellar template fitting is the limited wavelength range of the stellar template library chosen for fitting the LOSVD.  In this regard, the Indo-US Library is provides the largest optical range and highest resolution for currently available empirical stellar libraries.  However, if one chooses to use a different library of stellar templates, the LOSVD fitting range can be extended.\\
\indent In general, the quality of the fit to the LOSVD is S/N dependent.  We find that if if the continuum S/N$<10$, estimates of $v_*$ and $\sigma_*$ can have uncertainties $>50\%$, therefore, we allow the user to optionally disable fitting of the LOSVD and instead fit the stellar continuum with a single stellar population (SSP) model generated using the MILES Tune Stellar Libraries Webtool \citep{Vazdekis2010}, which is initialized with a metallicity $[\rm{M}/\rm{H}]=0.0$, age of 10.0 Gyr, and dispersion of 100 \kms to match the depth of stellar absorption features typically seen in SDSS galaxies.  The SSP template is normalized at 5500 \angstrom\; and is scaled by a multiplicative factor that is a free-parameter during the fitting process.  We also include alternative MILES SSP models with ages ranging from 0.1 Gyr to 14.0 Gyr which can be used as optional substitutes.  We show a range of MILES SSP models which can be used by BADASS in Figure \ref{fig:BADASS_MILES_ssp_models}.\\
\indent Due to the limited range of the MILES stellar library, if the fitting range is outside $(3525\;\text{\angstrom}\leq\lambda\leq7500\;\text{\angstrom})$, SSP models from \citet{Maraston2009} are used instead, which have a coverage of $(1150\;\text{\angstrom}\leq\lambda\leq25000\;\text{\angstrom})$, however have much larger dispersions that do not match strong stellar absorption features in SDSS spectra. \\
\indent We note that dispersions measured with pPXF already take into account the instrumental dispersion of the SDSS, since it first convolves input templates to the resolution of the SDSS before the fitting process.  We nonetheless place a lower limit on the allowed values for $\sigma_*$ to be the velocity scale (in units of km s$^{-1}$ pix$^{-1}$) of the input spectra.

\subsection{Fitting Procedure}

\subsubsection{Determination of Initial Parameter Values}

As is true for all fitting algorithms, the number of required MCMC iterations required for convergence on a solution is sensitive to the initial parameter values.  Ideally, one should initialize parameters as close as possible to their actual posterior values in order to minimize the number of iterations used for searching parameter spaces and maximize the number of posterior sampling iterations.  To do this, we employ maximum likelihood estimation of all parameters using the SciPy \citep{scipy} \texttt{scipy.optimize.minimize} function to find the \emph{negative} maximum (minimum) of the log-likelihood function, which we derive as follows.\\
\indent We assume that each datum of the spectrum can be approximated as a normally distributed random variable of mean $y_{\rm{data},i}$ and standard deviation $\sigma_i$.  The likelihood of the data given the model $y_{\rm{model}}$ is given by

\begin{equation}\label{eq:likelihood}
    L = \prod_{i=1}^N \frac{1}{(2\pi\sigma^2)^{1/2}}\exp\left[-\frac{\left(y_{\rm{data},i}-y_{\rm{model},i}\right)^2}{\sigma_i^2}\right]
\end{equation}

Since Equation \ref{eq:likelihood} can result in very large values, which often times exceeds the numerical precision of most computing machines, it is easier to use the natural log of the likelihood, i.e., the \emph{log-likelihood}:

\begin{equation}
    \mathcal{L} = \log(L) = -\frac{1}{2}\sum_{i=1}^N\log(2\pi\sigma_i^2)+\frac{\left(y_{\rm{data},i}-y_{\rm{model},i}\right)^2}{\sigma_i^2},
\end{equation}

The constant terms in this sum that do not change from one iteration to the next, including $\log(2\pi\sigma_i^2)$, can be dropped, as they do not play a role in determination of the minimum.  The log-likelihood is therefore given by

\begin{equation}\label{eq:loglikelihood}
    \mathcal{L} = \sum_{i=1}^N\frac{\left(y_{\rm{data},i}-y_{\rm{model},i}\right)^2}{\sigma_i^2},
\end{equation}
where the sum is performed over each spectral channel $i$ for each datum $y_{\rm{data},i}$, $y_{\rm{model},i}$ is the value of the model at each $i$, and $\sigma_i$ is the $1\sigma$ uncertainty at each $i$ determined from the SDSS inverse variance of the spectrum.\\
\indent The \textit{scipy.optimize.minimize} function, which employs the built-in Sequential Least SQuares Programming (SLSQP; \citet{slsqp}) method, is used to include bounds and constraints on all parameters.  Parameter bounds, which are the minimum and maximum values for each parameter, are determined from the data.  For example, the emission line amplitudes must be non-negative, and can have a maximum value of the data in the fitting region.  These simple boundary conditions, which are later used by \textit{emcee}, are effective in limiting the parameter space for timely convergence.  Constraints on parameters are used for the testing for possible blueshifted wing components in [\ion{O}{iii}] (see Section \ref{sec:outflow_tests}).\\
\indent By default, BADASS only performs one maximum likelihood fit to obtain initial parameter values in the interest of time and a single fit is sufficient for initializing parameters for MCMC fitting.  However, if one chooses not to perform MCMC fitting, or if one desires more robust initial parameter values, BADASS can optionally perform multiple iterations of maximum likelihood fitting by resampling the spectra with random normally-distributed noise from the spectral variance and re-fitting the spectrum, i.e., Monte Carlo ``bootstrapping".

\subsubsection{Testing for Presence of Outflows in Narrow Emission Lines}\label{sec:outflow_tests}

Additional ``blue wing" components in narrow emission lines, indicative of possible outflowing gas from the central BH, are known to be commonplace in AGN-host galaxies \citep{Nelson1996,Mullaney2013,Woo2016,Zakamska2014,Rakshit2018,DiPompeo2018,Davies2020}.  If present, failure to account for the blue excess in narrow line emission can lead to significant difference in measured line quantities, especially if one uses narrow line width as a proxy for $\sigma_*$ \citep{Woo2006,Komossa2007,Sexton2019,Bennert2018}.  \\
\indent Blue wings are most visibly obvious in the narrow [\ion{O}{iii}] emission line in type 1 AGNs because [\ion{O}{iii}] is not significantly contaminated by nearby broad lines.  To determine if blue wings are present, BADASS can optionally perform preliminary single-Gaussian and double-Gaussian fits to the H$\beta$/[\ion{O}{iii}] or H$\alpha$/[\ion{N}{ii}]/[\ion{S}{ii}] narrow line complexes to test if an additional Gaussian component in the model is justified.  The test for outflows is identical to the process used for fitting initial parameter values using maximum likelihood estimation (i.e., Monte Carlo bootstrapping).  The double-Gaussian fit makes the assumption that outflows are present by including a narrower ``core'' and a broader ``outflow'' component for the narrow emission lines.  Monte Carlo bootstrapping for a user-defined set of iterations is then used to obtain uncertainties on core and outflow components to assess the quality of the fit.  During the fitting process the FWHM of the outflow component is constrained to be greater than the core component, and the amplitude of the outflow component is constrained to be less than the core component, following what is typically seen in the literature.  However, since outflows are not necessarily always blueshifted, but sometimes at equal or redshifted velocities with respect to the core component (albeit in rare occurrences), we do not constrain the velocity offset of either component during the fitting process.  This feature is useful for detecting outflow components found in star forming galaxies, which are known to be less offset from the core component causing a more symmetric [\ion{O}{iii}] line profile \citep{Cicone2016,Davies2019,Manzano-King2019}. \\
\indent To quantify the presence of outflows in [\ion{O}{iii}], we visually identify 63 objects from a sample of 173 known type 1 AGN that both (1) exhibit the characteristic line profile asymmetry commonly seen in the literature, and (2) have a measurable non-gravitational component in [\ion{O}{iii}] relative to the systemic (stellar) velocity dispersion (this is discussed in detail in Section \ref{sec:sample_selection}) and derive empirical relationships between measurable parameters that recover these objects. After BADASS performs fits for both the single-Gaussian (no-outflow) and double-Gaussian (outflow) models, the following empirical diagnostics are used to determine if a secondary outflow component is justified in the model:

\begin{equation}\label{eq:amp_metric}
\textrm{Amplitude metric:}\quad\cfrac{A_{\rm{outflow}}}{\left(\sigma^2_{\rm{noise}} + \delta A^2_{\rm{outflow}}\right)^{1/2}} > 3.0
\end{equation}
\begin{equation}\label{eq:fwhm_metric}
\textrm{Width metric:}\quad\cfrac{\sigma_{\rm{outflow}}- \sigma_{\rm{core}}}{\left(\delta \sigma^2_{\rm{outflow}}+\delta \sigma^2_{\rm{core}}\right)^{1/2}} > 1.0
\end{equation}
\begin{equation}\label{eq:voff_metric}
\textrm{Velocity metric:}\quad\cfrac{v_{\rm{core}}- v_{\rm{outflow}}}{\left(\delta v^2_{\rm{core}}+\delta v^2_{\rm{outflow}}\right)^{1/2}} > 1.0
\end{equation}
\begin{equation}\label{eq:f_test}
\textrm{$F$-statistic:}\quad\cfrac{\left(\cfrac{\textrm{RSS}_{\rm{no\;outflow}}-\textrm{RSS}_{\rm{outflow}}}{k_2-k_1}\right)}{\left(\cfrac{\textrm{RSS}_{\rm{outflow}}}{N-k_2}\right)}
\end{equation}

where $A$ is the line amplitude (in units of $10^{-17}$ erg cm$^{-2}$ s$^{-1}$ \angstrom$^{-1}$), $\sigma$ is the Gaussian dispersion ($\rm{FWHM}/2.355$; in units of km s$^{-1}$), $v$ is the velocity offset of the line relative to the rest frame of the overall spectra (in units of km s$^{-1}$).  The quantity RSS is the sum-of-squares of the residuals within $\pm3\sigma$ of the full (core + outflow) [\ion{O}{iii}] line profile, $k_1=3$ is the number of degrees of freedom in the single-Gaussian model, $k_2=6$ is the number of degrees of freedom in the double-Gaussian model, and $N$ is the size of the sample used to calculate RSS.  If parameters of the core or outflow models do not adhere to their bounds or approach to the limits of their constraints, which in turn violates the number of degrees of freedom for each model, BADASS flags the relevant parameters and defaults to a single-Gaussian (no-outflow) model. \\
\indent Equation \ref{eq:amp_metric} is a measure of the amplitude of outflow component above the noise, while Equations \ref{eq:fwhm_metric} and \ref{eq:voff_metric} are a measure of how much we can significantly detect measurable differences between the core and outflow FWHM and velocity offsets, respectively. Another way of interpreting Equations \ref{eq:amp_metric}, \ref{eq:fwhm_metric}, and \ref{eq:voff_metric} is the uncertainty overlap between parameter values, which signifies how well BADASS can separate core and outflow components in parameter space.  For example, a value of 2 for the width metric indicates that there is 2$\sigma$ separation between the best-fit values of the core and outflow FWHM.\\
\indent Equation \ref{eq:f_test} is statistical $F$-test for model comparison between the single- and double-Gaussian models.  The $F$-statistic in this context calculates unexplained variance between the outflow and no-outflow models as a fraction of the unexplained variance in the outflow model alone.  The $F$-statistic is then used to calculate a $p$-value, which if less than a critical value (by default, $\alpha=0.05$), indicates that we can reject the null hypothesis that there is no significant difference between the single- and double-Gaussian models, and that the difference is greater than that which could be attributed to random chance.  We express our confidence in the outflow model by calculating $1-\alpha$.  For example, $\alpha=0.05$ indicates a 95\% confidence that a double-Gaussian model explains the variance in fitting the [\ion{O}{iii}] profile significantly better than a single-Gaussian model.\\
\indent All of these criteria can be toggled on or off, and the significance and confidence thresholds for each can be changed to meet the user's specific needs.  We find that the above criteria provides a satisfactory method in finding objects with strong blueshifted excess in [\ion{O}{iii}] with a success rate of $>90\%$ compared to visual identification, and recommend this method to determine if strong outflows are present or if the [\ion{O}{iii}] core dispersion needed as a surrogate for $\sigma_*$ when it cannot be otherwise measured.  We note that while these empirical criteria are successful in describing the types of objects in our sample, more sophisticated statistical modelling and cross-validation techniques with a larger sample will be required to improve the capability of BADASS to identify objects with outflows. \\
\indent If the above criteria are met, the final set of parameters that are fit with \textit{emcee} will include a second Gaussian component for all narrow lines in the fitting region, otherwise, the final model will only fit a single Gaussian component to each narrow line.  It is important to note that any emission line, whether it exhibits blue excess or not, can be fit with more than one component and produce a better fit.  However, if one is using the core component of [\ion{O}{iii}] as a surrogate for stellar velocity dispersion, using a two component fit when only one component is justified by the data can significantly underestimate the stellar velocity dispersion.  Likewise, not correcting for a strong and clearly visible outflow component can significantly overestimate the stellar velocity dispersion if the core component is used as a surrogate. \\
\indent While BADASS can perform tests for outflows on either the H$\beta$ or H$\alpha$ region, it is recommended that if one wants to fit outflows in both the H$\beta$ and H$\alpha$ regions that it be done simultaneously, in which case BADASS uses the [\ion{O}{iii}]$\lambda5007$ outflow component to constrain the properties of the outflow components for H$\alpha$, [\ion{N}{ii}], and [\ion{S}{ii}].  This constraint is activated because even if a broad line is not present in H$\alpha$, narrow H$\alpha$ and [\ion{N}{ii}] can still be severely blended due to the resolution of the SDSS.  However, if one chooses to fit outflows in H$\alpha$ independently from H$\beta$, it is still possible at the user's discretion.\\
\indent As an aside, there is no single definition or quantification of what constitutes an ``outflow'' in regards to emission lines.  BADASS, admittedly, can only detect significantly broad and offset [\ion{O}{iii}] wings that are commonly seen in the literature, but outflows can generally produce a wide range of emission line profiles.  For example, \citet{Bae2016} modeled biconical outflows in 3D and found that the emission line profile of outflows is strongly dependent on orientation and dust extinction, resulting in sometimes redshifted and non-Gaussian profiles.  Some models also indicate that a high-velocity outflow can be present without exhibiting an broad blue excess due simply to bicone orientation. \\
\indent The criteria we present here constitute a \emph{preliminary} means of filtering out objects with strong outflows which may cause significant residuals if not taken into account, and can make no claims as to whether an outflow is present if the secondary outflow component is close to the velocity offset or width of the primary core component of the emission line.  The best method of determining if a secondary component is necessary is to perform a full fit with the double-Gaussian model with \textit{emcee}, as well as fitting the LOSVD to estimate the gravitational influence of the NLR, and examining the individual parameter chains to ensure that they are not degenerate and have converged on a stable solution.

\subsubsection{Final Parameter Fitting}\label{sec:emcee}

\begin{figure*}
 \includegraphics[width=\textwidth]{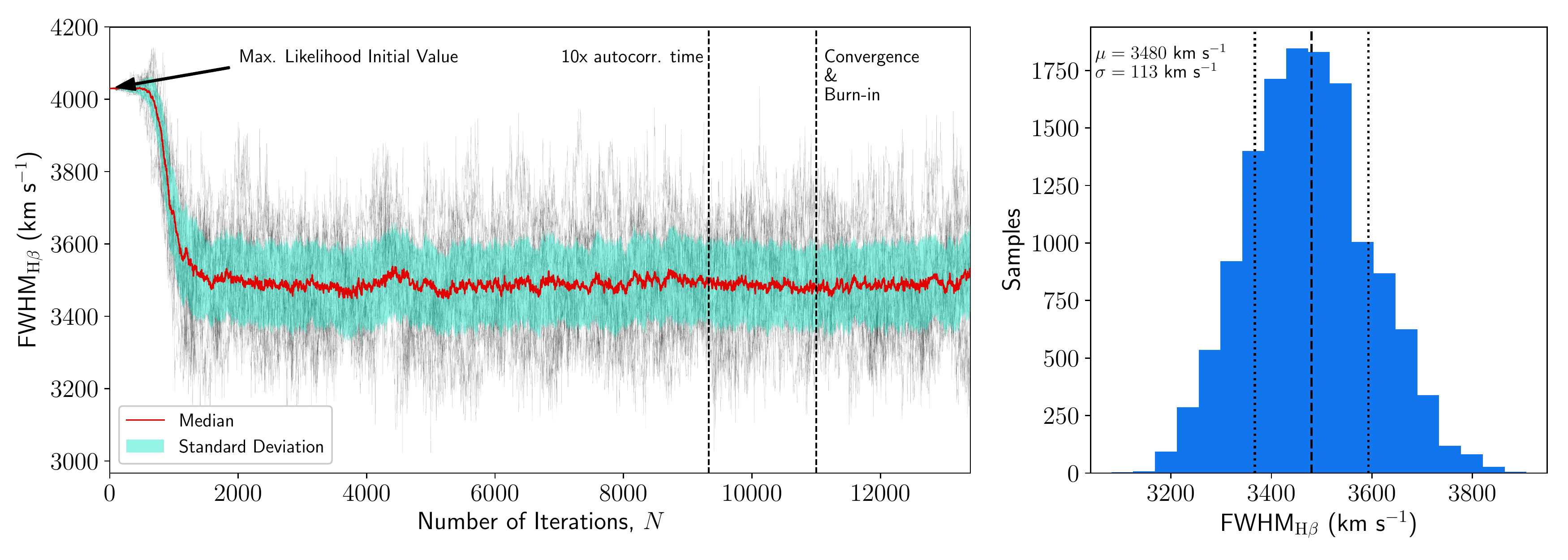}
 \caption{
\emph{Left}: The FWHM parameter MCMC chain for the H$\beta$ FWHM from the above example performed using 100 walkers. The initial starting position, estimated using maximum likelihood fitting, overestimates the final width of the line by more than 500 km s$^{-1}$.  As other parameters are fit, the value H$\beta$ FWHM decreases and settles into a stable solution by $\sim2000$ iterations.  Convergence is reached at $\sim11000$ iterations (for this example, when 10 times the autocorrelation time per parameter at 10\% tolerance per parameter is achieved), and the burn-in is chosen to be the final $2500$ iterations after all other parameters has been achieved. \emph{Right}: A histogram of the last 2500 iterations of all 100 walkers.  
}
 \label{fig:BADASS_chain_example}
\end{figure*}

\begin{figure*}
 \includegraphics[width=\textwidth,trim={1.5cm 0cm 1.5cm 1.5cm},clip]{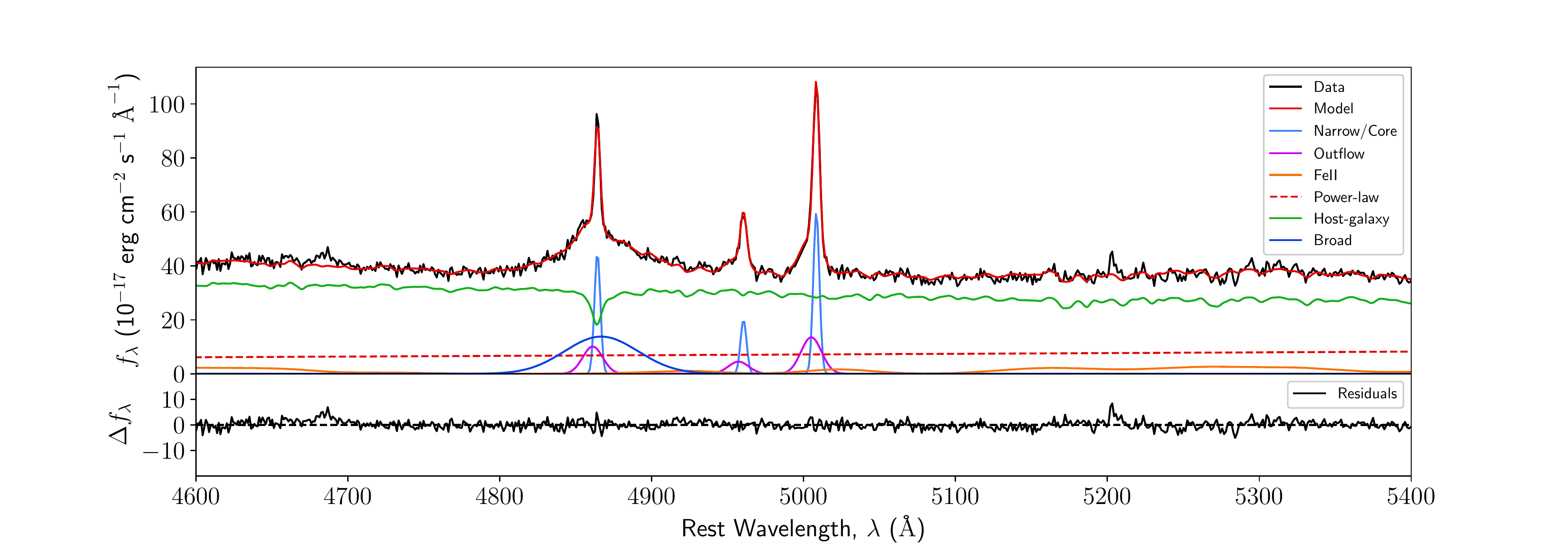}
 \caption{
An example of the best-fit final model output of a spectrum fit in the H$\beta$/[\ion{O}{iii}] region with individual fitting components designated by different colors.  This particular object exhibits the blueshifted [\ion{O}{iii}] outflow components (magenta) which BADASS is designed to detect and fit.
 }
 \label{fig:BADASS_fitting_example}
\end{figure*}

\indent Final parameter fitting performed using MCMC begins by initializing each parameter at its maximum likelihood value obtained from the initial fit.  We use Equation \ref{eq:loglikelihood} as the likelihood probability and initialize each parameter with a flat prior with lower and upper bounds determined by the data.  If the model contains outflow components, constraints on outflow parameters (see Section \ref{sec:outflow_tests}) are included in their respective flat priors.  We place an additional constraint on broad line components, whose widths must be greater than narrow line widths, if broad lines are included in the fit.  Fitting is then performed iteratively via MCMC until each component's parameters have converged. Each parameter space is randomly sampled using the affine invariant MCMC sampler \textit{emcee} until a user-defined number of iterations is reached or if autocorrelation analysis (recommended) has determined that parameter convergence has been sufficiently reached (see Section \ref{sec:autocorrelation}).  Fitting using the \textit{emcee} package is advantageous since the use of multiple simultaneous ``walkers'' efficiently explores each parameter space in parallel, all of which form MCMC parameter ``chains'' from which the final posterior distribution is estimated for each parameter, as shown in Figure \ref{fig:BADASS_chain_example}.  \\
\indent The values of parameters estimated using the initial maximum likelihood routine (\textit{scipy.optimize.minimize}) can differ substantially from parameters estimated using emcee, as shown in Figure \ref{fig:BADASS_chain_example}.  This large difference is attributed to the stellar continuum model used in each fit.  The initial fit uses only a single SSP template to estimate the contribution to the stellar continuum.  The \textit{scipy.optimize.minimize} algorithm, while relatively fast, is not sensitive enough to fit the LOSVD with stellar templates in addition to many other component parameters.  Fitting the LOSVD requires many iterations due to the fact that even moderate changes in stellar velocity or stellar velocity dispersion need not drastically change the shape of the resulting stellar continuum model, or drastically change the value of the calculated likelihood, which is partly due to the number and diversity of template stars used by BADASS.  Instead, the \textit{scipy.optimize.minimize} algorithm prioritizes the fitting of components that have the greatest effect on the calculated likelihood, such as emission lines and  continuum amplitudes.  In short, the limitations of the \textit{scipy.optimize.minimize} routine are due to a combination of (1) degeneracies inherent to the stellar template fitting process, and (2) the likelihood threshold requirements needed for \textit{scipy.optimize.minimize} to achieve a solution  As a result, the initial fitting routine in BADASS only fits a single SSP template for the stellar continuum.\\
\indent The advantage of MCMC fitting with \textit{emcee} allows for prolonged simultaneous fitting of parameters even after a likelihood threshold has been achieved.  Components that are fit very easily (such as emission lines and continuum amplitudes) converge on solutions very quickly, while components that are less sensitive to even moderate changes (such as stellar templates) can continue to converge on a solution, even if the calculated maximum likelihood does not vary considerably.  The advantage of MCMC sampling in this context allows BADASS to explore the LOSVD parameter space even after a maximum likelihood has been reached, which allows for greater variation in template stars used for achieving a best fit.  In other words, parameter convergence is more heavily dependent on parameter variation (autocorrelation; see Section \ref{sec:autocorrelation}) rather than achieving some maximum likelihood threshold.  In the case of the FWHM$_{\rm{H}\beta}$ parameter chain shown in Figure \ref{fig:BADASS_chain_example}, the value of FWHM$_{\rm{H}\beta}$ slowly adjusts to changing stellar templates of varying stellar H$\beta$ absorption.  These slow adjustments do not result in significant changes in the likelihood that the \textit{scipy.optimize.minimize} are sensitive to, however, over a prolonged number of fitting iterations, a compromise is eventually reached between the value of FWHM$_{\rm{H}\beta}$ and the stellar continuum absorption estimated from individual stellar templates. \\
\indent The MCMC fitting process uses the same likelihood function, priors, and constraints used in the initial fitting procedure, but instead produces parameter distributions from which best-fit values and uncertainties can be reliably estimated.  The best-fit values are then used to construct a final model, an example of which is shown in Figure \ref{fig:BADASS_fitting_example}.

\subsubsection{Autocorrelation Analysis}\label{sec:autocorrelation}

\begin{figure*}
 \includegraphics[width=\textwidth]{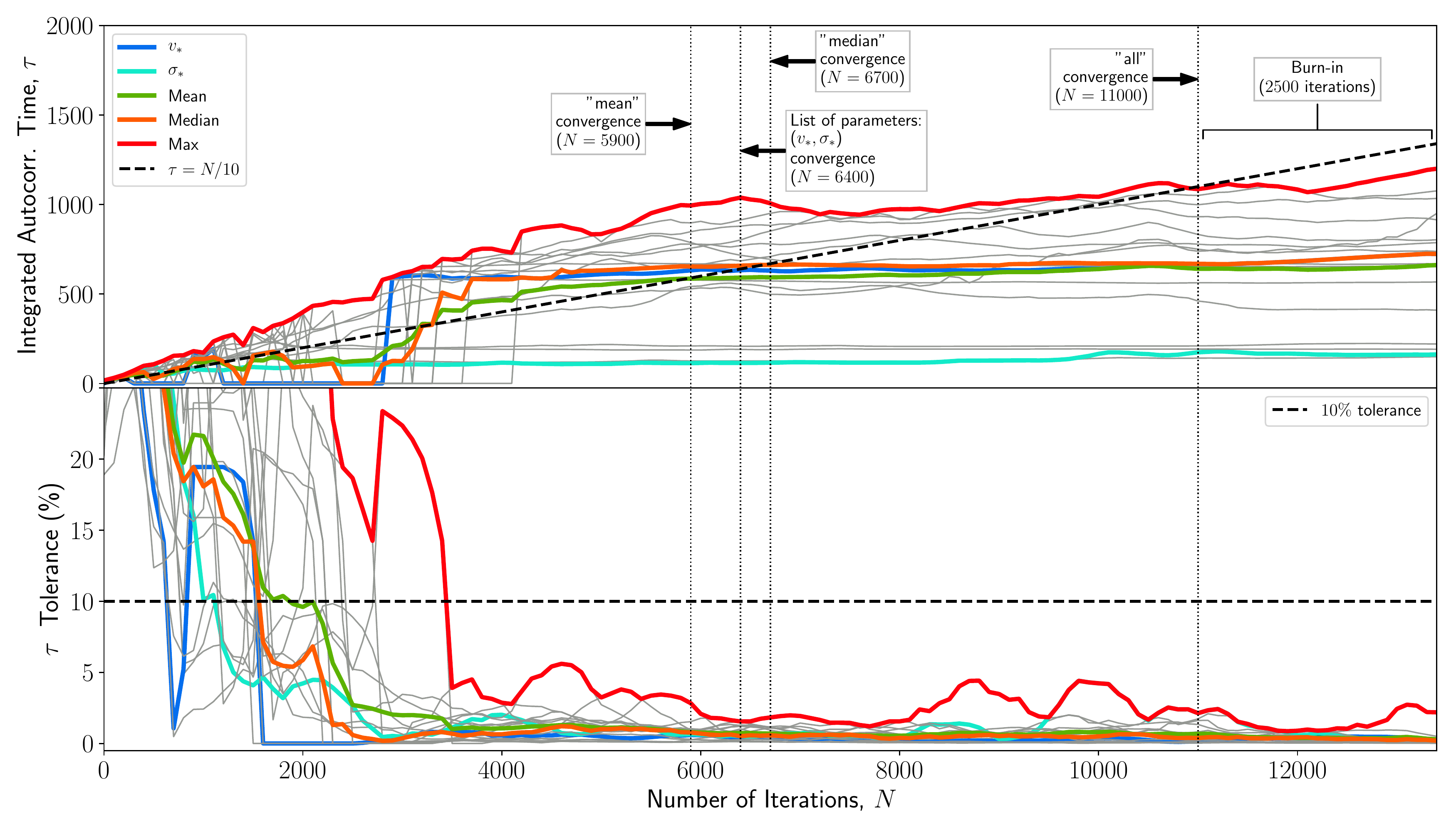}
 \caption{An example of different modes of autocorrelation convergence using a model spectrum with 17 free parameters.  The ``median'' mode of convergence typically converges the fastest because it tends to omit outlier $\tau$ values, which are typically values for which there are too few iterations to determine an accurate integrated autocorrelation time.  By specifying the LOSVD parameters ($v_*$ and $\sigma_*$), convergence is reached in 6400 iterations, however many other parameters have not yet converged.  The ``mean'' convergence, is a more strict convergence criteria because all $\tau$ estimates are weighted equally.  Requiring ``all'' parameters to converge requires the highest number of iterations and does not guarantee convergence if some parameters cannot converge within the maximum number of iterations.
 }
 \label{fig:BADASS_autocorrelation_plot}
\end{figure*}

\indent Over a number of iterations, the values of a parameter can fluctuate significantly as each parameter space is explored.  Since some parameters can be strongly correlated (for instance, the AGN power-law continuum component and the stellar continuum component amplitudes), there can exist parameter degeneracies that are not taken into account when using conventional fitting techniques.  Parameter fluctuations and degeneracies can also vary as a function of signal-to-noise.  The nature and diversity of galaxy and AGN data makes determination of parameter convergence a non-trivial issue, especially if the goal is to fit a large number of objects that vary intrinsically or in data quality.  Fitting algorithms typically address this by setting the number of fitting iterations to an arbitrarily high number or setting a minimum tolerance in the change in the likelihood value.  While these methods of convergence are generally good, they only guarantee convergence in the overall fit to the data, and not on the convergence of individual parameters. \\
\indent To address this, BADASS employs autocorrelation analysis to assess parameter convergence.  Autocorrelation analysis functions are built into \textit{emcee} (see \citet{emcee}), however, we tailor these functions for the purposes of spectral fitting.  The integrated autocorrelation time, which is the number of iterations required for a parameter chain to produce an independent sample, is calculated for all parameters at incremental fitting iterations (by default, every 100 fitting iterations), and then the convergence criteria are checked to see if they are satisfied by the most current fit parameters.  There are two criteria that must be satisfied that define convergence:  (1) if calculated autocorrelation times, which are multiplied by some multiplicative factor (by default, 10.0), exceeds the number of performed fitting iterations, and (2) if the difference between the current and previous calculated autocorrelation times is less than a specified percent change (by default, 10\%).  \\
\indent In practice, some parameters never reach adequate convergence, usually due to strong degeneracies with other model components (such as \ion{Fe}{ii}).  To accommodate these instances, BADASS offers the user four modes of convergence in terms of the integrated autocorrelation time: (1) mean, (2) median, (3) user-specified parameters, and (4) all parameters.  The first two options calculate the mean or median autocorrelation time of \emph{all} free-parameters to determine when an overall solution is reached, however it does not guarantee that all parameters reach convergence.  To guarantee the convergence of specific parameters of interest, option (3) allows the user to indicate which specific parameters are considered for autocorrelation analysis, which is useful for ignoring components which have high autocorrelation times (poorly constrained or highly degenerate components) or parameters of low importance. Finally, option (4) allows the user to specify that all parameters must converge on a solution, for which BADASS runs until all parameters satisfy the autocorrelation conditions or BADASS reaches the user-defined maximum number of iterations.\\
\indent Figure \ref{fig:BADASS_autocorrelation_plot} shows an example of different autocorrelation modes and the required number of iterations for a 17-parameter model.  We recommend that the user either select the specific parameters of interest for convergence or simply choose the ``mean'' criteria.  The ``median'' criteria is less sensitive to outlier parameters (parameters with large autocorrelation times) and will generally perform fewer iterations for convergence.  The ``all'' convergence mode is the most strict type of convergence, however, there is no guarantee that convergence can be reached for all parameters within the set maximum number of iterations, especially if the data has low S/N or there are degenerate parameters.\\
\indent Once the convergence criteria are met, BADASS continues to fit for a set number of iterations (by default, 2500), which is ultimately used for the posterior distribution to determine the best-fit parameter values and uncertainties.  The iteration at which convergence is achieved defines the ``burn-in'' for the parameter chains, after which all following iterations contribute to the final parameter estimation (i.e, the iterations that contribute to the histogram in Figure \ref{fig:BADASS_chain_example}).  Note that this is only true if autocorrelation analysis is used to assess convergence (\verb|auto_stop=True|), otherwise, BADASS runs for the maximum number of iterations using the burn-in defined by the user.  If for any reason convergence criteria are met and then subsequently violated, BADASS resets the burn-in and continues to sample until convergence is met again. This ensures that convergence is maintained at all times after the burn-in and that a best-fit is not achieved prematurely.

\subsection{Performance Tests}\label{sec:performance}

Since one of the primary goals of BADASS is the recovery of the LOSVD, and specifically the stellar velocity dispersion $\sigma_*$ in AGN host galaxies, we perform a series of performance tests as a function of different components and parameters which may affect measurements of $\sigma_*$ in the optical \ion{Mg}{ib}/\ion{Fe}{ii} region from 4400 \angstrom\; to 5500 \angstrom.  We note that the results of these tests are not strictly limited to BADASS, but also general fitting techniques concerned with fitting $\sigma_*$ in AGN host galaxies.

\subsubsection{Recovery of $\sigma_*$ as a function of S/N}

\begin{figure*}
 \includegraphics[width=\textwidth]{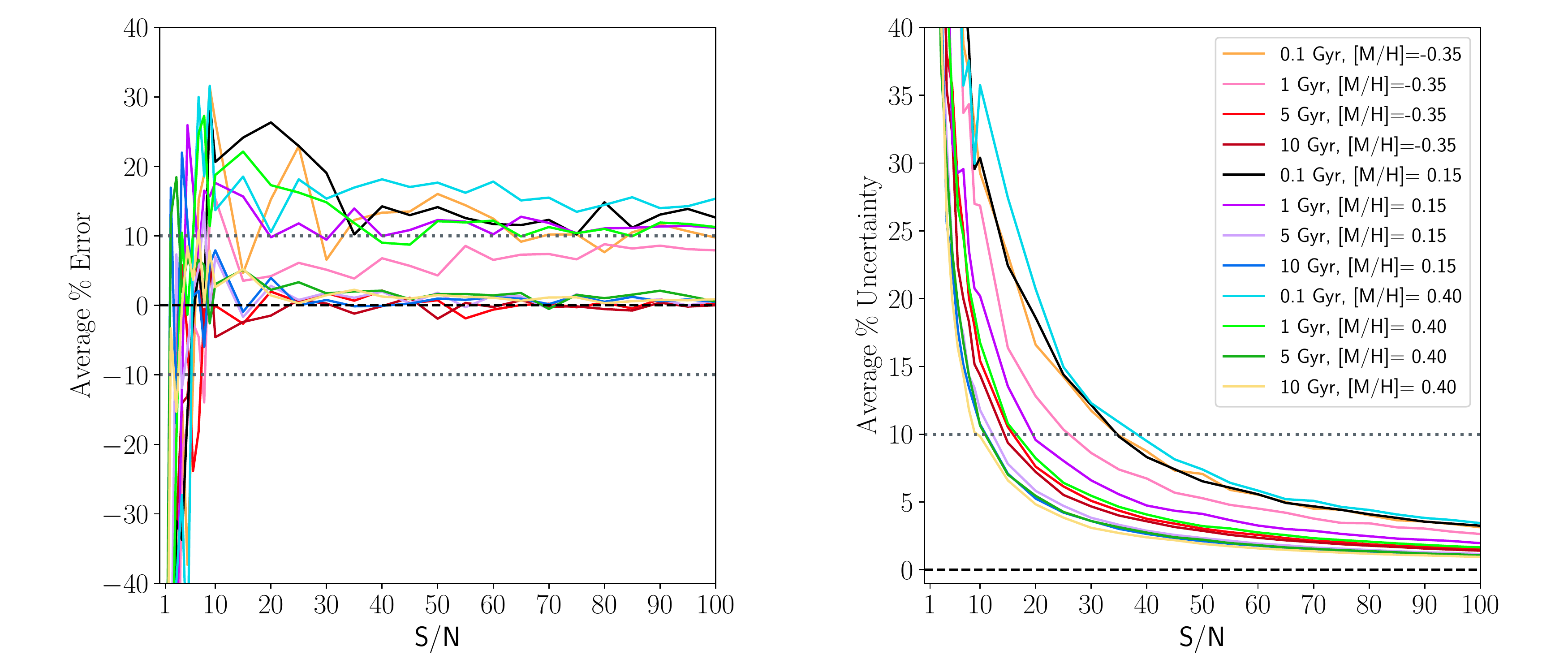}
 \caption{
 Recovery of $\sigma_*$ as a function of S/N for different fitted MILES SSP models.  The figure on the left shows the percent error relative to the actual value of $\sigma_*$, which exceeds 10\% (green dotted line) below S/N$\sim$15 and for the youngest (0.1 Gyr and 1 Gyr) SSP models.  The figure on the right shows the percent uncertainty, which increases with decreasing S/N.  This provides a minimum S/N for which LOSVD measurements can be reliably recovered.
 }
 \label{fig:ssp_model_test}
\end{figure*}

To investigate the effects of S/N level on the recovery of $\sigma_*$, we generate a series single stellar population model using the MILES Tune SSP model spectra webtool \citep{Vazdekis2010}.  We limit the stellar population ages to 0.1, 1, 5, and 10 Gyr and metallicities [M/H]$=-0.35$, 0.15, and 0.40, and initialize the simulated spectrum at a stellar velocity dispersion $\sigma_*=90$ \kms, taking into account the wavelength-dependent dispersion of the SDSS.  We then artificially add normally-distributed random noise at various S/N ratios to simulate real observations, and 10 mock spectra are generated per S/N level.  The S/N is measured relative to the value of the data in each spectral channel.  No other spectral components are added to these SSP models, and only $v_*$ and $\sigma_*$ are fit.  \\
\indent Figure \ref{fig:ssp_model_test} shows the results of S/N tests of various SSP models, where we calculate the percent error (deviation of the best-fit value from the actual value) and the percent uncertainty in $\sigma_*$.  The results of these tests provide a lower limit to the S/N of SDSS spectra for which $\sigma_*$ can be reliably measured.  Below a S/N of $\sim$15, the best-fit measurement of $\sigma_*$ begins to exceed the actual value by more than 10\%, and becomes increasingly unreliable at lower S/N.  We find a similar result for the average uncertainties of $\sigma_*$.  We therefore do not recommend measuring the LOSVD at S/N $<20$ if the scientific goal is to report accurate stellar kinematics. \\
\indent There is also a clear offset for the youngest (0.1 Gyr and 1 Gyr) SSP models, which can be explained by the lack of younger stellar templates included with BADASS, since these stellar types are considerably more rare.  In cases where BADASS is used for fitting active star forming galaxies, we recommend one include more O- and B-type templates for fitting the LOSVD and/or disabling the power-law component.

\subsubsection{Recovery of $\sigma_*$ as a function of \ion{Fe}{ii} Emission}

\begin{figure*}
 \includegraphics[width=\textwidth]{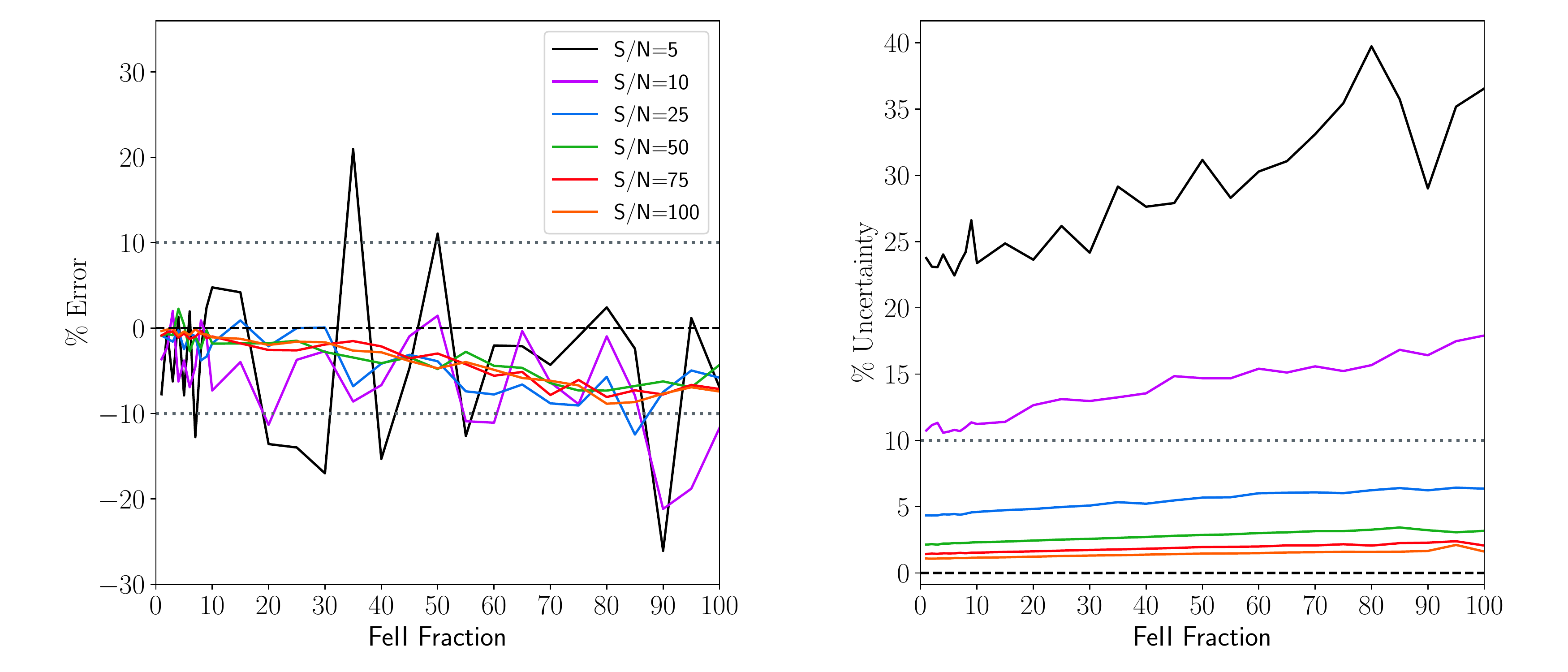}
 \caption{
 Recovery of $\sigma_*$ as a function of \ion{Fe}{ii} fraction.  There is a weak dependence on the value of $\sigma_*$ due to \ion{Fe}{ii} fraction, and a stronger dependence on S/N. Even in the most extreme cases in which the \ion{Fe}{ii} fraction fraction exceeds 50\%, \ion{Fe}{ii} emission does not significantly affect stellar template fitting.
 }
 \label{fig:feii_test}
\end{figure*}

To test the effects of \ion{Fe}{ii} emission on the measurement of $\sigma_*$, we hold the amplitude of the 10 Gyr, [M/H]=0.15 MILES SPP model constant and incrementally add broad and narrow \ion{Fe}{ii} at an increasing amplitudes, and fit $\sigma_*$ at S/N levels of 5, 10, 25, 50, 75, and 100.  We generate and fit 10 mock spectra per \ion{Fe}{ii} amplitude and S/N level.  We define the \ion{Fe}{ii} fraction as a function of stellar continuum amplitude, such that when \ion{Fe}{ii} amplitude is equal to the stellar continuum amplitude, the \ion{Fe}{ii} fraction is 100\%.  We assume that stellar absorption features are at the same velocity (redshift) as \ion{Fe}{ii} features, and note that subtle differences in velocity may make recovery of $\sigma_*$ more difficult, however, since narrow and broad \ion{Fe}{ii} features are typically blended due to the resolution of SDSS, we can only reliably measure the relative amplitude of \ion{Fe}{ii} emission.  At the very least, recovery of $\sigma_*$ as a function of \ion{Fe}{ii} fraction gives insight as to how stellar template fitting can recover the underlying stellar continuum when broad and narrow emission line features are superimposed on them. \\
\indent  The results of our tests in the recovery of $\sigma_*$ as a function of \ion{Fe}{ii} fraction are shown in Figure \ref{fig:feii_test}.  There is a weak dependence of the best-fit value of $\sigma_*$ as \ion{Fe}{ii} fraction increases, and the variance in $\sigma_*$ increases with decreasing S/N.  Even in the most extreme cases, where \ion{Fe}{ii} fraction exceeds 50\%, $\sigma_*$ can be reliably recovered.  Similarly for the uncertainty in $\sigma_*$, we find that \ion{Fe}{ii} fraction has no discernible effect on the measured uncertainties, and are more dependent on S/N.  While the effects of differing velocities between stellar absorption and \ion{Fe}{ii} features are not taken into consideration, these tests indicate that even extreme fractions of \ion{Fe}{ii} will not significantly affect stellar template matching of the underlying stellar continuum, as long as absorption features are not significantly diluted by the AGN continuum. 

\subsubsection{Recovery of $\sigma_*$ as a function of AGN Continuum Dilution}

\begin{figure*}
 \includegraphics[width=\textwidth,trim={0cm 0cm 0cm 0cm},clip]{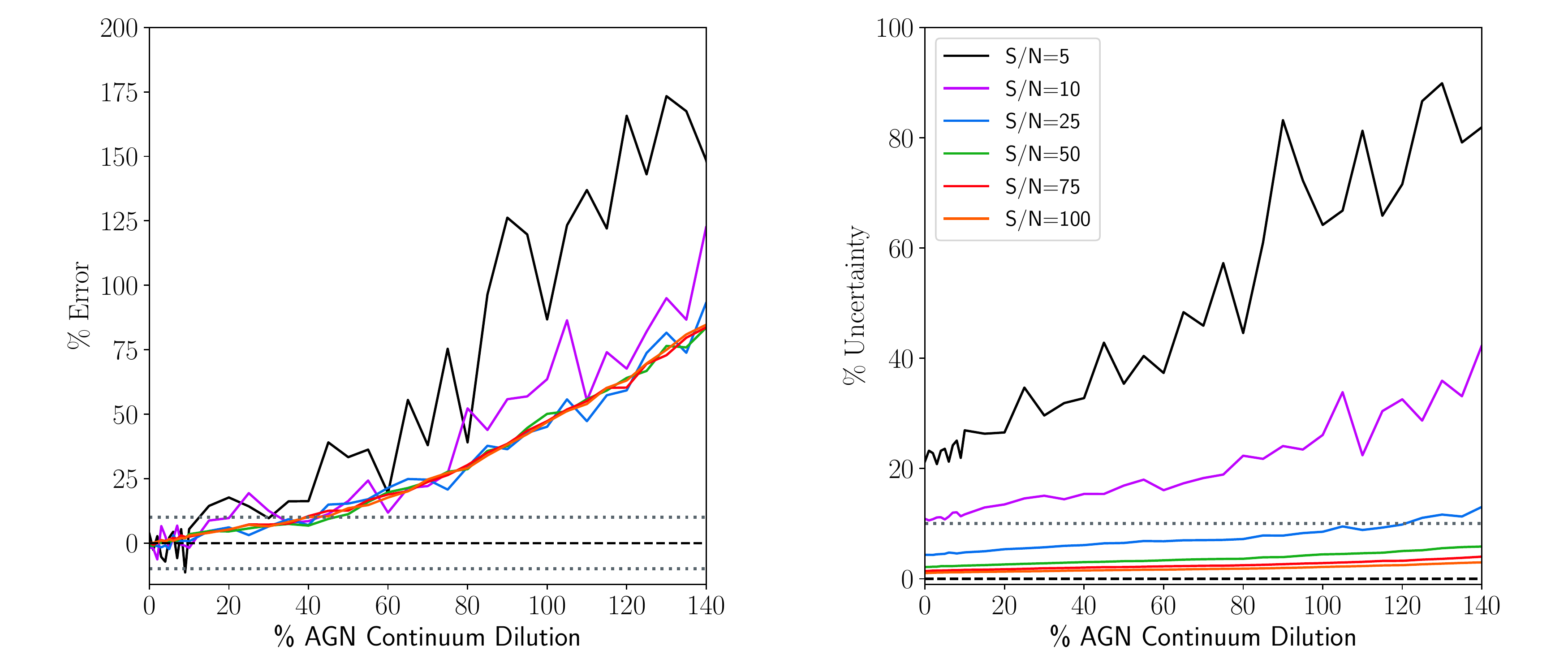}
 \caption{
 The effects of AGN continuum dilution on the recovery of $\sigma_*$ when a power-law continuum model component is not included in the fit.  At all S/N levels, the measured $\sigma_*$ can be biased to larger values at even low levels of continuum dilution.  Inclusion of a power-law component in the fit to the LOSVD completely resolves the percent error attributed to continuum dilution, and reduces the error to that attributed to S/N alone. 
 }
 \label{fig:cont_dil_test}
\end{figure*}

To simulate the effects of AGN continuum dilution, we construct a model spectrum using the MILES SSP 10 Gyr [M/H]$=0.15$ template held at a constant amplitude, followed by a flat ($\alpha_\lambda = 0.0$) continuum of increasing amplitude before re-normalizing the model.  Because young stellar types have a similar continuum shape as an AGN power-law continuum, we do not want the effects of template mismatch to skew measurements of $\sigma_*$ as a function of dilution of stellar absorption features, therefore we initialize the continuum to be flat.\\
\indent We define the percent continuum dilution as the ratio of the amplitude of the AGN continuum to the constant amplitude of the stellar continuum, and generate model spectra of percent continuum dilution ranging from 0\% to 140\%.  We then fit $\sigma_*$ at S/N levels of 5, 10, 25, 50, 75, and 100.  For each S/N level and continuum dilution level, 10 mock spectra are generated to determine a mean percent error from the true value of $\sigma_*$ and mean uncertainty.  For the fit, we only fit the LOSVD and do not include a power-law continuum model component, as we wish to determine the effect of dilution on the measurement of $\sigma_*$. \\
\indent Figure \ref{fig:cont_dil_test} shows the effects of AGN continuum dilution on the recovery of $\sigma_*$.  We find that continuum dilution can have significant effects at both low levels of dilution and is independent of S/N.  Because absorption features are nearly Gaussian (best approximated using Gauss-Hermite polynomials for high-resolution spectra), the correlation between amplitude and width causes the fitting algorithm to fit broader widths to absorption features with shallower depths, caused by the inclusion of the AGN continuum.  The effect of this dilution can be seen as the positive trend in \% error with increasing \% dilution in Figure \ref{fig:cont_dil_test}.  The error in measured $\sigma_*$ will exceed 10\% with only 40\% continuum dilution, and exceeds 50\% at 100\% dilution (when AGN continuum and stellar continuum amplitudes are equal).  We find similar trend for the uncertainties in $\sigma_*$, although the rate of increase is not as dramatic.\\
\indent Fortunately, the inclusion of a power-law continuum model to the fit completely solves the problem of continuum dilution, assuming that the power-law model accurately represents the observed continuum of observed Type 1 AGNs.  The power-law spectrum allows \textsc{pPXF} to accurately recover the true value of $\sigma_*$, at even the highest levels of continuum dilution, and reduces the error to that attributed to S/N alone.  Therefore, we highly recommend including the power-law model if fitting the LOSVD, for both Type 1 and Type 2 AGNs, since dilution need not be accompanied by a strong power-law slope typically observed in Type 1 AGNs. \\
\indent At extremely high levels of continuum dilution, the amplitude of absorption features becomes consistent with the amplitude of noise and other variations in the stellar continuum, and recovery of $\sigma_*$ becomes impossible.  This is observed in Type 1 AGNs with strong power-law continuum component, whose absorption features are typically significantly diluted or not observable.  \\
\indent Additionally, we tested the recovery of $\sigma_*$ as a function of the power-law continuum slope to investigate the effects of possible stellar template mismatch, for example, a steep AGN power-law slope resembling the steep stellar continuum of an O- or B-type star, however, we did not find any significant difficulties in the recovery of $\sigma_*$.\\
\indent For objects that exhibit strong \ion{Fe}{ii}, such as NLS1 \citep{Veron-Cetty2001,Xu2012} or Broad Absorption Line (BAL) objects \citep{Boroson1992,Zhang2010}, we find that strong \ion{Fe}{ii} is usually accompanied by a steep power-law continuum, indicating very strong AGN continuum fraction and thus dilution.  Our tests indicate that it isn't necessarily the presence of \ion{Fe}{ii} or a steep power-law slope, but the presence of strong continuum dilution that makes it nearly impossible to recover the LOSVD in NLS1 or BAL objects, and extra caution should be used when interpreting LOSVD fitting results from these types of objects.\\

% \begin{figure*}
%  \includegraphics[width=\textwidth]{BADASS_cont_slope_tests.pdf}
%  \caption{
% d 
%  }
%  \label{fig:cont_slope_test}
% \end{figure*}

\section{Application of BADASS: Correlation Analysis of [\ion{O}{iii}] outflows}\label{sec:Application}

\subsection{Motivation}\label{sec:motivation}

The emergence of scaling relations between supermassive black holes (henceforth, BHs) and their respective host galaxies implies that there is a fundamental mechanism that regulates their co-evolution \citep{Kormendy2013,DeGraf2015}, however the source of this mechanism remains poorly understood. Large statistical studies of galaxies have since established that galactic-scale outflows are commonplace in galaxies that harbor AGNs, hinting that AGN-driven outflows are strong candidates as the feedback messengers between BHs and their host galaxies \citep{Woo2016,Rakshit2018,Wang2018,DiPompeo2018}.  There is some observational evidence and theoretical arguments that point to the AGN as the central engine powering galactic scale outflows \citep{King2015,Fabian2012}.  Additionally, some numerical simulations indicate that AGN feedback can act to disrupt gas cooling and subsequent star formation on galactic scales \citep{Croton2006a,Dubois2013,Costa2020}, which could give rise to the scaling relations we observe today.  \\
\indent Evidence of such feedback is believed to manifest itself at optical wavelengths as a broad flux-excess in the base or wings of ionized gas emission lines.  The flux-excess, which is most easily identified as extended emission in [\ion{O}{iii}]$\lambda5007$, is typically found to be blueshifted with respect the core component of the line, resulting in a significantly asymmetric line profile \citep{Woo2016,Komossa2018}.
These so-called ``blue wing'' outflow components, which have widths ranging from a few hundred to a few thousand kilometers per second \citep{Harrison2014,Zakamska2016,Manzano-King2019}, can be interpreted as outflowing ionized gas that is no longer gravitationally bound to the narrow-line region (NLR) of the galaxy.  The absence of a ``red wing'' in the profile could also indicate significant dust attenuation of outflowing gas moving radially along the line of sight, possibly due to the presence of a galactic disk or AGN torus structure \citep{Bae2016}. \\
\indent Ionized outflows in narrow forbidden emission lines were first identified in early studies of individual radio sources \citep{Grandi1977,Afanasev1980} and it was soon found that signatures of blueshifted outflows were common in larger samples of Seyfert and radio galaxies \citep{Heckman1981,DeRobertis1984,Whittle1985}.  It was \citet{Heckman1981} that first suggested that because the source of radio emission is due to a compact non-thermal central radio source, the outflow emission must originate along the line of sight between the observer and nuclear region of the galaxy.  Later, \citet{Heckman1984} confirmed a relatively strong correlation between radio emission and the presence of outflows which holds true to this day \citep{Jackson1991,Veilleux1991,Brotherton1996,Mullaney2013,Zakamska2014}.  The interest in ionized gas outflows has accelerated within recent years to include extensive IFU observations \citep{MullerSanchez2011,Bae2017,Freitas2018,Wylezalek2020} and hydrodynamical simulations \citep{Melioli2015,Costa2020}. \\
\indent The kinematic properties of ionized gas outflows in relation to other galaxy and AGN properties have also been explored in detail since their discovery.  \citet{Nelson1996} first studied the relationship between the bulge and NLR stellar and gas kinematics, showing that the stellar velocity dispersion is relatively correlated with the [\ion{O}{iii}] gas dispersion, largely due to the gravitational potential of the bulge.  However, they noted that [\ion{O}{iii}] lines with blue wings do not correlate as well with stellar velocity dispersion, indicating the presence of a strong non-gravitational component \citep{Nelson1996,Mullaney2013,Woo2016,Rakshit2018,Wang2018,DiPompeo2018}.  When the blue wing outflow component is properly removed from [\ion{O}{iii}] line profile, and if any possible \ion{Fe}{ii} contamination is accounted for, there is better agreement with stellar velocity dispersion \citep{Boroson2003,Greene2005}.  Although the correlation cannot be used on an object-to-object basis, correcting [\ion{O}{iii}] for outflow components \ion{Fe}{ii} emission provides a means to estimate stellar velocity dispersion for scaling relations such as the $M_{\rm{BH}}-\sigma_*$ relation for larger, higher-redshift statistical samples for which stellar absorption features cannot easily be measured \citep{Wang2001,Boroson2003, Woo2006,Komossa2007,Bennert2018, Sexton2019}.\\
\indent Correlations between the kinematics of the [\ion{O}{iii}] profile and properties of the AGN also exist.  For instance, by studying the combined (core+outflow) [\ion{O}{iii}] profile of large samples of type 2 and type 1 SDSS AGNs, \citet{Woo2016} and \citet{Rakshit2018} found that the launching velocity of outflows increases with AGN luminosity.  Although these studies examined the combined flux-weighted kinematics of the [\ion{O}{iii}] profile, the results of \citet{Bennert2018} imply that the core component of the [\ion{O}{iii}] profile can be independently used to estimate stellar velocity dispersion once the outflow component has been removed.  This invites inquiry as to whether or not the core or outflow components \emph{independently} may exhibit other relationships with each other or host galaxy properties. \\
\indent Since detection and fitting of outflow components in [\ion{O}{iii}] is a feature specifically implemented in BADASS, it presents an opportunity to examine both the core and outflow component kinematics of the [\ion{O}{iii}] line to investigate all correlations related to ionized gas outflows and the host galaxy to further understand the physical interpretation of the emission line profile.  Furthermore, while the use of powerful techniques such as integral field spectroscopy are becoming mainstream for studying the spatially resolved kinematics of AGN host galaxies, these studies are limited to nearby objects.  Our objective here is to study the emission line profile of [\ion{O}{iii}] in local AGNs with outflows to better understand the relationships between outflows, AGNs, and their host galaxies, and apply our knowledge in future studies to objects in the non-local universe for which spatially-resolved observations are not possible.\\
\indent Throughout the following sections, we refer to individual objects in our sample using their truncated object ID, for example, J001335.  We commonly refer to different components of the double-Gaussian ``outflow'' model of the [\ion{O}{iii}] profile as ``core'' and ``outflow'' when referring to specific quantities of each, such as $\sigma_{\rm{core}}$ for the core component velocity dispersion. 

\subsection{Sample Selection}\label{sec:sample_selection}

\begin{figure}
 \includegraphics[width=\columnwidth]{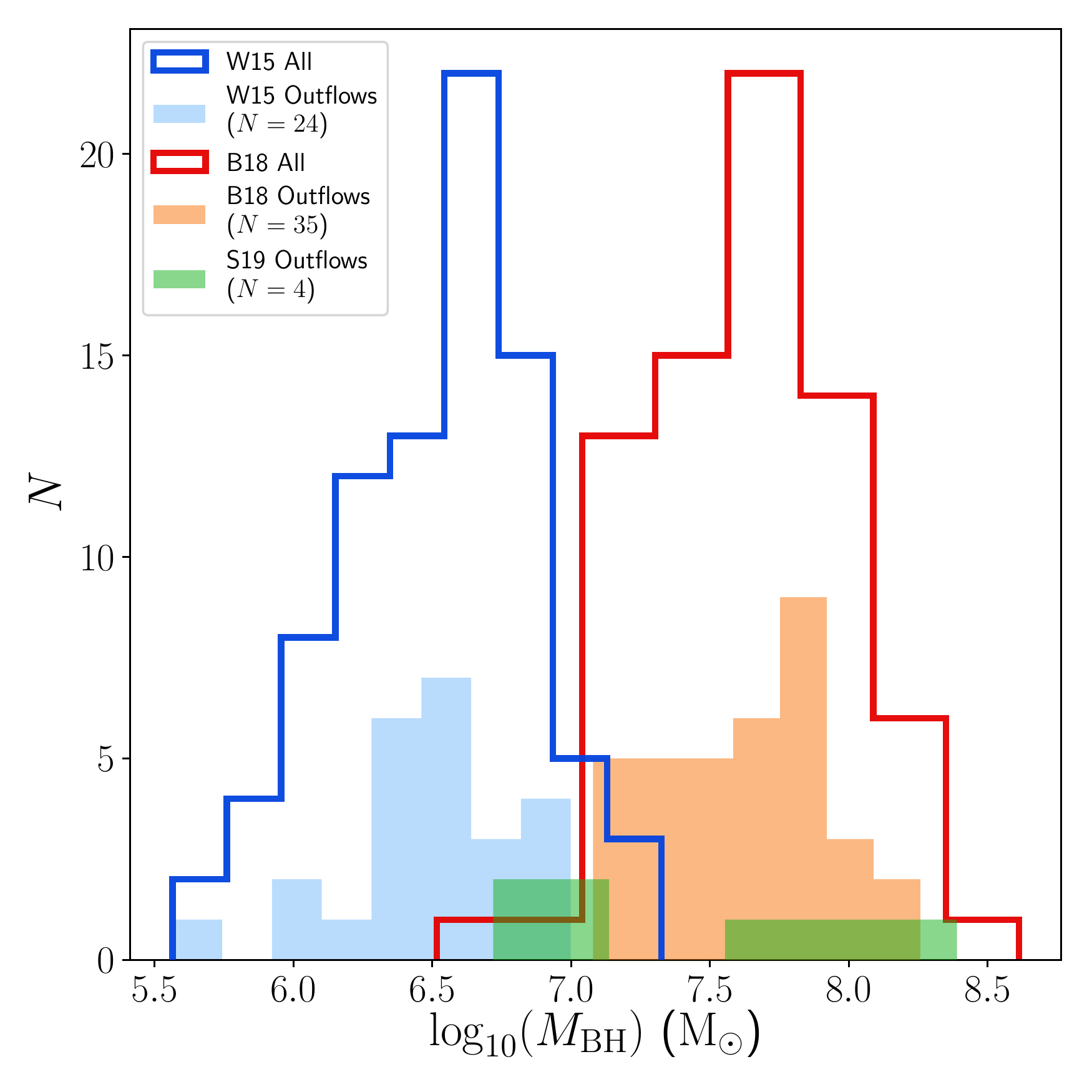}
 \caption{
Histogram of BH mass for the individual samples from B18, W15, and S19 used for our sample of 162 type 1 AGN.  The shaded regions indicate the 63 objects for which we have detected significant non-gravitational outflow signatures in the [\ion{O}{iii}] profile.
 }
 \label{fig:sample_hist}
\end{figure}

Since we wish to investigate the relationships between emission line outflow properties of [\ion{O}{iii}] and both the AGN and host galaxy, we select previously studied nearby type 1 AGNs with SDSS spectroscopy for which BH mass can be measured from the broad H$\beta$ emission and fit these objects using BADASS.  For this we included 81 type 1 AGNs studied by \citet{Bennert2018} (henceforth B18) originally selected from SDSS Data Release 6 (DR6), which selected BH masses $(6.6\leq \log_{10}(M_{\rm{BH}}/M_{\odot}) \leq 8.7)$ at redshifts $(0.02\leq z \leq 0.10)$.  The study from B18 performed detailed follow-up observations with Keck/LRIS and performed a detailed decomposition of the [\ion{O}{iii}] profile to study how different line decompositions affect the $M_{\rm{BH}}-\sigma_*$ relation, however we fit only the SDSS spectra from B18 here for the purposes of benchmarking the capabilities of BADASS.  \\
\indent To extend the sample to lower-mass BHs we include NLS1 objects from \citet{Woo2015} (henceforth W15), which have a BH mass range of $(5.6\leq \log_{10}(M_{\rm{BH}}/M_{\odot}) \leq 7.4)$ and redshift range $(0.01\leq z \leq 0.10)$.  The W15 sample was selected from SDSS DR7 by sequentially selecting objects with ($500 ~\rm{km~s}^{-1} < \rm{FWHM}_{\rm{Br.H}\beta} \leq 2000 ~\rm{km~s}^{-1}$), ($800 ~\rm{km~s}^{-1} < \rm{FWHM}_{\rm{Br.H}\alpha} \leq 2200 ~\rm{km~s}^{-1}$), and a line flux ratio of [\ion{O}{iii}]/H$\beta<3$, resulting in a final sample of 93 NLS1s. \\
\indent  Finally, we include 5 objects from \citet{Sexton2019} (henceforth S19) for which there is sufficient S/N to adequately measure $\sigma_*$ in the SDSS spectra and have previously determined outflow signatures in the [\ion{O}{iii}] profile.  The S19 sample consists of 22 type 1 AGNs observed with Keck-I LRIS comprised of both BLS1s and NLS1s.  The S19 sample has a BH mass range of ($6.3\leq \log_{10}(M_{\rm{BH}}/M_{\odot}) \leq 8.3$) and is comprised of objects in a broad range of redshifts ($0.03\leq z\leq0.57$) used to study evolution in the $M_{\rm{BH}}-\sigma_*$ relation in the non-local universe.  The 5 objects we include here consists of 4 BLS1 objects and one NLS1 object, which have a BH mass range of ($6.9\leq \log_{10}(M_{\rm{BH}}/M_{\odot}) \leq 8.2$) and range in redshift from ($0.09\leq z\leq0.43$) as reported by S19.\\
\indent We removed 16 objects (8 from B18, 8 from W15) for which we could not fit a broad H$\beta$ line, i.e., are type 2 AGNs or have significant host galaxy absorption that makes fitting the broad line highly uncertain.  We also removed one object (J112229) listed twice in Table 1 of W15 after confirming there were no nearby neighbors.  \\
%Since taking into account the stellar absorption features can considerably change previously measured broad H$\beta$ widths, classifications for NLS1s ($500 ~\rm{km~s}^{-1} < \rm{FWHM}_{\rm{Br.H}\beta} \leq 2000  ~\rm{km~s}^{-1}$) and Broad Line Seyfert 1s (BLS1s; $\rm{FWHM}_{\rm{Br.H}\beta} > 2000 ~\rm{km~s}^{-1})$  have also changed.  By this arbitrary definition, the combined sample includes only 61 NLS1s, and 101 BLS1s.  \\
\indent The final sample of 162 objects span a BH mass range of $(5.6\leq \log_{10}(M_{\rm{BH}}/M_{\odot}) \leq 8.7)$ and a average redshift of $z=0.06$.  Of these 162 objects, 76 contain measurable outflows in the [\ion{O}{iii}] line profile as determined using the BADASS outflow criteria given in Section \ref{sec:outflow_tests}.  We also included an additional 6 objects which have some visually identifiable asymmetry in the [\ion{O}{iii}] line profile, which may be attributable to outflows, bringing the total number of objects with outflows to 82.   We plot the distribution of BH mass for all 162 objects in Figure \ref{fig:sample_hist}.

\subsection{Methods}

\subsubsection{Spectral Fitting with BADASS} \label{sec:spectrum_fitting}

\begin{figure}
 \includegraphics[width=\columnwidth]{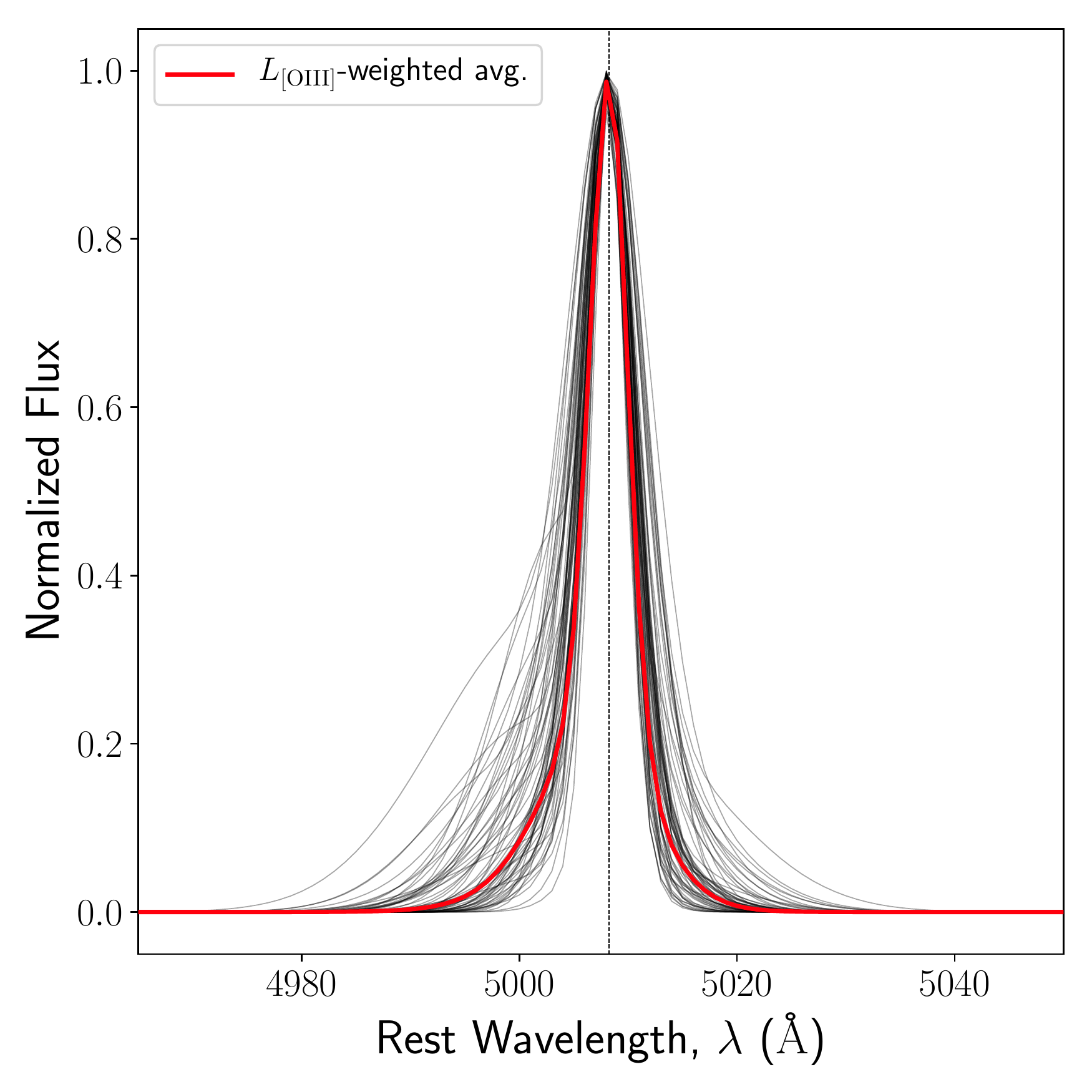}
 \caption{
 Superimposed [\ion{O}{iii}] profiles of all 63 outflow objects in our sample shown in black.  Individual [\ion{O}{iii}] profiles are aligned at the rest frame velocity of the core component and normalized by the maximum amplitude of the full [\ion{O}{iii}] profile.  The luminosity-weighted average is show by the red profile.
 }
 \label{fig:outflow_stack}
\end{figure}

All 162 objects are re-fit with BADASS in two different ways.  The first fit is forced to include outflow components in [\ion{O}{iii}] even if they do not satisfy the outflow criteria used by BADASS.  The second fit is forced to not include outflow components in [\ion{O}{iii}].  Because we wish to investigate how the non-gravitational component of the [\ion{O}{iii}] profile correlates with other galaxy properties, both single- and double-Gaussian profiles must be fit to determine which decomposition produces better agreement with $\sigma_*$.  In both fits we include all other model components, i.e., broad line H$\beta$, narrow and broad \ion{Fe}{ii}, power-law continuum, and the LOSVD, in the wavelength range $(4400\leq\lambda\leq 5800)$ following the methods from S19.  This fitting region is ideal, not only because it contains the emission lines we want to study, but is also large enough to adequately constrain the amplitude of \ion{Fe}{ii} emission such that it can be distinguished from the stellar absorption features near \ion{Mg}{ib} used to estimate the LOSVD.  \\
\indent We allow BADASS to fit for a minimum of 2500 iterations with 100 walkers until the LOSVD parameters (stellar velocity and velocity dispersion), and emission line parameters (amplitude,width, and velocity offset) for the broad H$\beta$, [\ion{O}{iii}] core, and [\ion{O}{iii}] outflow components have achieved convergence at a minimum of 10 times the autocorrelation time and within a 10\% autocorrelation tolerance, with a post-convergence burn-in of 2500 iterations.  We set a maximum iteration ceiling of $50,000$ iterations, however, objects with high S/N and clearly visible outflow profiles in [\ion{O}{iii}] typically converge by $\sim$12,000 iterations, which is actually $\sim$5-10 times the autocorrelation time for the parameters we consider for convergence. \\
\indent Using both the BADASS criteria given in Section \ref{sec:outflow_tests} and by visually inspecting the fits of both the outflow and no-outflow models, we determine that 82 of the 162 objects have outflow components with significant width and offset differences from their core components. \\ 
\indent Finally, since we wish to examine the effects of the non-gravitational outflow component of the [\ion{O}{iii}] profile and compare them to the stronger gravitational component $\sigma_*$, we remove 19 objects for which we do not see an improvement in agreement between the $\sigma_{\rm{core}}$ and $\sigma_*$, which is necessary to remove objects for which BADASS potentially overfit with a double-Gaussian profile and thus no strong non-gravitational outflow component is present.\\
\indent The final sample includes 63 objects with strong non-gravitational kinematics in the [\ion{O}{iii}]$\lambda 5007$ emission line, of which 55 outflow components are blueshifted and 8 outflow components are redshifted relative to their core components.  The 63 objects with strong outflows are shown in the shaded regions of \ref{fig:sample_hist} and listed in Table \ref{tab:measurements} with their relevant measurements. \\
\indent To visualize the diversity of the 63 strong [\ion{O}{iii}] outflows in our sample, we align their full profiles by shifting them to the rest frame velocity of the core component and normalize them by the amplitude of the full profile, as shown in Figure \ref{fig:outflow_stack}, with the luminosity-weighted average shown by the red profile.\\

\subsubsection{Correcting $\sigma_*$ for Disk Inclination}\label{sec:disk_inclination}

The relatively large $3''$ diameter SDSS fiber can cover a significant fraction of the host galaxy, introducing contamination from non-bulge components.  This is of particular concern for $\sigma_*$ measurements on the $M_{\rm{BH}}-\sigma_*$ relation since BH mass does not correlate with the stellar velocity dispersion of disks (see \citet{Kormendy2013} for a review of all BH mass correlations).\\
\indent As such, a significant number of objects in our sample contain disks, which at high inclinations, can artificially increase the measured stellar velocity dispersion, and overestimate values by as much as 25\% \citep{Hartmann2014}.  Using $N$-body simulations, \citet{Bellovary2014} derived a prescription to correct measured $\sigma_*$ to face-on ($i=0$) values using common observables, which depend significantly on the inclination $i$ and rotational velocity $v_{\rm{rot}}$ of the disk.\\
\indent To correct the measured velocity dispersions in our sample, we first obtain disk inclinations and disk scale lengths from \citep{Simard2011}, who performed bulge+disk decompositions of over 1.1 million SDSS galaxies, from which we obtain measurements for 57 objects from our sample of 63.  We then estimate the disk rotational velocities from scale lengths using the SDSS $RV$ Relation from \citet{Hall2012}.  The prescription from \citet{Bellovary2014} depends on the ratio $(v/\sigma)_{\rm{spec}})$, for which we assume a value of 0.6 for a fast-rotating late-type galaxy \citep{FalconBorroso2017} following the same procedure from S19.  We note that varying values of  $(v/\sigma)_{\rm{spec}})$ do not significantly change the correction factor as much as values for $i$ and $v_{\rm{rot}}$.  We propagate all uncertainties in quadrature and assume an additional 10\% uncertainty in correction prescription.  The average change in $\sigma_*$ due to this correction for all of our objects is only 12 km s$^{-1}$, but can be as high as 39 km s$^{-1}$ for the highest of inclinations.  We determine that despite this correction, the overall scatter for $\sigma_*$ in our sample does not change, and that this correction will have a negligible effect on our results.

\subsection{Results}

\begin{figure*}
 \includegraphics[width=\textwidth]{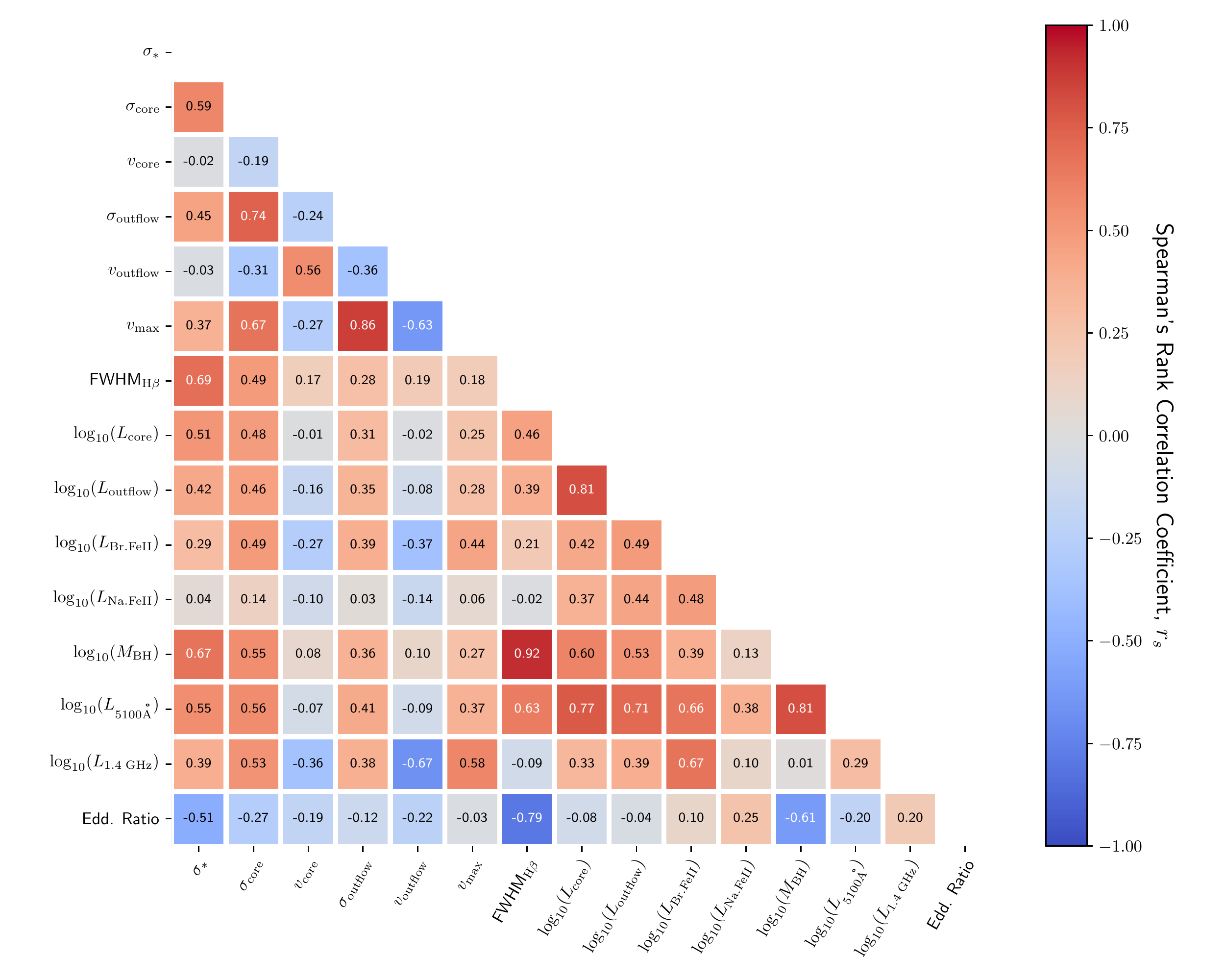}
 \caption{
Correlation matrix of relevant quantities measured with BADASS.  The colorscale represents the absolute Spearman's rank correlation coefficient $r_s$. The average uncertainty for all calculated values of $r_s$ is 0.04. 
}
 \label{fig:corr_matrix}
\end{figure*}

The relevant measurements obtained from spectral fitting with BADASS for the 63 objects in our sample are presented in Table \ref{tab:measurements}.  Calculated AGN luminosities at $5100$ \angstrom\; are obtained via the empirical relation between the luminosity of the broad H$\beta$ emission line and $\lambda L_{\text{5100\angstrom}}$ from \citet{Greene2005}, and $M_{\rm{BH}}$ is calculated using the relation from S19 based on the mass recalibration from reverberation mapping measurements from \citet{Woo2015}.  Black hole masses for the 63 outflow objects span nearly three orders of magnitude from ($5.6\leq \log_{10}(M_{\rm{BH}}/M_{\odot})\leq8.4$).  To quantify the maximal velocity of the outflows, we adopt the relation from \citet{Harrison2014} given by

\begin{equation}\label{eq:v_max}
    v_{\rm{max}}=\Delta v_0+\frac{W_{80}}{2}
\end{equation}

where $v_0$ is the velocity offset of the outflow component measured with respect to the velocity offset of the core component, and $W_{80}=1.09\;\rm{FWHM}$, which represents the width containing 80\% of the Gaussian flux of the outflow component.  Values of $v_{\rm{max}}$ listed in Table \ref{tab:measurements} are negative if the outflow is blueshifted with respect to the core component, and positive if redshifted with respect to the core component.  Velocities for the core and outflow components, $v_{\rm{core}}$ and $v_{\rm{outflow}}$, are reported as velocities with respect to the systemic (stellar) velocity.  All reported dispersion are corrected for the SDSS redshift-dependent instrumental dispersion during the fitting process by BADASS.  The vast majority of our objects ($N=55$) have blueshifted outflow components with respect to their core component.\\
\indent We use these measurements to further investigate correlations of outflows with their host galaxy and AGN to understand their relationship, if any.  For reference, we plot a heatmap of the Spearman's rank correlation coefficient $r_s$ for all relevant and possibly interesting quantities measured with BADASS in Figure \ref{fig:corr_matrix}.  We discuss the most notable correlations in detail in the following subsections.  In the following figures we adopt a consistent colorscale shown in Figure \ref{fig:vvd_diagram}, which represents the absolute value of $v_{\rm{max}}$.

\begin{landscape}
\begin{table}
\caption{BADASS measurements of relevant quantities for outflow, host galaxy, and AGN properties.  Column 1: SDSS object designation.  Column 2: reference; (1) B18, (2) W15, (3) S19.  Column 3: systemic redshift determined using stellar kinematics.  Column 4: stellar velocity dispersion.  Column 5: [\ion{O}{iii}] core component systemic velocity.  Column 6: [\ion{O}{iii}] core component velocity dispersion.  Column 7: [\ion{O}{iii}] core component luminosity.  Column 8:  [\ion{O}{iii}] outflow component systemic velocity.  Column 9: [\ion{O}{iii}] outflow component velocity dispersion.  Column 10: [\ion{O}{iii}] outflow component luminosity.  Column 11: maximal outflow velocity measured using $W_{80}$.  Column 12: FWHM of the broad H$\beta$ emission line.  Column 13: AGN luminosity at 5100 \angstrom\; measured using the relation from \citet{Greene2005}.  Column 14: BH mass estimated using relation from S19. 
}
\label{tab:measurements}
\begin{tabular}{llcccccccccccc}
\hline
Object & Ref. & $z$ & $\sigma_*$ & $v_{\rm{core}}$ & $\sigma_{\rm{core}}$ & $L_{\rm{core}}$ & $v_{\rm{outflow}}$  & $\sigma_{\rm{outflow}}$ & $L_{\rm{outflow}}$ & $v_{\text{max}}$ & FWHM$_{\text{H}\beta}$ & $\lambda L_{\text{5100\angstrom}}$ & $\log(M_{\rm{BH}})$ \\

(SDSS) & & & (km s$^{-1}$) & (km s$^{-1}$) & (km s$^{-1}$) & (erg s$^{-1}$) & (km s$^{-1}$) & (km s$^{-1}$) & (erg s$^{-1}$) & (km s$^{-1}$) & (km s$^{-1}$) & (erg s$^{-1}$) & ($M_{\odot}$)\\
\hline
J000338.94+160220.6		&	3	&	0.11668		&	$91_{-13}^{+13}$	&	$-129_{-11}^{+11}$	& 	$107_{-4}^{+4}$		&	$41.01_{-0.02}^{+0.02}$		&	$-331_{-26}^{+24}$	&	$397_{-21}^{+21}$	&	$40.90_{-0.03}^{+0.03}$		&	$-712_{-36}^{+35}$		&	$3513_{-114}^{+116}$	&	$43.38_{-0.01}^{+0.01}$		&	$7.63_{-0.13}^{+0.19}$ \\
J001335.38-095120.9		&	1	&	0.06196		&	$55_{-18}^{+18}$	&	$-38_{-13}^{+14}$	& 	$127_{-11}^{+11}$	&	$40.25_{-0.05}^{+0.05}$		&	$-284_{-59}^{+45}$	&	$392_{-36}^{+40}$	&	$40.28_{-0.06}^{+0.05}$		&	$-749_{-75}^{+64}$		&	$3587_{-50}^{+53}$		&	$43.34_{-0.01}^{+0.01}$		&	$7.62_{-0.13}^{+0.19}$ \\
J010939.01+005950.4		&	1	&	0.09376		&	$105_{-14}^{+15}$	&	$-198_{-11}^{+11}$	& 	$119_{-3}^{+3}$		&	$41.26_{-0.01}^{+0.01}$		&	$-398_{-16}^{+16}$	&	$393_{-12}^{+13}$	&	$41.17_{-0.01}^{+0.01}$		&	$-705_{-20}^{+20}$		&	$3279_{-96}^{+102}$		&	$43.23_{-0.02}^{+0.02}$		&	$7.47_{-0.12}^{+0.19}$ \\
J012159.81-010224.4		&	1	&	0.05484		&	$116_{-12}^{+12}$	&	$-130_{-12}^{+12}$	& 	$117_{-3}^{+3}$		&	$41.38_{-0.02}^{+0.02}$		&	$-303_{-13}^{+12}$	&	$272_{-3}^{+3}$		&	$41.50_{-0.01}^{+0.01}$		&	$-521_{-7}^{+7}$		&	$4210_{-47}^{+51}$		&	$43.53_{-0.00}^{+0.00}$		&	$7.86_{-0.13}^{+0.20}$ \\
J021257.59+140610.0		&	1	&	0.06228		&	$120_{-9}^{+10}$	&	$-88_{-5}^{+5}$		& 	$138_{-3}^{+3}$		&	$40.83_{-0.02}^{+0.01}$		&	$-290_{-27}^{+25}$	&	$413_{-30}^{+35}$	&	$40.51_{-0.03}^{+0.03}$		&	$-732_{-46}^{+46}$		&	$4221_{-89}^{+90}$		&	$43.14_{-0.01}^{+0.01}$		&	$7.65_{-0.13}^{+0.18}$ \\
J024912.86-081525.7		&	2	&	0.02983		&	$20_{-11}^{+11}$	&	$-72_{-5}^{+5}$		& 	$62	_{-4}^{+4}$		&	$39.60_{-0.02}^{+0.02}$		&	$-212_{-24}^{+22}$	&	$303_{-20}^{+23}$	&	$39.42_{-0.03}^{+0.03}$		&	$-529_{-35}^{+34}$		&	$891_{-38}^{+38}$		&	$41.76_{-0.03}^{+0.02}$		&	$5.56_{-0.13}^{+0.22}$ \\
J030124.26+011022.8		&	1	&	0.07216		&	$94_{-11}^{+11}$	&	$-172_{-10}^{+10}$	& 	$120_{-6}^{+6}$		&	$40.50_{-0.02}^{+0.02}$		&	$-560_{-41}^{+37}$	&	$487_{-30}^{+32}$	&	$40.51_{-0.03}^{+0.03}$		&	$-1013_{-56}^{+53}$		&	$3263_{-66}^{+65}$		&	$43.22_{-0.02}^{+0.02}$		&	$7.47_{-0.13}^{+0.19}$ \\
J030144.19+011530.8		&	1	&	0.07558		&	$98_{-10}^{+10}$	&	$-208_{-8}^{+8}$	& 	$131_{-5}^{+5}$		&	$40.89_{-0.02}^{+0.02}$		&	$-517_{-18}^{+16}$	&	$345_{-8}^{+8}$		&	$41.00_{-0.02}^{+0.02}$		&	$-752_{-21}^{+19}$		&	$3622_{-45}^{+47}$		&	$43.42_{-0.01}^{+0.01}$		&	$7.68_{-0.14}^{+0.21}$ \\
J030417.78+002827.2		&	2	&	0.04488		&	$55_{-8}^{+8}$		&	$-81_{-5}^{+5}$		& 	$63	_{-4}^{+4}$		&	$40.27_{-0.03}^{+0.03}$		&	$-172_{-11}^{+10}$	&	$166_{-6}^{+6}$		&	$40.26_{-0.03}^{+0.03}$		&	$-304_{-13}^{+12}$		&	$1505_{-27}^{+27}$		&	$42.80_{-0.01}^{+0.01}$		&	$6.58_{-0.14}^{+0.22}$ \\
J073106.86+392644.5		&	2	&	0.04894		&	$32_{-10}^{+10}$	&	$-134_{-6}^{+6}$	& 	$107_{-3}^{+3}$		&	$40.17_{-0.01}^{+0.01}$		&	$-363_{-10}^{+10}$	&	$318_{-5}^{+5}$		&	$40.28_{-0.01}^{+0.01}$		&	$-637_{-11}^{+11}$		&	$1492_{-34}^{+35}$		&	$42.38_{-0.01}^{+0.01}$		&	$6.36_{-0.14}^{+0.22}$ \\
J073505.65+423545.7		&	3	&	0.08644		&	$45_{-13}^{+12}$	&	$-55_{-7}^{+7}$		& 	$76	_{-6}^{+6}$		&	$40.59_{-0.05}^{+0.05}$		&	$-156_{-12}^{+11}$	&	$174_{-6}^{+7}$		&	$40.70_{-0.04}^{+0.04}$		&	$-324_{-14}^{+13}$		&	$1712_{-60}^{+61}$		&	$42.85_{-0.02}^{+0.02}$		&	$6.72_{-0.13}^{+0.19}$ \\
J073703.28+424414.6		&	1	&	0.08861		&	$102_{-10}^{+10}$	&	$-46_{-9}^{+8}$		& 	$135_{-3}^{+3}$		&	$41.27_{-0.02}^{+0.01}$		&	$-249_{-27}^{+23}$	&	$318_{-22}^{+24}$	&	$40.83_{-0.04}^{+0.04}$		&	$-610_{-38}^{+36}$		&	$4004_{-65}^{+63}$		&	$43.33_{-0.01}^{+0.01}$		&	$7.71_{-0.14}^{+0.21}$ \\
J073714.28+292634.1		&	2	&	0.08029		&	$83_{-10}^{+9}$		&	$-160_{-10}^{+10}$	& 	$108_{-12}^{+9}$	&	$40.41_{-0.07}^{+0.06}$		&	$-353_{-103}^{+69}$	&	$231_{-42}^{+41}$	&	$40.07_{-0.15}^{+0.17}$		&	$-488_{-117}^{+88}$		&	$2395_{-189}^{+218}$	&	$42.62_{-0.04}^{+0.05}$		&	$6.89_{-0.15}^{+0.21}$ \\
J080243.40+310403.3		&	1	&	0.04130		&	$99_{-8}^{+8}$		&	$-26_{-6}^{+6}$		& 	$104_{-3}^{+3}$		&	$40.69_{-0.02}^{+0.01}$		&	$-199_{-44}^{+36}$	&	$301_{-41}^{+48}$	&	$40.07_{-0.06}^{+0.06}$		&	$-559_{-69}^{+64}$		&	$5511_{-80}^{+81}$		&	$43.22_{-0.01}^{+0.01}$		&	$7.93_{-0.14}^{+0.22}$ \\
J081718.55+520147.7		&	2	&	0.03911		&	$42_{-13}^{+12}$	&	$-156_{-7}^{+7}$	& 	$59	_{-4}^{+3}$		&	$40.20_{-0.02}^{+0.02}$		&	$-101_{-20}^{+25}$	&	$205_{-24}^{+28}$	&	$39.63_{-0.07}^{+0.09}$		&	$318_{-36}^{+39}$		&	$1941_{-55}^{+58}$		&	$42.39_{-0.01}^{+0.01}$		&	$6.59_{-0.14}^{+0.19}$ \\
J082912.68+500652.3		&	2	&	0.04373		&	$75_{-7}^{+8}$		&	$-56_{-6}^{+6}$		& 	$77	_{-1}^{+1}$		&	$40.79_{-0.00}^{+0.00}$		&	$-225_{-10}^{+10}$	&	$358_{-9}^{+10}$	&	$40.38_{-0.01}^{+0.01}$		&	$-627_{-14}^{+14}$		&	$1017_{-22}^{+23}$		&	$42.54_{-0.01}^{+0.01}$		&	$6.10_{-0.14}^{+0.22}$ \\
J085504.16+525248.3		&	2	&	0.08994		&	$76_{-9}^{+9}$		&	$-171_{-13}^{+15}$	& 	$206_{-22}^{+18}$	&	$40.60_{-0.09}^{+0.07}$		&	$-309_{-42}^{+32}$	&	$522_{-68}^{+80}$	&	$40.60_{-0.07}^{+0.08}$		&	$-808_{-98}^{+94}$		&	$2092_{-93}^{+104}$		&	$42.96_{-0.02}^{+0.02}$		&	$6.95_{-0.14}^{+0.22}$ \\
J090902.35+133019.4		&	1	&	0.05005		&	$88_{-10}^{+10}$	&	$-69_{-10}^{+10}$	& 	$92	_{-13}^{+15}$	&	$39.72_{-0.07}^{+0.10}$		&	$4_{-19}^{+24}$		&	$249_{-19}^{+27}$	&	$39.89_{-0.07}^{+0.05}$		&	$392_{-31}^{+34}$		&	$3524_{-134}^{+147}$	&	$42.49_{-0.02}^{+0.02}$		&	$7.15_{-0.13}^{+0.20}$ \\
J092343.00+225432.7		&	1	&	0.03357		&	$137_{-6}^{+7}$		&	$-157_{-8}^{+8}$	& 	$146_{-3}^{+3}$		&	$41.16_{-0.01}^{+0.01}$		&	$-233_{-8}^{+8}$	&	$417_{-6}^{+7}$		&	$41.37_{-0.01}^{+0.01}$		&	$-612_{-9}^{+9}$		&	$3598_{-26}^{+26}$		&	$43.69_{-0.00}^{+0.00}$		&	$7.81_{-0.14}^{+0.21}$ \\
J093259.60+040506.0		&	1	&	0.05990		&	$103_{-5}^{+4}$		&	$-133_{-5}^{+5}$	& 	$89	_{-4}^{+4}$		&	$40.43_{-0.03}^{+0.03}$		&	$-282_{-25}^{+20}$	&	$235_{-20}^{+22}$	&	$40.20_{-0.05}^{+0.05}$		&	$-450_{-36}^{+33}$		&	$4829_{-194}^{+214}$	&	$42.74_{-0.02}^{+0.02}$		&	$7.56_{-0.14}^{+0.22}$ \\
J094057.19+032401.2		&	2	&	0.06122		&	$63_{-12}^{+12}$	&	$-134_{-9}^{+9}$	& 	$90	_{-9}^{+9}$		&	$40.39_{-0.04}^{+0.04}$		&	$-291_{-38}^{+32}$	&	$335_{-28}^{+32}$	&	$40.34_{-0.05}^{+0.05}$		&	$-587_{-52}^{+48}$		&	$1577_{-99}^{+101}$		&	$42.69_{-0.02}^{+0.02}$		&	$6.56_{-0.14}^{+0.22}$ \\
J094529.36+093610.4		&	2	&	0.01394		&	$80_{-6}^{+6}$		&	$-167_{-4}^{+4}$	& 	$114_{-2}^{+2}$		&	$40.15_{-0.01}^{+0.01}$		&	$-247_{-6}^{+6}$	&	$300_{-7}^{+7}$		&	$39.95_{-0.02}^{+0.02}$		&	$-465_{-10}^{+10}$		&	$2084_{-50}^{+51}$		&	$41.95_{-0.02}^{+0.01}$		&	$6.41_{-0.15}^{+0.22}$ \\
J094838.43+403043.5		&	1	&	0.04771		&	$92_{-11}^{+11}$	&	$-176_{-7}^{+7}$	& 	$103_{-3}^{+2}$		&	$40.78_{-0.01}^{+0.01}$		&	$-304_{-49}^{+40}$	&	$457_{-56}^{+63}$	&	$40.12_{-0.05}^{+0.05}$		&	$-714_{-86}^{+82}$		&	$3374_{-64}^{+66}$		&	$43.07_{-0.01}^{+0.01}$		&	$7.43_{-0.14}^{+0.18}$ \\
J104925.39+245123.7		&	1	&	0.05543		&	$105_{-11}^{+11}$	&	$-81_{-8}^{+8}$		& 	$98	_{-2}^{+2}$		&	$41.15_{-0.01}^{+0.01}$		&	$-166_{-13}^{+13}$	&	$281_{-13}^{+15}$	&	$40.72_{-0.03}^{+0.03}$		&	$-445_{-20}^{+19}$		&	$5072_{-45}^{+47}$		&	$43.48_{-0.00}^{+0.00}$		&	$8.00_{-0.14}^{+0.20}$ \\
J110016.03+461615.2		&	2	&	0.03257		&	$63_{-5}^{+5}$		&	$-156_{-4}^{+4}$	& 	$79	_{-3}^{+3}$		&	$40.30_{-0.02}^{+0.02}$		&	$-256_{-8}^{+7}$	&	$213_{-5}^{+6}$		&	$40.23_{-0.02}^{+0.02}$		&	$-374_{-10}^{+9}$		&	$1433_{-52}^{+53}$		&	$42.17_{-0.02}^{+0.02}$		&	$6.20_{-0.14}^{+0.21}$ \\
J110101.78+110248.8		&	1	&	0.03596		&	$104_{-9}^{+9}$		&	$-69_{-6}^{+6}$		& 	$115_{-3}^{+3}$		&	$40.93_{-0.01}^{+0.01}$		&	$-24_{-8}^{+7}$		&	$320_{-9}^{+9}$		&	$40.80_{-0.02}^{+0.02}$		&	$455_{-12}^{+12}$		&	$6092_{-82}^{+80}$		&	$43.13_{-0.01}^{+0.01}$		&	$7.97_{-0.14}^{+0.20}$ \\
J110456.03+433409.1		&	1	&	0.04952		&	$70_{-7}^{+7}$		&	$-40_{-5}^{+5}$		& 	$71	_{-4}^{+4}$		&	$40.57_{-0.02}^{+0.02}$		&	$44	_{-10}^{+12}$	&	$218_{-12}^{+13}$	&	$40.34_{-0.04}^{+0.04}$		&	$363_{-19}^{+19}$		&	$4031_{-237}^{+253}$	&	$42.46_{-0.03}^{+0.04}$		&	$7.25_{-0.14}^{+0.23}$ \\
J112526.51+022039.0		&	2	&	0.04897		&	$76_{-11}^{+11}$	&	$-57_{-7}^{+7}$		& 	$75	_{-5}^{+5}$		&	$40.31_{-0.04}^{+0.03}$		&	$-99_{-22}^{+18}$	&	$229_{-24}^{+29}$	&	$39.97_{-0.07}^{+0.07}$		&	$-336_{-38}^{+35}$		&	$1618_{-116}^{+121}$	&	$42.31_{-0.03}^{+0.03}$		&	$6.38_{-0.15}^{+0.23}$ \\
J114545.18+554759.6		&	1	&	0.05419		&	$96_{-12}^{+12}$	&	$-159_{-9}^{+8}$	&	$79_{-11}^{+11}$	&	$40.18_{-0.07}^{+0.08}$		&	$-122_{-12}^{+13}$	&	$248_{-22}^{+28}$	&	$40.30_{-0.05}^{+0.05}$		&	$356_{-30}^{+30}$		&	$3765_{-160}^{+168}$	&	$42.67_{-0.03}^{+0.03}$		&	$7.31_{-0.14}^{+0.19}$ \\											
J115333.22+095408.4		&	2	&	0.06965		&	$99_{-9}^{+9}$		&	$-134_{-7}^{+7}$	& 	$123_{-3}^{+2}$		&	$41.19_{-0.01}^{+0.01}$		&	$-260_{-18}^{+16}$	&	$345_{-18}^{+19}$	&	$40.74_{-0.03}^{+0.03}$		&	$-569_{-29}^{+27}$		&	$1937_{-66}^{+69}$		&	$42.90_{-0.02}^{+0.02}$		&	$6.86_{-0.13}^{+0.19}$ \\
J120556.01+495956.4		&	1	&	0.06376		&	$120_{-7}^{+8}$		&	$-170_{-6}^{+6}$	& 	$148_{-2}^{+2}$		&	$41.67_{-0.01}^{+0.01}$		&	$-208_{-9}^{+9}$	&	$390_{-17}^{+18}$	&	$41.05_{-0.03}^{+0.03}$		&	$-539_{-22}^{+22}$		&	$7451_{-166}^{+177}$	&	$43.28_{-0.01}^{+0.01}$		&	$8.24_{-0.13}^{+0.19}$ \\
J120626.29+424426.1		&	1	&	0.05234		&	$119_{-8}^{+8}$		&	$-100_{-7}^{+7}$	& 	$110_{-4}^{+4}$		&	$40.47_{-0.01}^{+0.01}$		&	$-290_{-45}^{+42}$	&	$544_{-57}^{+65}$	&	$40.14_{-0.04}^{+0.03}$		&	$-889_{-86}^{+84}$		&	$3819_{-58}^{+62}$		&	$43.12_{-0.01}^{+0.01}$		&	$7.57_{-0.14}^{+0.17}$ \\
J121044.27+382010.3		&	1	&	0.02319		&	$97_{-6}^{+7}$		&	$-75_{-6}^{+6}$		& 	$103_{-4}^{+4}$		&	$40.70_{-0.02}^{+0.02}$		&	$-63_{-8}^{+8}$		&	$266_{-12}^{+13}$	&	$40.52_{-0.03}^{+0.03}$		&	$354_{-17}^{+17}$		&	$6302_{-98}^{+108}$		&	$43.03_{-0.01}^{+0.01}$		&	$7.96_{-0.14}^{+0.18}$ \\
J123152.04+450442.9		&	1	&	0.06276		&	$140_{-11}^{+11}$	&	$-195_{-11}^{+12}$	& 	$208_{-9}^{+8}$		&	$40.71_{-0.04}^{+0.03}$		&	$-749_{-117}^{+118}$&	$442_{-76}^{+72}$	&	$40.29_{-0.09}^{+0.11}$		&	$-1121_{-152}^{+153}$	&	$2708_{-83}^{+90}$		&	$42.89_{-0.02}^{+0.02}$		&	$7.14_{-0.14}^{+0.18}$ \\
J123228.08+141558.7		&	3	&	0.42747		&	$90_{-38}^{+39}$	&	$-183_{-27}^{+28}$	& 	$154_{-7}^{+7}$		&	$41.97_{-0.03}^{+0.03}$		&	$-225_{-31}^{+32}$	&	$484_{-30}^{+34}$	&	$41.95_{-0.03}^{+0.03}$		&	$-663_{-42}^{+42}$		&	$5760_{-281}^{+301}$	&	$44.01_{-0.02}^{+0.02}$		&	$8.39_{-0.13}^{+0.18}$ \\
J123455.90+153356.2		&	3	&	0.04625		&	$98_{-7}^{+7}$		&	$-47_{-6}^{+6}$		& 	$90	_{-2}^{+2}$		&	$40.83_{-0.01}^{+0.01}$		&	$-240_{-15}^{+14}$	&	$235_{-9}^{+9}$		&	$40.51_{-0.02}^{+0.03}$		&	$-494_{-18}^{+17}$		&	$2514_{-53}^{+55}$		&	$42.88_{-0.02}^{+0.02}$		&	$7.07_{-0.14}^{+0.19}$ \\
J123651.17+453904.1		&	2	&	0.03079		&	$98_{-6}^{+6}$		&	$-130_{-6}^{+5}$	& 	$65	_{-3}^{+3}$		&	$40.12_{-0.01}^{+0.02}$		&	$-186_{-6}^{+6}$	&	$339_{-5}^{+5}$		&	$40.55_{-0.01}^{+0.01}$		&	$-491_{-7}^{+7}$		&	$1964_{-46}^{+47}$		&	$42.50_{-0.01}^{+0.01}$		&	$6.66_{-0.14}^{+0.17}$ \\
\hline
\end{tabular}
\end{table}
\end{landscape}

\begin{landscape}
\begin{table}
\contcaption{}
\begin{tabular}{llcccccccccccc}
\hline
Object & Ref. & $z$ & $\sigma_*$ & $v_{\rm{core}}$ & $\sigma_{\rm{core}}$ & $L_{\rm{core}}$ & $v_{\rm{outflow}}$  & $\sigma_{\rm{outflow}}$ & $L_{\rm{outflow}}$ & $v_{\text{max}}$ & FWHM$_{\text{H}\beta}$ & $\lambda L_{\text{5100\angstrom}}$ & $\log(M_{\rm{BH}})$ \\

(SDSS) & & & (km s$^{-1}$) & (km s$^{-1}$) & (km s$^{-1}$) & (erg s$^{-1}$) & (km s$^{-1}$) & (km s$^{-1}$) & (erg s$^{-1}$) & (km s$^{-1}$) & (km s$^{-1}$) & (erg s$^{-1}$) & ($M_{\odot}$)\\
\hline
J123932.59+342221.3		&	2	&	0.08516		&	$78_{-8}^{+8}$		&	$-184_{-7}^{+7}$	& 	$70	_{-6}^{+7}$		&	$40.27_{-0.03}^{+0.03}$		&	$-450_{-19}^{+16}$	&	$321_{-12}^{+12}$	&	$40.68_{-0.02}^{+0.02}$		&	$-678_{-24}^{+22}$		&	$2198_{-114}^{+120}$	&	$42.85_{-0.04}^{+0.04}$		&	$6.94_{-0.14}^{+0.18}$ \\
J124035.82-002919.4		&	2	&	0.08154		&	$94_{-18}^{+19}$	&	$-133_{-15}^{+15}$	& 	$76	_{-2}^{+2}$		&	$41.28_{-0.01}^{+0.01}$		&	$-171_{-16}^{+16}$	&	$251_{-8}^{+8}$		&	$41.03_{-0.02}^{+0.02}$		&	$-360_{-11}^{+11}$		&	$1553_{-66}^{+69}$		&	$42.81_{-0.02}^{+0.02}$		&	$6.62_{-0.13}^{+0.20}$ \\
J124129.42+372201.9		&	1	&	0.06363		&	$119_{-12}^{+12}$	&	$-41_{-7}^{+7}$		& 	$116_{-3}^{+3}$		&	$41.13_{-0.01}^{+0.01}$		&	$-169_{-25}^{+23}$	&	$390_{-28}^{+30}$	&	$40.66_{-0.03}^{+0.03}$		&	$-629_{-43}^{+42}$		&	$4412_{-90}^{+94}$		&	$43.22_{-0.01}^{+0.01}$		&	$7.74_{-0.13}^{+0.19}$ \\
J132310.39+270140.4		&	1	&	0.05618		&	$70_{-10}^{+10}$	&	$-13_{-6}^{+6}$		& 	$91	_{-6}^{+6}$		&	$40.40_{-0.04}^{+0.04}$		&	$-54_{-8}^{+8}$		&	$251_{-10}^{+11}$	&	$40.54_{-0.03}^{+0.03}$		&	$-362_{-15}^{+15}$		&	$4176_{-198}^{+217}$	&	$42.59_{-0.02}^{+0.02}$		&	$7.34_{-0.13}^{+0.21}$ \\
J135345.93+395101.6		&	1	&	0.06330		&	$134_{-6}^{+6}$		&	$-117_{-6}^{+6}$	& 	$126_{-6}^{+5}$		&	$40.57_{-0.03}^{+0.02}$		&	$-240_{-33}^{+27}$	&	$466_{-64}^{+73}$	&	$40.30_{-0.04}^{+0.04}$		&	$-721_{-88}^{+86}$		&	$6036_{-318}^{+334}$	&	$42.78_{-0.03}^{+0.03}$		&	$7.78_{-0.13}^{+0.20}$ \\
J140514.86-025901.2		&	1	&	0.05460		&	$107_{-12}^{+12}$	&	$-67_{-8}^{+8}$		& 	$116_{-9}^{+9}$		&	$40.31_{-0.05}^{+0.04}$		&	$-175_{-34}^{+26}$	&	$320_{-25}^{+28}$	&	$40.20_{-0.06}^{+0.06}$		&	$-519_{-47}^{+41}$		&	$3689_{-98}^{+100}$		&	$42.95_{-0.01}^{+0.01}$		&	$7.44_{-0.13}^{+0.18}$ \\
J141630.82+013707.9		&	1	&	0.05436		&	$115_{-9}^{+9}$		&	$-137_{-7}^{+7}$	& 	$145_{-8}^{+8}$		&	$40.45_{-0.04}^{+0.04}$		&	$-231_{-17}^{+15}$	&	$411_{-21}^{+24}$	&	$40.55_{-0.03}^{+0.03}$		&	$-620_{-32}^{+31}$		&	$3667_{-156}^{+173}$	&	$42.66_{-0.03}^{+0.03}$		&	$7.28_{-0.14}^{+0.19}$ \\
J141908.30+075449.6		&	1	&	0.05634		&	$168_{-10}^{+10}$	&	$-6	_{-7}^{+7}$		& 	$195_{-4}^{+3}$		&	$41.18_{-0.01}^{+0.01}$		&	$-329_{-34}^{+27}$	&	$449_{-16}^{+16}$	&	$40.80_{-0.03}^{+0.03}$		&	$-898_{-39}^{+33}$		&	$6065_{-294}^{+299}$	&	$42.92_{-0.02}^{+0.02}$		&	$7.86_{-0.14}^{+0.19}$ \\
J143452.45+483942.8		&	1	&	0.03669		&	$110_{-9}^{+9}$		&	$-54_{-7}^{+7}$		& 	$108_{-2}^{+2}$		&	$40.99_{-0.01}^{+0.01}$		&	$53_{-18}^{+20}$	&	$336_{-23}^{+26}$	&	$40.40_{-0.04}^{+0.04}$		&	$538_{-34}^{+35}$		&	$4855_{-47}^{+44}$		&	$43.40_{-0.00}^{+0.00}$		&	$7.93_{-0.14}^{+0.16}$ \\
J152209.56+451124.0		&	2	&	0.06593		&	$95_{-11}^{+11}$	&	$-58_{-11}^{+10}$	& 	$167_{-22}^{+18}$	&	$40.42_{-0.11}^{+0.09}$		&	$-92_{-56}^{+29}$	&	$418_{-71}^{+99}$	&	$40.27_{-0.11}^{+0.14}$		&	$-570_{-107}^{+96}$		&	$2083_{-90}^{+102}$		&	$42.66_{-0.03}^{+0.03}$		&	$6.79_{-0.13}^{+0.19}$ \\
J152324.42+551855.3		&	2	&	0.03987		&	$86_{-10}^{+11}$	&	$-217_{-6}^{+6}$	& 	$86	_{-5}^{+5}$		&	$40.12_{-0.03}^{+0.03}$		&	$-435_{-63}^{+43}$	&	$234_{-31}^{+23}$	&	$39.66_{-0.10}^{+0.08}$		&	$-519_{-75}^{+59}$		&	$2459_{-143}^{+158}$	&	$42.15_{-0.03}^{+0.03}$		&	$6.66_{-0.13}^{+0.21}$ \\
J152940.58+302909.3		&	2	&	0.03641		&	$93_{-5}^{+5}$		&	$-98_{-6}^{+6}$		& 	$97	_{-5}^{+4}$		&	$40.48_{-0.03}^{+0.03}$		&	$-231_{-31}^{+23}$	&	$228_{-20}^{+22}$	&	$40.15_{-0.06}^{+0.07}$		&	$-426_{-40}^{+35}$		&	$2412_{-40}^{+40}$		&	$42.98_{-0.01}^{+0.01}$		&	$7.08_{-0.13}^{+0.18}$ \\
J153552.40+575409.3		&	1	&	0.03077		&	$128_{-12}^{+13}$	&	$-93_{-10}^{+10}$	& 	$130_{-2}^{+2}$		&	$41.44_{-0.01}^{+0.01}$		&	$-171_{-13}^{+12}$	&	$273_{-9}^{+10}$	&	$41.01_{-0.04}^{+0.04}$		&	$-429_{-14}^{+14}$		&	$4447_{-39}^{+38}$		&	$43.57_{-0.00}^{+0.00}$		&	$7.93_{-0.13}^{+0.18}$ \\
J154351.49+363136.7		&	1	&	0.06794		&	$78_{-9}^{+9}$		&	$-163_{-7}^{+6}$	& 	$108_{-3}^{+3}$		&	$41.27_{-0.02}^{+0.02}$		&	$-323_{-10}^{+10}$	&	$267_{-6}^{+6}$		&	$41.18_{-0.02}^{+0.02}$		&	$-502_{-12}^{+12}$		&	$2898_{-44}^{+44}$		&	$43.39_{-0.01}^{+0.01}$		&	$7.46_{-0.13}^{+0.19}$ \\
J154507.53+170951.1		&	1	&	0.04837		&	$119_{-9}^{+10}$	&	$8_{-6}^{+6}$		& 	$103_{-1}^{+1}$		&	$41.12_{-0.01}^{+0.01}$		&	$-119_{-9}^{+9}$	&	$311_{-8}^{+8}$		&	$40.72_{-0.01}^{+0.01}$		&	$-527_{-12}^{+12}$		&	$5501_{-115}^{+112}$	&	$42.99_{-0.01}^{+0.01}$		&	$7.80_{-0.13}^{+0.18}$ \\
J160746.00+345048.9		&	2	&	0.05478		&	$86_{-8}^{+8}$		&	$-200_{-8}^{+7}$	& 	$108_{-4}^{+4}$		&	$40.61_{-0.02}^{+0.02}$		&	$-411_{-11}^{+11}$	&	$424_{-8}^{+9}$		&	$40.84_{-0.01}^{+0.01}$		&	$-755_{-14}^{+14}$		&	$1651_{-44}^{+47}$		&	$42.75_{-0.01}^{+0.01}$		&	$6.63_{-0.13}^{+0.18}$ \\
J161156.30+521116.8		&	1	&	0.04149		&	$108_{-7}^{+7}$		&	$-44_{-6}^{+6}$		& 	$131_{-4}^{+4}$		&	$40.50_{-0.02}^{+0.02}$		&	$-368_{-31}^{+29}$	&	$426_{-21}^{+21}$	&	$40.33_{-0.03}^{+0.03}$		&	$-870_{-41}^{+39}$		&	$3727_{-135}^{+141}$	&	$42.72_{-0.02}^{+0.02}$		&	$7.33_{-0.13}^{+0.19}$ \\
J163159.59+243740.2		&	2	&	0.04384		&	$72_{-6}^{+6}$		&	$-58	_{-6}^{+6}$	& 	$94	_{-2}^{+2}$		&	$40.65_{-0.01}^{+0.01}$		&	$-207_{-24}^{+21}$	&	$278_{-16}^{+16}$	&	$39.97_{-0.04}^{+0.04}$		&	$-506_{-31}^{+28}$		&	$1065_{-44}^{+48}$		&	$42.37_{-0.02}^{+0.02}$		&	$6.07_{-0.14}^{+0.18}$ \\
J163501.46+305412.1		&	2	&	0.05460		&	$95_{-14}^{+15}$	&	$-108_{-9}^{+9}$	& 	$122_{-6}^{+6}$		&	$40.79_{-0.02}^{+0.03}$		&	$-223_{-12}^{+11}$	&	$393_{-11}^{+13}$	&	$40.97_{-0.02}^{+0.01}$		&	$-620_{-17}^{+17}$		&	$2333_{-110}^{+120}$	&	$42.73_{-0.02}^{+0.02}$		&	$6.94_{-0.14}^{+0.18}$ \\
J170859.15+215308.1		&	1	&	0.07277		&	$123_{-12}^{+12}$	&	$-85_{-9}^{+9}$		& 	$164_{-4}^{+4}$		&	$40.94_{-0.01}^{+0.01}$		&	$2_{-15}^{+15}$		&	$551_{-22}^{+24}$	&	$40.86_{-0.02}^{+0.02}$		&	$795_{-31}^{+31}$		&	$6325_{-115}^{+122}$	&	$43.37_{-0.01}^{+0.01}$		&	$8.13_{-0.13}^{+0.18}$ \\
J172759.14+542147.0		&	2	&	0.09989		&	$40_{-16}^{+19}$	&	$-89_{-13}^{+12}$	& 	$73	_{-3}^{+3}$		&	$40.80_{-0.02}^{+0.02}$		&	$-185_{-34}^{+29}$	&	$241_{-31}^{+36}$	&	$40.26_{-0.06}^{+0.07}$		&	$-406_{-51}^{+47}$		&	$1295_{-78}^{+78}$		&	$42.66_{-0.02}^{+0.02}$		&	$6.38_{-0.13}^{+0.20}$ \\
J205822.14-065004.3		&	2	&	0.07413		&	$35_{-13}^{+13}$	&	$-59_{-8}^{+8}$		& 	$96	_{-2}^{+2}$		&	$41.05_{-0.01}^{+0.01}$		&	$-225_{-13}^{+12}$	&	$287_{-8}^{+8}$		&	$40.78_{-0.02}^{+0.02}$		&	$-533_{-14}^{+14}$		&	$1302_{-32}^{+32}$		&	$42.87_{-0.01}^{+0.01}$		&	$6.48_{-0.13}^{+0.19}$ \\
J210226.54+000702.3		&	2	&	0.05222		&	$73_{-8}^{+9}$		&	$-92_{-9}^{+8}$		& 	$77	_{-8}^{+7}$		&	$39.93_{-0.05}^{+0.05}$		&	$-275_{-49}^{+39}$	&	$199_{-28}^{+31}$	&	$39.73_{-0.09}^{+0.09}$		&	$-438_{-60}^{+53}$		&	$1984_{-116}^{+129}$	&	$42.19_{-0.03}^{+0.03}$		&	$6.50_{-0.14}^{+0.20}$ \\
J222246.61-081943.9		&	1	&	0.08312		&	$111_{-8}^{+8}$		&	$-222_{-7}^{+7}$	& 	$172_{-4}^{+4}$		&	$41.16_{-0.01}^{+0.01}$		&	$-599_{-13}^{+13}$	&	$570_{-8}^{+8}$		&	$41.35_{-0.01}^{+0.01}$		&	$-1109_{-16}^{+15}$		&	$3933_{-105}^{+110}$	&	$43.27_{-0.01}^{+0.01}$		&	$7.67_{-0.13}^{+0.18}$ \\
J223338.42+131243.5		&	1	&	0.09438		&	$123_{-15}^{+16}$	&	$-203_{-9}^{+10}$	& 	$150_{-4}^{+4}$		&	$41.40_{-0.02}^{+0.02}$		&	$-258_{-13}^{+12}$	&	$483_{-20}^{+20}$	&	$41.29_{-0.02}^{+0.02}$		&	$-675_{-28}^{+28}$		&	$4326_{-51}^{+55}$		&	$43.66_{-0.00}^{+0.00}$		&	$7.96_{-0.13}^{+0.19}$ \\
J235128.75+155259.1		& 	1	&	0.09675		&	$136_{-11}^{+12}$	&	$-23_{-9}^{+8}$		& 	$98	_{-4}^{+4}$		&	$41.21_{-0.01}^{+0.01}$		&	$-135_{-8}^{+8}$	&	$249_{-2}^{+2}$		&	$41.55_{-0.00}^{+0.00}$		&	$-431_{-4}^{+4}$		&	$7236_{-175}^{+180}$	&	$43.43_{-0.01}^{+0.01}$		&	$8.28_{-0.13}^{+0.19}$ \\		
\hline
\end{tabular}
\end{table}
\end{landscape}

\subsubsection{Correlations with Velocity} \label{sec:vel_correlations}

\begin{figure*}
 \includegraphics[width=\textwidth]{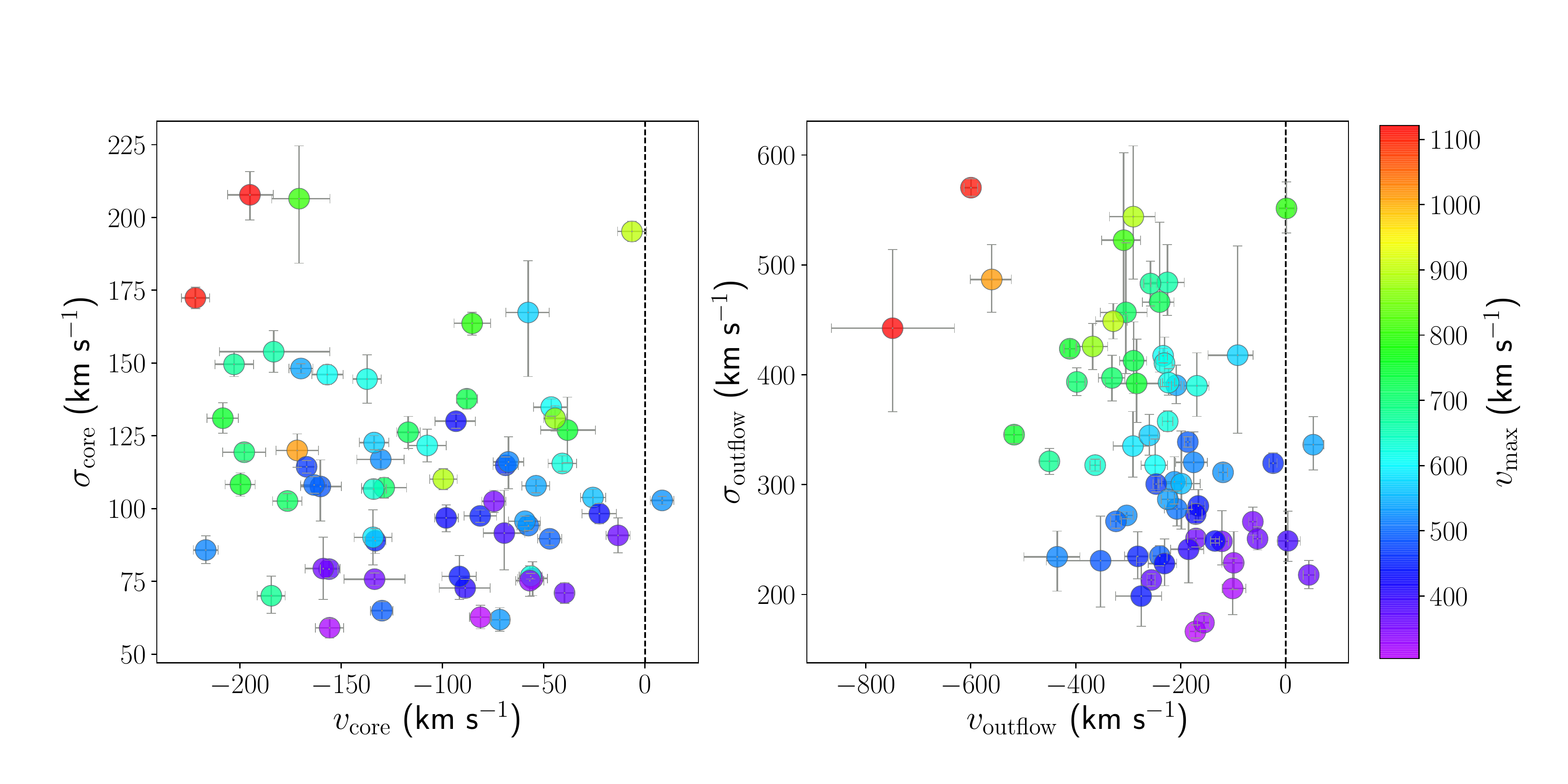}
 \caption{
VVD diagrams for the [\ion{O}{iii}] core and outflow components.  The black dashed line indicates zero velocity offset with respect to the systemic (stellar) velocity. (\emph{Left}):  There is no correlation between $\sigma_{\rm{core}}$ and $v_{\rm{core}}$, however the majority of velocities are blueshifted with respect to the systemic velocity.  (\emph{Right}): There is a significant correlation between $\sigma_{\rm{outflow}}$ and $v_{\rm{outflow}}$, characterized by the distinct ``fan'' shape described by \citet{Woo2016}, caused by increasing extinction with increasing outflow velocity.
}
 \label{fig:vvd_diagram}
\end{figure*}

\begin{figure}
 \includegraphics[width=\columnwidth]{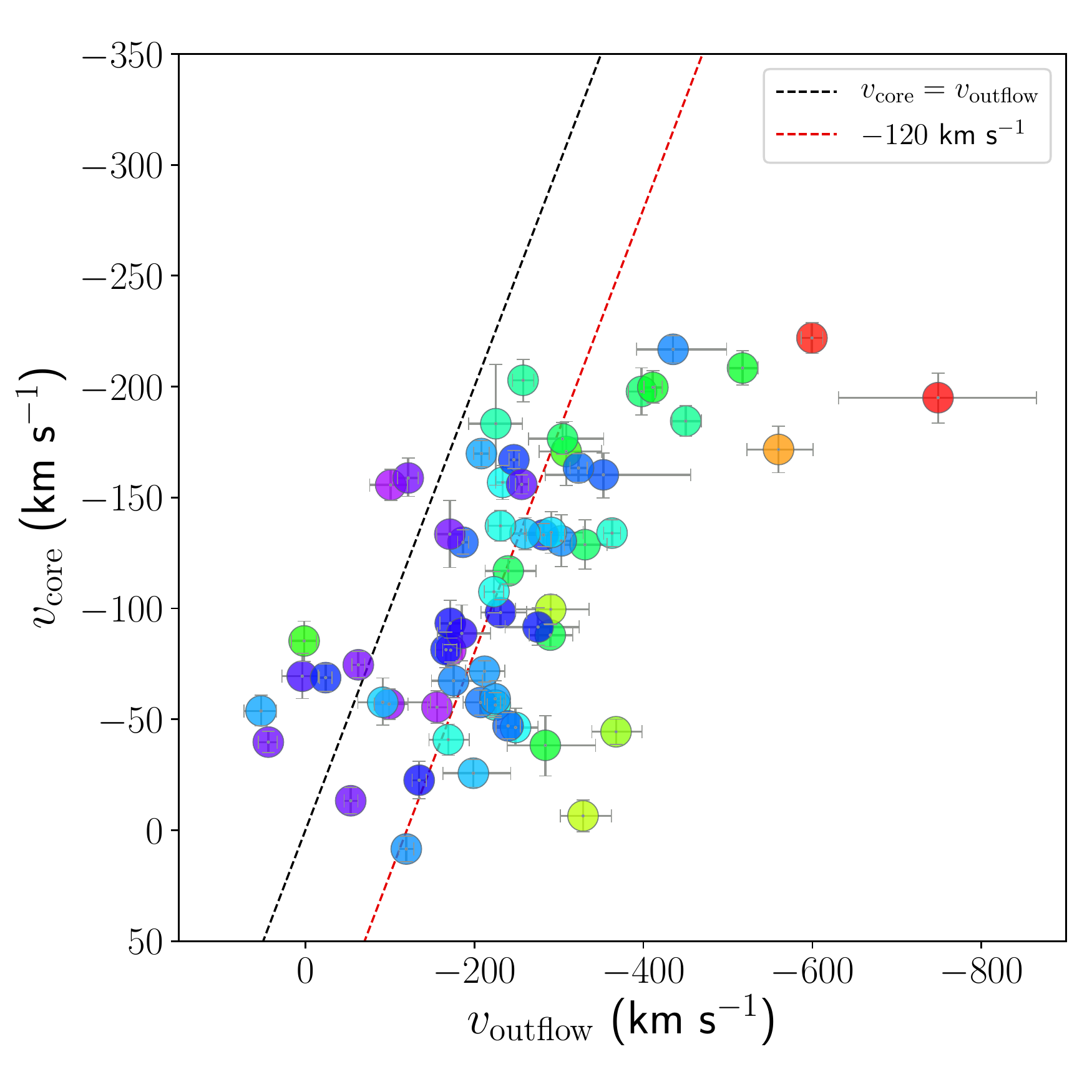}
 \caption{
The strong correlation between $v_{\rm{core}}$ and $v_{\rm{outflow}}$ measured with respect to the systemic (stellar) velocity.  The black dashed line indicates the perfect correlation $v_{\rm{core}}=v_{\rm{outflow}}$, and the red dashed line indicates an average -120 km s$^{-1}$ offset from the perfect correlation for blueshifted outflows.  Objects with larger $v_{\rm{max}}$ appear to deviate from this correlation, however a larger sample of objects with $v_{\rm{max}}>900$ km s$^{-1}$ is needed to conclusively determine if the correlation holes true, and likewise for redshifted outflows.
}
 \label{fig:vel_line_lock}
\end{figure}

We first investigate correlations with the velocities of the [\ion{O}{iii}] core and outflow components.  Following \citet{Woo2016}, we plot Velocity-Velocity Dispersion (VVD) diagrams of the core and outflow components in Figure \ref{fig:vvd_diagram}. The Spearman correlation coefficient for the [\ion{O}{iii}] core VVD diagram is $r_s=-0.19$, indicating no significant correlation, while the [\ion{O}{iii}] outflow VVD diagram shows stronger correlation of $r_s=-0.36$.  The [\ion{O}{iii}] outflow VVD diagram also exhibits the same ``fan'' shape characterized by \citet{Woo2016}, which according to 3D biconical outflow models \citep{Bae2016}, is caused by an increasing extinction due to the presence of an obscuring dust plane.  In theory, if there exists a dust plane that bisects a biconical outflow for which one of the cones points toward the observer along the LOS (the blueshifted cone), the dust plane will obscure the cone on the far side (the redshifted cone) causing an observed blueshifted flux excess.  The fan-shaped VVD diagram for outflows does not appear to extend to the core component.  We also confirm the result from \citet{Rakshit2018} that blueshifted outflows are significantly more common than redshifted outflows for type 1 AGN.  \\
\indent There does appear to be strong correlation ($r_s=0.56$) between $v_{\rm{core}}$ and $v_{\rm{outflow}}$ measured with respect to the systemic (stellar) velocity, as shown in Figure \ref{fig:vel_line_lock}, which shows $v_{\rm{core}}$ scales linearly with $v_{\rm{outflow}}$.  Our data suggest that for blueshifted outflows (right of the dashed line in Figure \ref{fig:vel_line_lock}) there is an average offset of 120 km s$^{-1}$ between the core and outflow components, and there appears to be an increasing offset from $v_{\rm{core}}$ with $v_{\rm{max}}$. A larger sample with  $v_{\rm{max}}>900$ km s$^{-1}$ and objects with redshifted outflows is needed to conclusively determine whether this trend holds, or if there is a value of  $v_{\rm{max}}$ for which this trend no longer holds true.

\subsubsection{Correlations with Dispersion} \label{sec:disp_correlations}

\begin{figure*}
 \includegraphics[width=\textwidth,trim={0cm 0cm 0cm 0cm},clip]{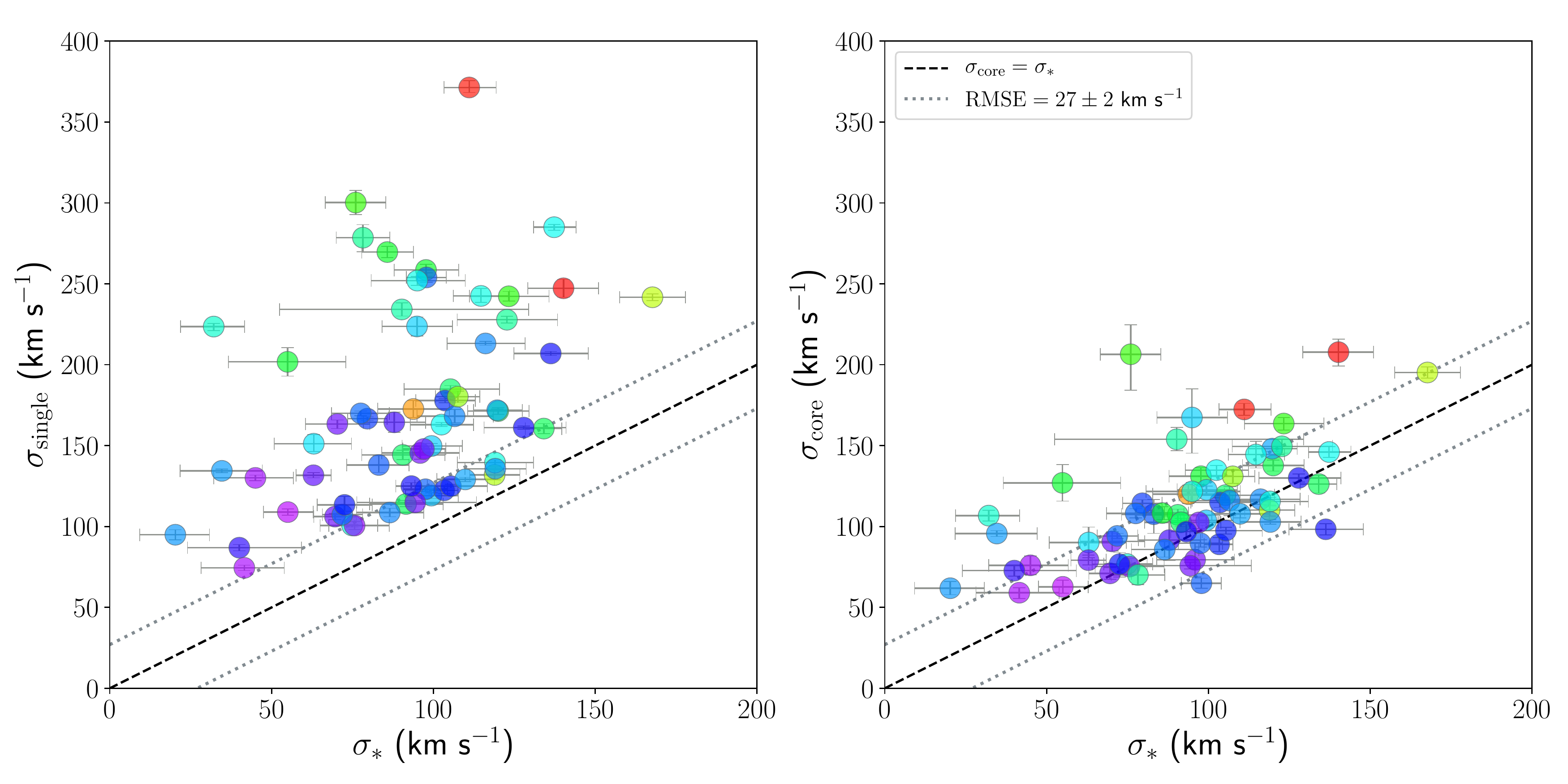}
 \caption{
(\emph{Left}): The single gaussian no-outflow model [\ion{O}{iii}] dispersion $\sigma_{\rm{single}}$ as a function of stellar velocity dispersion $\sigma_*$.  There is a clear offset in objects which exhibit strong outflows with large $v_{\rm{max}}$.  (\emph{Right}):  The double-Gaussian outflow model [\ion{O}{iii}] dispersion as a function of $\sigma_*$.  The gray dotted lines in both plots give the scatter of the $\sigma_{\rm{core}}-\sigma_*$ relation for comparison.  The colorscale is the same as in Figure \ref{fig:vvd_diagram}.  The black dashed line represents perfect correlation, i.e., $\sigma_{\rm{core}}=\sigma_{\rm{*}}$.
 }
 \label{fig:core_vs_stellar}
\end{figure*}

\begin{figure}
 \includegraphics[width=\columnwidth,trim={0cm 0cm 0cm 0cm},clip]{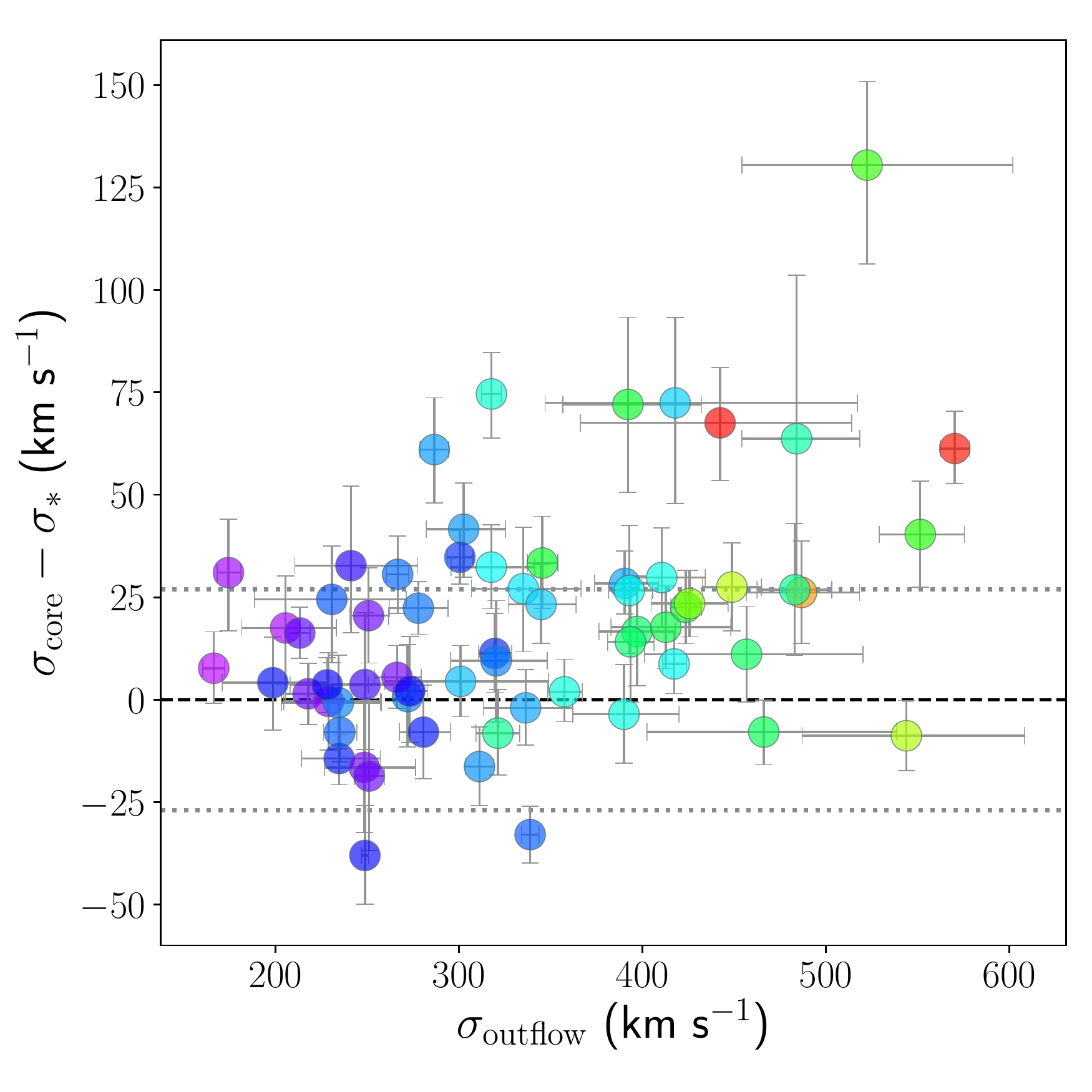}
 \caption{
The difference between the decomposed $\sigma_{\rm{core}}$ and $\sigma_*$ as a function of $\sigma_{\rm{outflow}}$.  There is a clear dependence on how well $\sigma_{\rm{core}}$ traces the gravitational component $\sigma_*$, which appears to scale with $\sigma_{\rm{outflow}}$.
}
 \label{fig:delta_core_stellar_vs_outflow}
\end{figure}

In Figure \ref{fig:core_vs_stellar}, we plot the single-Gaussian ``no-outflow'' model [\ion{O}{iii}] dispersion $\sigma_{\rm{single}}$ alongside the double-Gaussian ``outflow'' model core dispersion $\sigma_{\rm{core}}$, both as a function of $\sigma_*$.  We confirm the results from \citet{Bennert2018} that $\sigma_{\rm{core}}$ correlates more strongly with $\sigma_*$ once the secondary outflow component is accounted for.  Values of $\sigma_{\rm{core}}$ are scattered about the perfect correlation with $\sigma_*$, with a root-mean-square error (RMSE) of $27\pm2$ km s$^{-1}$.  The mean of this distribution of $\sigma_{\rm{core}}$ values is $18\pm2$ km s$^{-1}$, caused primarily by objects with $v_{\rm{max}}>700$ km s$^{-1}$.  \\
\indent In Figure \ref{fig:delta_core_stellar_vs_outflow} we plot the difference ($\sigma_{\rm{core}}-\sigma_*$) as a function of $\sigma_{\rm{outflow}}$ and find that there is some dependence on how well $\sigma_{\rm{core}}$ can recover the gravitational component $\sigma_*$ as a function of $\sigma_{\rm{outflow}}$ and $v_{\rm{max}}$, however a larger sample will also be needed to confirm this trend. 

\begin{figure}
 \includegraphics[width=\columnwidth,trim={0cm 0cm 0cm 0cm},clip]{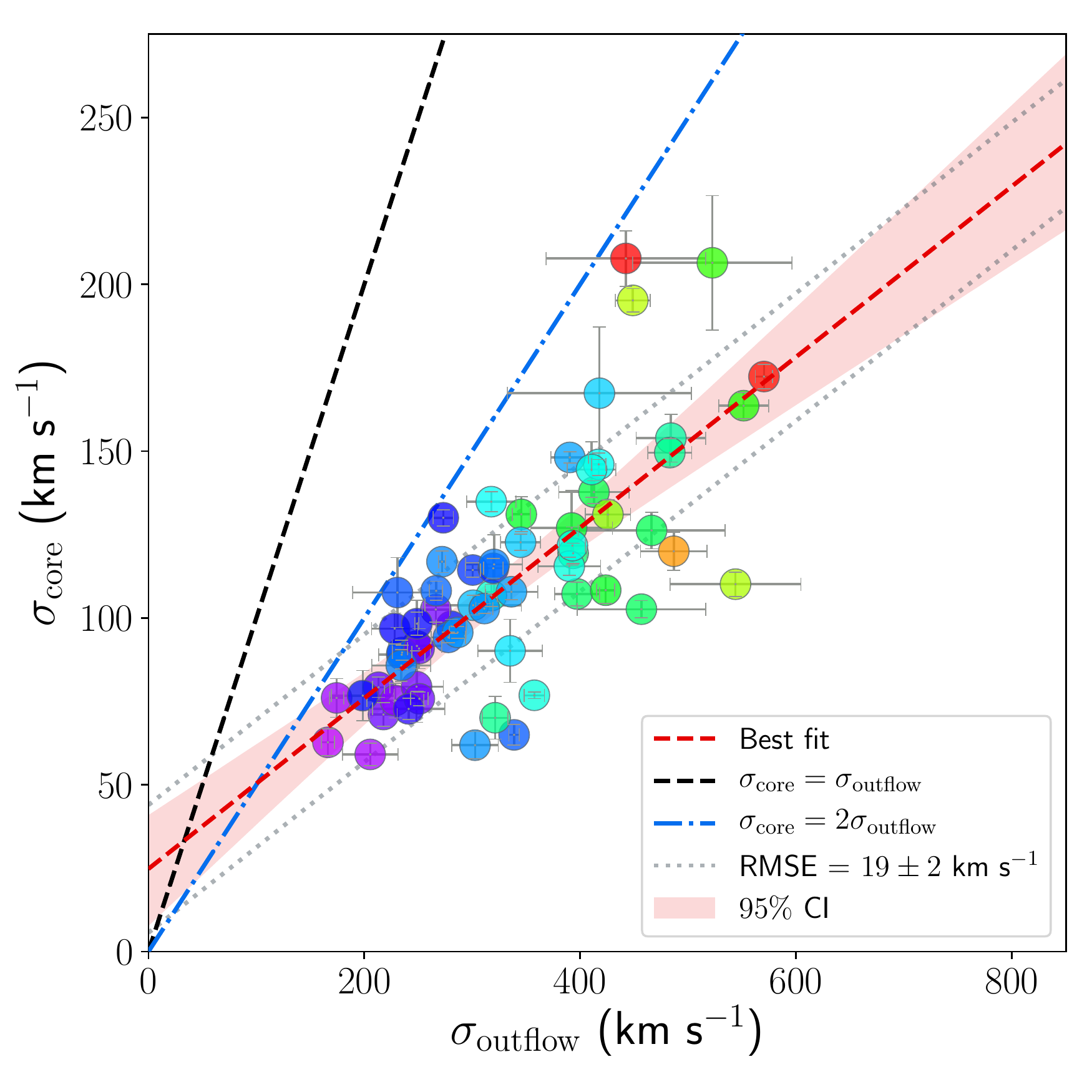}
 \caption{
The correlation between $\sigma_{\rm{core}}$  and $\sigma_{\rm{outflow}}$.  The best-fit regression line is given by the red dashed line, and the shaded red region corresponds to 95\% confidence interval.  The gray dotted lines correspond to the scatter in the relation.  The identity correlation ($\sigma_{\rm{core}}=\sigma_{\rm{outflow}}$) is shown by the black dashed line, and the $\sigma_{\rm{core}}=2\sigma_{\rm{outflow}}$ relation is shown by the blue dashed-dotted line for comparison.
}
 \label{fig:core_vs_outflow}
\end{figure}

\indent There also exists a strong correlation between the [\ion{O}{iii}] core component dispersion $\sigma_{\rm{core}}$ and the outflow component dispersion $\sigma_{\rm{outflow}}$ as first reported by \citet{Zhang2017}, which we plot in Figure \ref{fig:core_vs_outflow} for our sample.  This correlation has so far been largely overlooked, mostly due to the parameterization other studies have used to quantify outflows.  For instance, \citet{Woo2016} performed double-Gaussian decomposition of the [\ion{O}{iii}], but did not study individual dispersions and instead adopted a flux-weighted integrated dispersion for the full line (core+outflow) profile.  The Spearman rank correlation coefficient for this relation is $r_s=0.74$, implying a very strong correlation, and stronger than the $\sigma_{\rm{core}}-\sigma_*$ correlation ($r_s=0.59$).  We perform linear regression using \emph{emcee} following the same methods used in S19, and determine a best-fit slope of $m=0.26\pm0.03$, intercept of $b=25.00_{-8.89}^{+9.05}$ km s$^{-1}$, and intrinsic scatter of $f=19.11_{-2.08}^{+2.16}$, which we plot in Figure \ref{fig:core_vs_outflow}.\\  
\indent It is important to emphasize that the $\sigma_{\rm{core}}-\sigma_{\rm{outflow}}$ correlation is not a result of our definition of an ``outflow'' or our selection criteria for outflows given in Section \ref{sec:outflow_tests}.  Although we define an outflow to have $\sigma_{\rm{outflow}}>\sigma_{\rm{core}}$, which excludes objects above the dashed line ($\sigma_{\rm{core}}=\sigma_{\rm{outflow}}$) in Figure \ref{fig:core_vs_outflow} by design, this does not explain the tightness in the correlation below the dashed line. The outflow criterion for dispersion given in Section \ref{sec:outflow_tests} also does not select objects based on the ratio of $\sigma_{\rm{outflow}}$ and $\sigma_{\rm{core}}$, but by the ratio of the difference of $\sigma_{\rm{outflow}}$ and $\sigma_{\rm{core}}$ and their relative uncertainties.  Recall that 90\% of the objects with outflows were first identified visually from their strong asymmetric profile, implying that we are not overfitting [\ion{O}{iii}] profiles which do not require double-Gaussian decomposition.  We therefore are confident that the $\sigma_{\rm{core}}-\sigma_{\rm{outflow}}$ correlation is real and not an artifact of the fitting process.

\subsubsection{Correlations with Luminosity} \label{sec:lum_correlations}

\begin{figure*}
 \includegraphics[width=\textwidth,trim={0cm 0cm 0cm 0cm},clip]{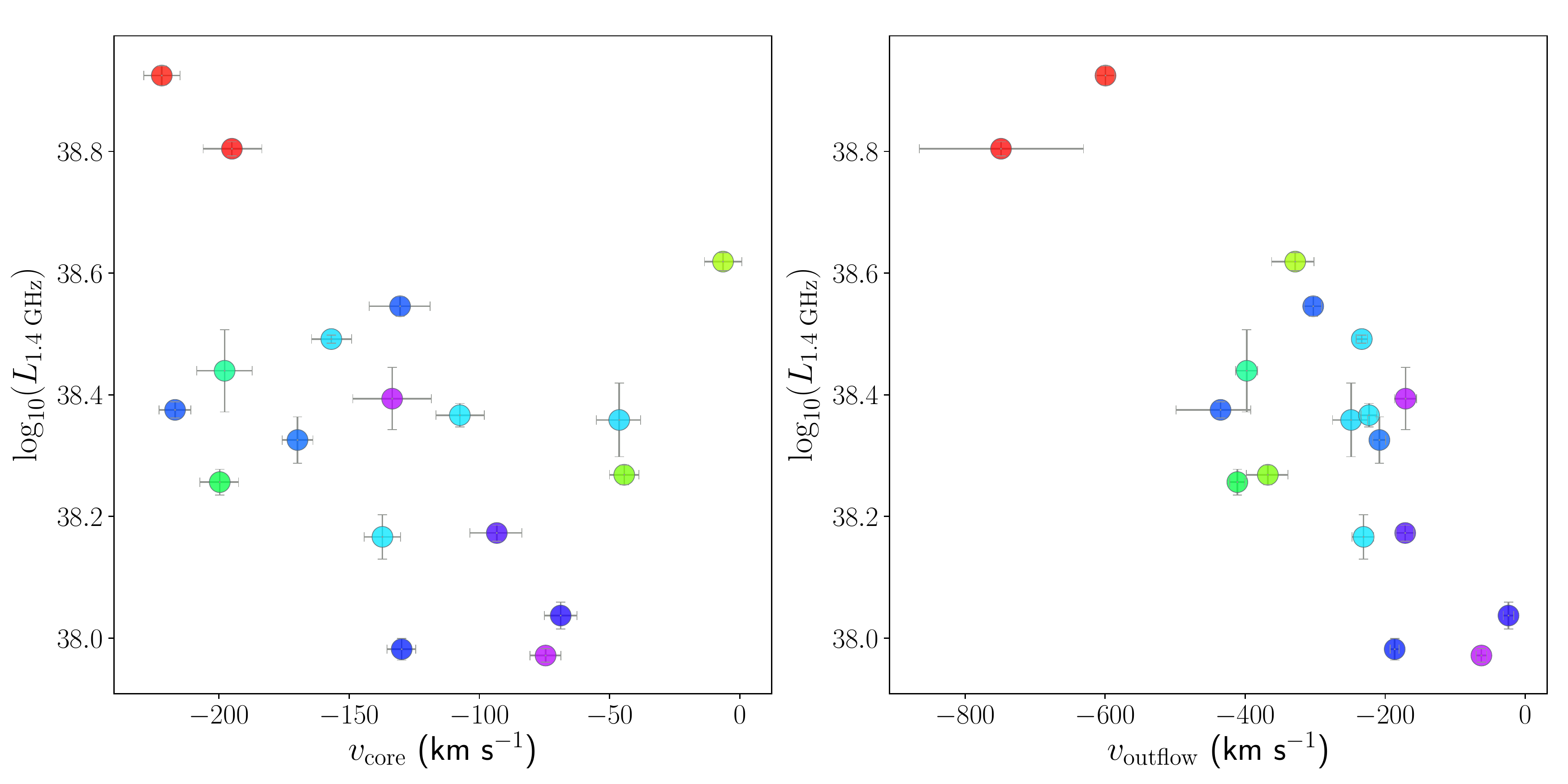}
 \caption{
 Correlations of $L_{\rm{1.4\;GHz}}$ versus $v_{\rm{core}}$ (left) and $v_{\rm{outflow}}$ (right) for 18 objects with available measurements.  The correlation correlation with $\sigma_{\rm{outflow}}$ ($r_s=-0.67$) is nearly twice that of the correlation with $v_{\rm{core}}$.
}
 \label{fig:radio_lum_corr}
\end{figure*}

\indent There is strong correlation between the AGN luminosity at 5100 \angstrom\; ($L_{\text{5100 \angstrom}}$) and $L_{\rm{core}}$ ($r_s=0.77$), and a slightly weaker correlation for $L_{\rm{outflow}}$ ($r_s=0.71$), although the weaker correlation with $L_{\rm{outflow}}$ is likely due to larger uncertainties.  Because we are estimating $L_{\text{5100\angstrom}}$ using the luminosity of the broad H$\beta$ emission line \citep{Greene2005}, correlations with $L_{\text{5100\angstrom}}$ presented here are comparable to correlations with $M_{\rm{BH}}$, which is estimated using both the luminosity and width of the broad H$\beta$ emission line.  However, the correlation between $L_{\text{5100 \angstrom}}$ and outflow kinematics are comparatively weaker. \\
\indent To investigate correlations with the radio luminosity at 1.4 GHz, we obtain $L_{\rm{1.4\;GHz}}$ measurements from the VLA First Survey Catalog \citep{White1997}, which covers 10,575 square degrees of sky for a total of 946,432 radio sources, from which 18 of our 63 outflow objects have measurements.  Referring to Figure \ref{fig:corr_matrix}, the correlations between $L_{\rm{1.4\;GHz}}$ and $\sigma_{\rm{core}}$ and $\sigma_{\rm{outflow}}$ are comparable to their correlations with $L_{\text{5100 \angstrom}}$.  However, when compared to systemic velocities, there is much stronger correlation between $L_{\rm{1.4\;GHz}}$ and $v_{\rm{outflow}}$ ($r_s=-0.67$) than for $v_{\rm{core}}$ ($r_s=-0.36$).  In Figure \ref{fig:radio_lum_corr}, we plot $L_{\rm{1.4\;GHz}}$ as a function of $v_{\rm{core}}$ and $v_{\rm{outflow}}$. \\
\indent The only other notable correlations found between luminosities are those with the luminosity of the broad \ion{Fe}{ii}.  Both $L_{\text{5100\angstrom}}$ and $L_{\rm{1.4\;GHz}}$ correlate strongly with $L_{\rm{Br.FeII}}$, and with nearly identical degrees of correlation of $r_s\sim0.67$. 

\subsubsection{The $M_{\rm{BH}}-\sigma_*$ Relation}

\begin{figure*}
 \includegraphics[width=\textwidth]{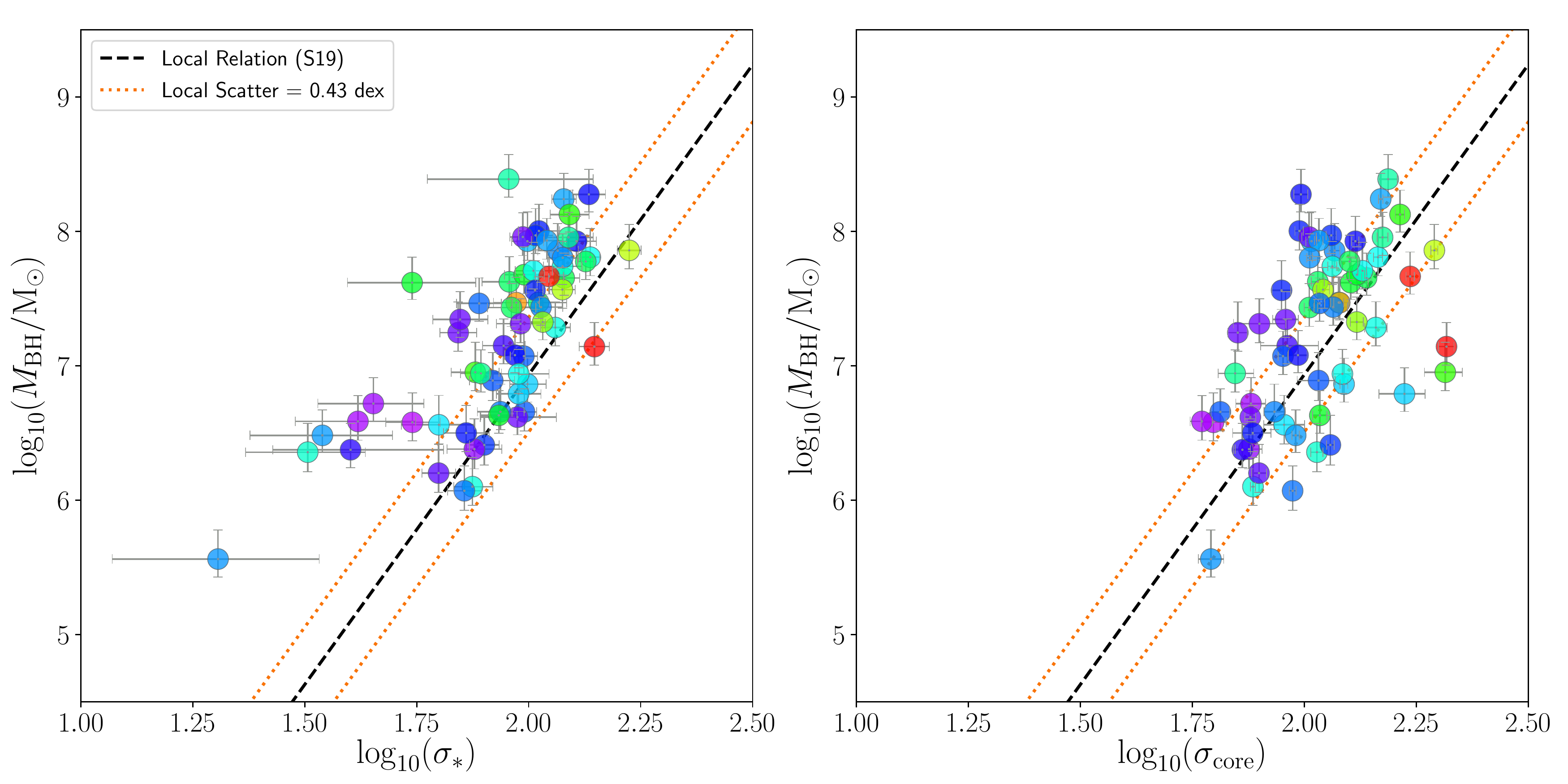}
 \caption{
The $M_{\rm{BH}}-\sigma$ relation as a function of $\sigma_*$ (left) and $\sigma_{\rm{core}}$ (right).  We plot the local relation calculated from S19 (black dashed line) as well as the local scatter (orange dotted lines) for comparison.  The majority of objects on the $M_{\rm{BH}}-\sigma_*$ relation agree with the local relation with some scatter ($f=0.54$ dex), but the mean of the distribution is 0.57 dex above the local relation, caused by significant outliers likely due to poorly understood systematics and data quality.  The $M_{\rm{BH}}-\sigma_{\rm{core}}$ relation also agrees with the local relation, with a comparable amount of scatter ($f=0.58$ dex), mostly caused by stratification in $\sigma_{\rm{core}}$, and a mean of only 0.17 dex above the local relation.  
}
 \label{fig:m_sigma_relation}
\end{figure*}

The ultimate goal of our analysis is to determine the effect - if any - of outflow kinematics on the $M_{\rm{BH}}-\sigma_*$ relation.  In Figure \ref{fig:m_sigma_relation} we plot the $M_{\rm{BH}}-\sigma$ relation using both $\sigma_*$ and $\sigma_{\rm{core}}$, and plot the local relation derived from S19 (black dashed line) and the 0.43 dex scatter (orange dotted line) for comparison. \\
\indent There is considerable scatter in the $M_{\rm{BH}}-\sigma_*$ relation (left of Figure \ref{fig:m_sigma_relation}) for our objects, however the majority of our sample falls within or close to the expected scatter of the relation, with the exception of some outliers above the relation by as much as $\sim1 $ dex.  The total scatter about the relation for our objects is 0.6 dex.  There are considerable uncertainties we cannot account for given the nature of SDSS data that may affect our measurements of $\sigma_*$.  Despite our efforts to correct for the effects of inclination, it is possible that the bulge+disk decomposition performed by \citet{Simard2011} resulted in a poor match to the image PSF, since the decompositions do not take into account the point-spread function (PSF) of the AGN, which would in turn affect measured disk quantities such as ellipticity $(b/a)$ and inclination.  As mentioned in Section \ref{sec:disk_inclination}, the $3''$ diameter SDSS fiber can cover a significant fraction of the host galaxy to include contamination from non-bulge components, which can bias measurements of $\sigma_*$, but also decrease the fraction of light from the AGN.  If the fraction of light from the AGN decreased due to significant host galaxy absorption, we would underestimate the amplitude and therefore overestimate the FWHM of the broad H$\beta$ emission line, leading to an overestimation of $M_{\rm{BH}}$.  Dependencies on AGN continuum dilution also play a role in how well $\sigma_*$ can be recovered from absorption features, as we showed in Figure \ref{fig:cont_dil_test}.  It remains that measurements $\sigma_*$ are one of the most uncertain measurements in BH scaling relations, due to both data limitations and poorly understood systematics.\\
\indent Since we are interested in using $\sigma_{\rm{core}}$ as a surrogate for $\sigma_*$ on the $M_{\rm{BH}}-\sigma_*$ relation, we plot $M_{\rm{BH}}-\sigma_{\rm{core}}$ on the right in Figure \ref{fig:m_sigma_relation}.
We find that the scatter of the $M_{\rm{BH}}-\sigma_*$ relation is 0.54 dex, slightly less than that of the $M_{\rm{BH}}-\sigma_{\rm{core}}$ relation with a scatter of 0.57 dex.  However, the mean of $M_{\rm{BH}}-\sigma_*$ relation is 0.57 dex above the local relation, driven by clear outliers between 1-2 dex above the local relation.  The mean of the $M_{\rm{BH}}-\sigma_{\rm{core}}$ is 0.17 and more-evenly distributed about the local relation.\\
\indent It is clear from Figure \ref{fig:m_sigma_relation} that the scatter in $M_{\rm{BH}}-\sigma_{\rm{core}}$ is due primarily to stratification in $\sigma_{\rm{core}}$, with $\sigma_{\rm{core}}<600\text{ km s}^{-1}$ primarily above the relation, and $\sigma_{\rm{core}}>900\text{ km s}^{-1}$ below the relation.  It is possible that this separation in dispersion across the local relation could be attributed to the core broadening as a function of $v_{\rm{max}}$ we see in Figure \ref{fig:delta_core_stellar_vs_outflow}.  It is also worthy to note that this stratification in $\sigma_*$ is not as obvious in the $M_{\rm{BH}}-\sigma_*$ relation, although there is similar trend for  $\sigma_{\rm{*}}<600\text{ km s}^{-1}$.

\subsection{Discussion}

In the following sections we discuss the different correlations and their possible interpretation.  We emphasize that although we can only speculate on the physical interpretation of these correlations, these observations represent observational constraints that should be considered when developing models that describe AGN-driven outflows.

\subsubsection{Correlations with Velocity}

The differences in the VVD diagrams shown in Figure \ref{fig:vvd_diagram} indicate that the core and outflow components of the NLR are kinematically distinct.  According 3D biconical outflow modelling from \citet{Bae2016}, the fan-shaped distribution of the $\sigma_{\rm{outflow}}-v_{\rm{outflow}}$ relation is due to a number of factors, the most important of which are bicone inclination, ejection velocity, and dust extinction along the line of sight.  For a symmetric biconical outflow in the absence of any dust extinction, we would expect to measure zero velocity offset along line of sight due to the cancelling of velocities in opposite directions.  With the addition of extinction effects, the obscuration of one side of the bicone would lead to a shift in observed velocity offset.  The large number of blueshifted outflows in our sample can be explained as varying obscuration of the receding (redshifted) cone.
One interpretation of the strong correlation in the outflow VVD diagram is evidence of collimation, that is, we expect to see an increase in $\sigma_{\rm{outflow}}$ with an increase in $v_{\rm{outlfow}}$ along the line of sight if the flow  subtends relatively small solid angle (such as a cone) and has a preferred inclination.  For example, we expect to see larger velocities as well as a larger velocity dispersion for a flow that is directed along the LOS, as opposed to a flow directed at some angle with respect to the LOS, which would produce a smaller observed velocity and thus smaller dispersion.  This interpretation is consistent with the model grids for 3D biconical outflow models from \citet{Bae2016}.\\
\indent We do not observe the same strong correlation for the $\sigma_{\rm{core}}-v_{\rm{core}}$ VVD relation. There is an overall blueshift of $v_{\rm{core}}$ which correlates with $v_{\rm{outflow}}$, as shown in Figure \ref{fig:vel_line_lock}, however there is a larger spread in $v_{\rm{core}}$ for a given value of $\sigma_{\rm{core}}$.  Given our above argument for the outflow component, the lack of correlation for the core component would imply that the source of the core gas emission is less collimated and more spherically symmetric.  This would agree with the interpretation that the core component represents the original NLR gas that is still strongly coupled to the gravitational potential.  The core VVD diagram clearly does not exhibit the same kinematic properties of outflows, and  should be treated as separate kinematic component when trying to model outflows in AGN. \\
\indent The linear increase between $v_{\rm{core}}$ and $v_{\rm{outflow}}$ and constant 120 km s$^{-1}$ offset in velocity shown in Figure \ref{fig:vel_line_lock} could indicate that the two components are locked in velocity, at least up to a certain value of $v_{\rm{max}}$.  We can only speculate the physical interpretation of this trend, but it could mean that the core NLR gas can be coupled to the outflowing gas below a certain velocity threshold, causing it to become entrained and expand with the outflow.  At the highest velocities, the core gas may decouple from the outflowing gas, causing this trend to plateau as shown in Figure \ref{fig:vel_line_lock}.  We would require a larger sample of objects with outflows with $v_{\rm{max}}>900\text{ km s}^{-1}$ to determine if this occurs.

\subsubsection{Correlations with Dispersion} 

\begin{figure*}
 \includegraphics[width=\textwidth]{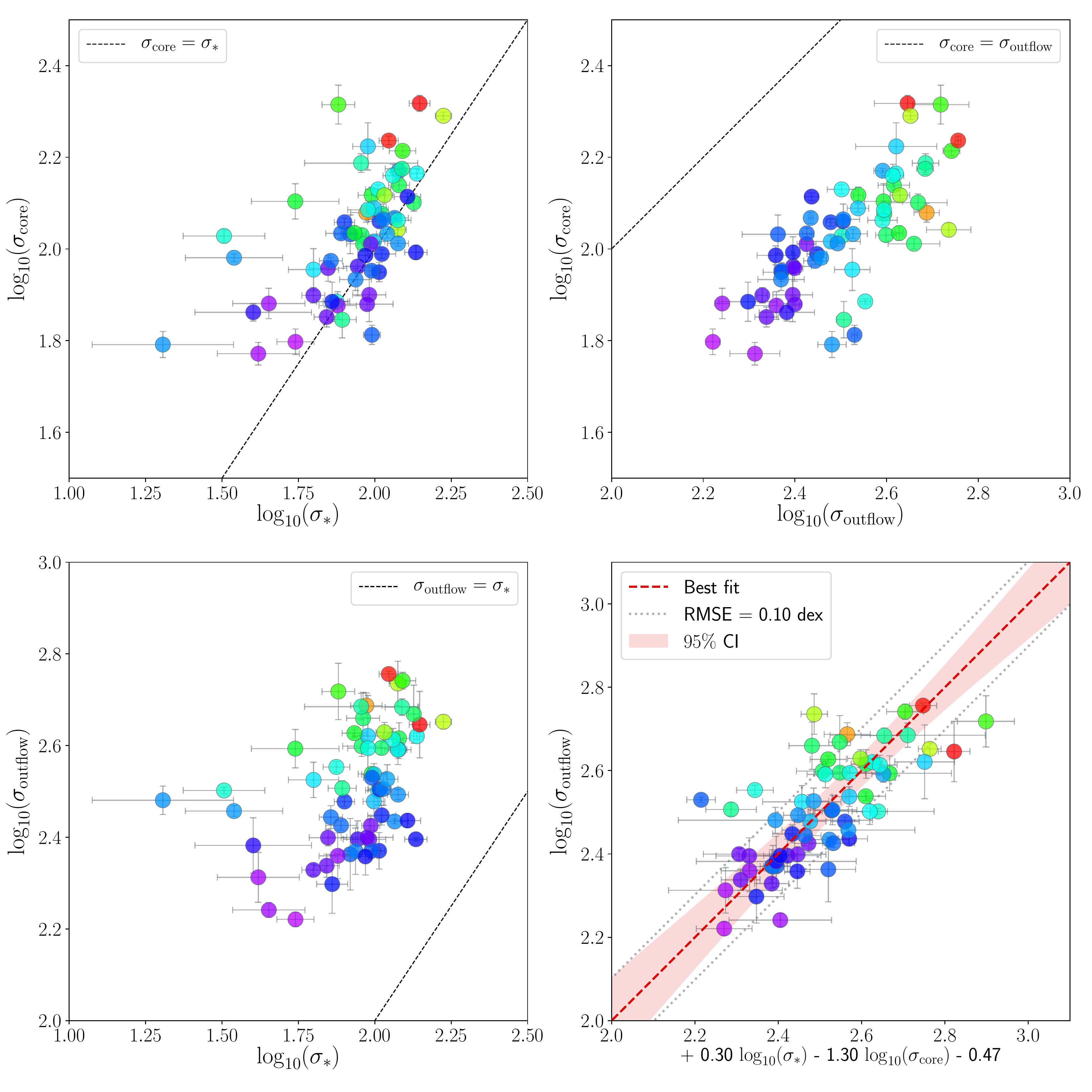}
 \caption{
The three projections of the $\sigma_{\rm{outflow}}-\sigma_{\rm{core}}-\sigma_*$ relation, and the best fit relation projected parallel to the best fit plane.  The identity correlations are given by the black dashed line in each plot.  The scatter about the best fit plane relation is $f=0.10$, which is considerably smaller than the scatter in the $\sigma_{\rm{core}}-\sigma_*$ ($f=0.17$ dex) and $\sigma_{\rm{outflow}}-\sigma_*$ ($f=0.19$ dex) relations, and comparable to the $\sigma_{\rm{core}}-\sigma_{\rm{outflow}}$ relation ($f=0.09$ dex).
}
 \label{fig:dispersion_plane_with_projections}
\end{figure*}

In an idealized gravitationally bound system, such as in an undisturbed NLR, we should expect the gas and stellar components to have similar velocity distributions.  If a secondary component in the same region as the source of NLR emission is present, and exhibits some collimation (increase in velocity and velocity dispersion), we would expect a distribution similar to that shown on the left of Figure \ref{fig:core_vs_stellar}.  We do recover the core NLR gas within some scatter about the perfect correlation with $\sigma_*$ after correction, as shown on the right in Figure \ref{fig:core_vs_stellar}.  Objects with $v_{\rm{max}}<600\text{ km s}^{-1}$ are more evenly distributed about the perfect correlation with $\sigma_*$ after correcting for $\sigma_{\rm{outflow}}$.  The majority of objects with $v_{\rm{max}}>600\text{ km s}^{-1}$ tend to fall above with the relation even after correcting for $\sigma_{\rm{outflow}}$, which may indicate that the presence of outflows may introduce additional non-gravitational broadening which may only be detected for the strongest cases. Some scatter is expected, as we cannot fully account for all non-gravitational interactions nor fully account for systematics involving the measurements of $\sigma_*$ given the nature of SDSS data, such as inclination, aperture effects, or merger history.  This correlation is enough to suggest that the core component of the [\ion{O}{iii}] profile traces the original NLR gas that is dominated by the gravitational potential of the stellar component.\\
\indent It is not unreasonable to suggest that if a secondary outflowing component arises from within the NLR, it must start out with the same velocity distribution as the core components.  If that velocity distribution then undergoes some interaction, we expect the original distribution to broaden.  We can interpret the $\sigma_{\rm{core}}-\sigma_{\rm{outflow}}$ correlation shown in Figure \ref{fig:core_vs_outflow} to be the broadening of the original NLR core gas due to the outflowing gas.  What is still puzzling is the linear rate at which the outflow dispersion grows with the core dispersion and its small scatter.  We can interpret this as the outflow component having a strong dependence on the original NLR gas from which it is believe to have originated, and the strong linear dependence describes the manner by which the flow propagates through the ambient medium.  Another possible interpretation is that the strength of the outflow component (parameterized by $v_{\rm{max}}$) causes a broadening of the core component, such that $\sigma_{\rm{core}}$ approaches its respective value of $\sigma_*$.  There is some correlation shown in Figure \ref{fig:delta_core_stellar_vs_outflow} that suggests that the core component broadens as a function of $\sigma_{\rm{outflow}}$ (and therefore $v_{\rm{max}}$), however a larger sample of objects with $v_{\rm{max}}>900$ is needed to determine if this trend is real or simply increased scatter.\\
\indent The $\sigma_{\rm{core}}-\sigma_*$ and $\sigma_{\rm{core}}-\sigma_{\rm{outflow}}$ imply that there is some connection between $\sigma_*$ and $\sigma_{\rm{outflow}}$.  Ideally, if there is a constant linear relationship between  $\sigma_{\rm{core}}$ and  $\sigma_{\rm{outflow}}$, and if  $\sigma_{\rm{core}}$ traces $\sigma_*$, then $\sigma_{\rm{outflow}}$ should also scale with $\sigma_*$ but positively offset by some constant.  We can fit the interdependence of the three dispersions as a plane of the form 
\begin{equation} \label{eq:plane}
    a\log_{10}(\sigma_*)+b\log_{10}(\sigma_{\rm{core}})+c\log_{10}(\sigma_{\rm{outflow}})+d=0
\end{equation}
% a = -2.388, -0.510, +0.539
% b = 10.255, -1.191, +0.997
% c = -7.906, -0.961, +1.119
% d = 3.631, -1.602, +1.460
We perform orthogonal regression using \emph{emcee} following the methods of S19 and obtain best fit coefficients of $a=-2.39_{-0.51}^{+0.54}$, $b=10.26_{-1.19}^{+1.00}$, and $c=-7.91_{-0.96}^{+1.12}$, $d=3.63_{-1.60}^{+1.46}$, and a scatter about the best fit plane of $f=0.10$ dex.  We plot the projections of the three dispersions, and the projection along the parallel axis of the plane in Figure \ref{fig:dispersion_plane_with_projections}.  Despite the decreased scatter, there is still a large uncertainty in $a$, i.e., the slope of the $\sigma_{\rm{outflow}}-\sigma_*$ relation, which is caused by large scatter.  Further study with a larger sample is needed to better constrain this slope before the functional form of \ref{eq:plane} can be used to calculate $\sigma_*$ using both $\sigma_{\rm{core}}$ and $\sigma_{\rm{outflow}}$.\\
\indent The physical interpretation of the plane relationship between the three dispersions does not necessarily imply that $\sigma_{\rm{outflow}}$ can somehow influence $\sigma_*$ or vice versa, neither does it answer the proverbial ``chicken or egg'' problem, that is, we do not know if outflows are the causal explanation for the broadening of $\sigma_{\rm{core}}$ or if $\sigma_{\rm{outflow}}$ correlates with $\sigma_{\rm{core}}$ because it originated from an already-broad gas velocity distribution.  A larger sample, along with integral field spectroscopy, to determine if these relationships hold true.

\subsubsection{Correlations with  Luminosity}

It has been known for some time that the incidence of [\ion{O}{iii}] outflows correlates with radio emission in both type 1 and type 2 AGN \citep{Wilson1980,Whittle1985,Whittle1988,Nelson1996}.  More recent studies suggest that the strongest correlation with luminosity is between the [\ion{O}{iii}] width and the radio luminosity at 1.4 GHz ($L_{\rm{1.4\;GHz}}$), especially in objects with high-velocity outflows and at much higher redshifts. \citep{Mullaney2013,Zakamska2014,Zakamska2016,Hwang2018,Perrotta2019}. \\
\indent As mentioned in Section \ref{sec:lum_correlations}, the strong correlation between the core and outflow components, $L_{\rm{core}}$ and $L_{\rm{outflow}}$, and the optical AGN luminosity $L_{\text{5100\angstrom}}$ is not surprising if the core and outflow components originate in close proximity to the ionizing source.  There is still some correlation with core and outflow dispersion, but even lesser so for the core and outflow velocities.  We see similar lack of correlation when we compare $\sigma_{\rm{core}}$ and $v_{\rm{core}}$ to $L_{\rm{1.4\;GHz}}$.  By far, the strongest correlation we find between any measured luminosities and kinematics is with $L_{\rm{1.4\;GHz}}$ and $v_{\rm{outflow}}$.\\
\indent Previous studies by \citet{Woo2016} and \citet{Rakshit2018} used a total (core+outflow) integrated [\ion{O}{iii}] velocity dispersion parameterization and normalized it by the stellar velocity dispersion to quantify non-gravitational kinematics to compare to radio luminosity, finding no strong correlations with radio activity.  In this study, the outflow component is designated as the only non-gravitational component, for which we do find strong correlation with radio luminosity, although with a much smaller sample size.  Our findings agree with \citet{Mullaney2013}, who similarly found strong correlation between objects with high $L_{\rm{1.4\;GHz}}$ and objects with the broadest [\ion{O}{iii}] profiles.\\

\subsubsection{The $M_{\rm{BH}}-\sigma_*$ Relation}

Figure \ref{fig:m_sigma_relation} shows that when corrected for the outflow component, $\sigma_{\rm{core}}$ can be used as a surrogate for $\sigma_*$ on the $M_{\rm{BH}}-\sigma_*$ relation with comparable scatter, and agree with the results found by \linebreak\citet{Bennert2018}.  However, if we are to use $\sigma_{\rm{core}}$ for studies on the non-local $M_{\rm{BH}}-\sigma_*$, we do not have the luxury of comparing it $\sigma_*$ to ensure we have evidence of non-gravitational kinematics as we have done in our sample.  Performing a double-Gaussian decomposition of the [\ion{O}{iii}] profile when there is no evidence of an additional non-gravitational component, while always producing a better fit, can cause one to measure a smaller $\sigma_{\rm{core}}$ than what $\sigma_*$ suggests, which can give the impression that one is measuring BHs that are overmassive relative to the local $M_{\rm{BH}}-\sigma_*$ relation.\\
\indent We advise that if $\sigma_{\rm{core}}$ is used as a surrogate for $\sigma_*$, that one always fit a double-Gaussian component and check that the object falls within the acceptable scatter of the $\sigma_{\rm{core}}-\sigma_{\rm{outflow}}$ relation.  Furthermore, for $\sigma_{\rm{outflow}}<200\text{ km s}^{-1}$, the scatter of the $\sigma_{\rm{core}}-\sigma_{\rm{outflow}}$ relation begins to intersect with that of the $\sigma_{\rm{outflow}}-\sigma_*$ relation, and it becomes increasingly unclear if there are additional non-gravitational kinematics present in the [\ion{O}{iii}] profile with respect to $\sigma_*$. 
Therefore, we recommend that for $\sigma_{\rm{outflow}}<200\text{ km s}^{-1}$ one does \emph{not} use a double-Gaussian decomposition for the risk of severely overfitting the [\ion{O}{iii}] profile and significantly underestimating $\sigma_*$. Likewise, if a single-Gaussian fit to the [\ion{O}{iii}] profile exceeds $\sim200$ km s$^{-1}$, it is recommended to perform a double-Gaussian decomposition and assess the quality of the fit.  In this regard, the outflow confidence calculated by BADASS by performing an $F$-statistic model comparison makes it clear when a double-Gaussian fit is warranted by the data.

\section{Conclusion}

To summarize, we have presented BADASS, a new, thoroughly-tested, and powerful fitting software for optical SDSS spectra that is open source and specialized for fitting AGN spectra.  Since BADASS can fit numerous components simultaneously, it can be generalized to fit not just AGN spectra, but non-AGN host galaxies as well.  The use of MCMC allows the user to fit objects with unprecedented detail, obtain robust uncertainties, and determine the quality of fits using a broad range of metrics and outputs.  BADASS also utilizes multiprocessing to efficiently fit large samples of objects without excessive memory overhead. \\
\indent Currently, BADASS is being used for a variety of research projects and number of collaborations.  For instance, BADASS is being run in a cluster environment to fit over 19,000 SDSS galaxies to determine the significance of outflows as a function of separation distance and as a function of environment.  BADASS will also be used to fit a larger sample of type 1 AGN to follow up on results presented here.\\
\indent Performance tests with BADASS we have presented here in the recovery of $\sigma_*$ can also be applied to other fitting routines which attempt to measure the LOSVD.  We summarize the results of these tests below:
\begin{enumerate}
    \item In non-AGN host galaxies and Type 2 AGN, where significant \ion{Fe}{ii} and AGN continuum dilution is absent, the LOSVD can be recovered in the \ion{Mg}{ib}/\ion{Fe}{ii} region (4400 \angstrom\;-5800 \angstrom) with less than 10\% error and uncertainty for $\text{S/N}>20$.  For objects which exhibit active star formation, the steep continuum from young stellar populations complicates measurements of $\sigma_*$, and we recommend to include more O- and B-type template stars and disable the AGN power-law component from the fit.
    \item Measurements of $\sigma_*$ are not significantly affected by the inclusion of \ion{Fe}{ii}, and are more affected by S/N level. 
    \item Dilution of stellar absorption features by strong AGN continuum contributes to the largest error and uncertainty in measuring $\sigma_*$.  We find that while that strong \ion{Fe}{ii} and steep AGN power-law slope can be indicative of strong continuum dilution, they are not the root cause.  Continuum dilution caused to a large fraction of continuum flux being dominated by the AGN is the root cause of large uncertainties in the estimation of $\sigma_*$, and extra caution should be given in the estimate of the LOSVD to objects which exhibit strong \ion{Fe}{ii} or steep power-law slope, such as NLS1 or BAL objects. 
\end{enumerate}
\indent As an application of BADASS, we fit a sample of 63 SDSS type 1 AGN with strong evidence of outflows in the [\ion{O}{iii}]$\lambda 5007$ emission line and performed a correlation analysis of kinematics to determine the relationships between outflows, the AGN, and the host galaxy, expanding upon previous similar studies.  We summarize our most important results below:
\begin{enumerate}
    \item By performing a double-Gaussian decomposition of the [\ion{O}{iii}]$\lambda5007$ emission line profile into separate core and outflow components, we find that the core dispersion of the [\ion{O}{iii}] profile ($\sigma_{\rm{core}}$) is a suitable surrogate for stellar velocity dispersion ($\sigma_*$) in a statistical context but should not be used on an object-to-object basis.  There is some evidence that the measured difference $\sigma_{\rm{core}}-\sigma_*$ scales with increasing outflow component dispersion ($\sigma_{\rm{outflow}}$), which may imply that there is some broadening of the NLR gas due to the presence of outflows, causing the scatter we see in the $\sigma_{\rm{core}}-\sigma_*$ relation. 
    \item Velocity-Velocity Dispersion (VVD) digrams of the outflow component resemble the ``fan-shaped'' VVD profiles exhibited by 3D biconical outflow models from \citet{Bae2016}, indicating possible orientation-dependent or collimated flow.  The core component does not exhibit the same VVD shape as the outflow component, indicating that it is a kinematically distinct component of the [\ion{O}{iii}] gas, more strongly coupled to the gravitational potential.
    \item There is a systematic broadening of the $\sigma_{\rm{core}}$ component which scales with $\sigma_{\rm{outflow}}$, resulting a tight correlation between $\sigma_{\rm{core}}$ and $\sigma_{\rm{outflow}}$.  This tight correlation implies a very specific relationship between outflow kinematics and the kinematics of the narrow line region, which could be used to constrain theoretical models of AGN outflows.
     \item We present a new planar relationship between $\sigma_*$, $\sigma_{\rm{core}}$ and $\sigma_{\rm{outflow}}$ with a scatter about the best-fit plane of 0.10 dex.  However, a larger sample is still needed to constrain the relationship between $\sigma_{\rm{outflow}}$ and $\sigma_*$ before it can be used to obtain values for $\sigma_*$. 
    \item We recover the strong correlation between $L_{\rm{1.4\;GHz}}$ and properties of outflows found in previous studies.  We do not observe strong correlations between outflow kinematics and the optical AGN luminosity $L_{\text{5100\angstrom}}$. 
    \item We find that $\sigma_{\rm{core}}$ is a suitable surrogate for $\sigma_*$ on the $M_{\rm{BH}}-\sigma_*$ relation with comparable scatter in a statistical context in agreement with \citet{Bennert2018}.  Additionally, we present recommendations and caveats for using $\sigma_{\rm{core}}$ for studies of the $M_{\rm{BH}}-\sigma_*$ relation in the non-local universe for which $\sigma_*$ cannot be measured.\\ 
\end{enumerate}

The correlations we have presented here showcase a number of observational constraints that theoretical models of AGN outflows should satisfy.  Further investigation into these correlations and their causes will be necessary with larger samples, and we have shown here that BADASS is capable of such detailed analyses.  As newer and larger surveys begin to come online, tools such as BADASS, which underscore the need for a generalized open-source framework for fitting a variety of objects with advanced statistical techniques, will be needed for increasingly-detailed analysis of astronomical spectra in the coming decade.\\
\indent The BADASS source code for both Python 2.7 and 3.6, as well as their documentation can be found at \url{https://github.com/remingtonsexton}.

\section*{Acknowledgments}
\indent We thank the anonymous referee for their careful reading of our manuscript, as well as their helpful suggestions in improving the manuscript and code.\\
\indent ROS acknowledges financial support from the NASA MIRO program through the Fellowships and Internships for Extremely Large Data Sets (FIELDS) in the form of a Graduate Student Fellowship.  ROS  personally thanks Dr. Bahram Mobasher for his generous support as a FIELDS graduate student fellow.  \\
\indent Partial support for this project was provided by the National Science Foundation, under grant No. AST 1817233.\\
\indent Funding for SDSS-III has been provided by the Alfred P. Sloan Foundation, the Participating Institutions, the National Science Foundation, and the U.S. Department of Energy Office of Science. The SDSS-III web site is \url{http://www.sdss3.org/}.\\
\indent SDSS-III is managed by the Astrophysical Research Consortium for the Participating Institutions of the SDSS-III Collaboration including the University of Arizona, the Brazilian Participation Group, Brookhaven National Laboratory, Carnegie Mellon University, University of Florida, the French Participation Group, the German Participation Group, Harvard University, the Instituto de Astrofisica de Canarias, the Michigan State/Notre Dame/JINA Participation Group, Johns Hopkins University, Lawrence Berkeley National Laboratory, Max Planck Institute for Astrophysics, Max Planck Institute for Extraterrestrial Physics, New Mexico State University, New York University, Ohio State University, Pennsylvania State University, University of Portsmouth, Princeton University, the Spanish Participation Group, University of Tokyo, University of Utah, Vanderbilt University, University of Virginia, University of Washington, and Yale University.

%%%%%%%%%%%%%%%% DATA AVAILABILITY STATEMENT%%%%%%

\section*{Data Availability Statement}

The BADASS code referenced throughout this manuscript is available at \url{https://github.com/remingtonsexton}.  The spectral data used in the preparation of this manuscript is publicly available at \url{http://www.sdss3.org/}.



\bibliographystyle{mnras}
\bibliography{mybib} % if your bibtex file is called example.bib

%%%%%%%%%%%%%%%%%%%%%%%%%%%%%%%%%%%%%%%%%%%%%%%%%%

%%%%%%%%%%%%%%%%% APPENDICES %%%%%%%%%%%%%%%%%%%%%

% \appendix 

% \section{Examples}

%%%%%%%%%%%%%%%%%%%%%%%%%%%%%%%%%%%%%%%%%%%%%%%%%%

% Don't change these lines
\bsp	% typesetting comment
\label{lastpage}

\end{document}